\documentclass{appolb}
\usepackage{graphicx}% graphicx package included for placing figures in the text
\usepackage{psfig}
\usepackage{lscape}
\usepackage{amssymb}
\usepackage{amsfonts}
\usepackage{mathrsfs}
\renewcommand{\thetable}{\Roman{table}}
\begin{document}
\title{
Reggeometry of deeply virtual Compton scattering (DVCS)\\%
and exclusive vector meson production (VMP) at HERA%
%\thanks{Presented at ...}%
}
\author{S.~Fazio\\[-0.4cm]
\address{Brookhaven National Laboratory, 11973 Upton NY, U.S.A.}
\\[0.4cm]
 R.~Fiore, A.~Lavorini\\[-0.3cm]
 \address{Dipartimento di Fisica, Universit\`a della Calabria, %\\%
 Istituto Nazionale di Fisica Nucleare, Gruppo collegato di Cosenza, %\\%
 I-87036 Arcavacata di Rende, Cosenza, Italy}
 \\[0.4cm]
 L.~Jenkovszky\footnote{jenk@bitp.kiev.ua} and A. Salii\footnote{saliy.andriy@gmail.com} \\[-0.3cm]
 \address{Bogolyubov Institute for Theoretical Physics,\\ National Academy of Sciences of Ukraine, %\\%
 UA-03680 Kiev, Ukraine}
}
\maketitle
\begin{abstract}
A Reggeometric (Regge+Geometry) model, based on the observed proportionality between the forward slope of the differential cross section and the interaction radius, the latter depending on virtuality $Q^2$ of the incoming virtual photon and on the mass $M^2$ of the produced particle, is constructed. 
The objective of this study is the dependence of the Regge-pole amplitude on the virtuality $Q^2$ and masses of the external particles, which remains an open problem for the theory.
The present analysis is based on the HERA data on Deeply Virtual Compton Scattering ({DVCS}) and exclusive diffractive Vector Meson Production ({VMP}). We treat each class of reactions separately, anticipating a further study \cite{FFJS} that will include both a soft and a hard component of the unique Pomeron.
\end{abstract}
  
\section{Introduction}

\begin{figure}[ht]
  \begin{center}
  \includegraphics[clip,scale=0.6]{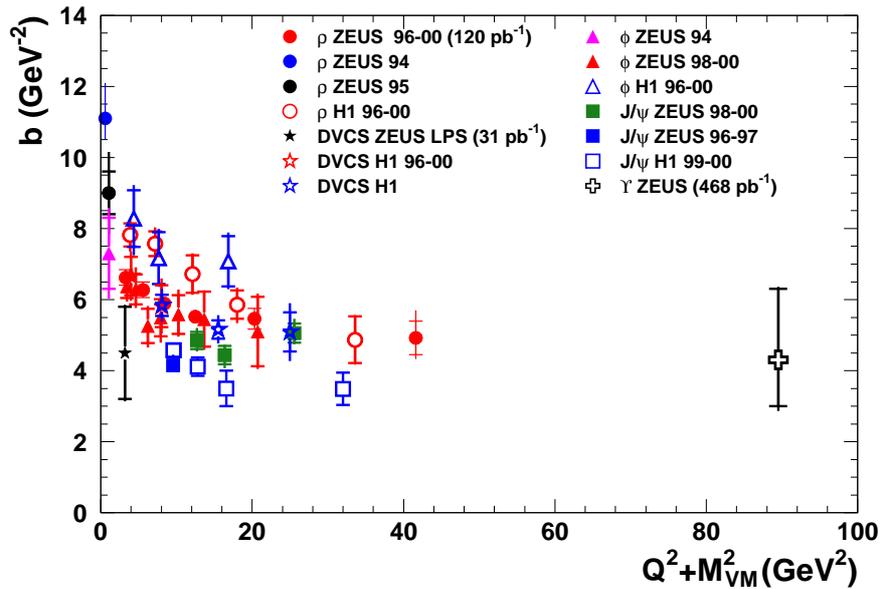}
  \end{center}
 %  \vspace{-1.3cm}
   \caption{The forward slope of differential cross sections as function of $\widetilde Q^2=Q^2+M_V^2$. Compilation of the data for DVCS and VMP measured by the ZEUS and H1 Collaborations see~\cite{Nikolaev}.}
 \label{fig:slope}
\end{figure}

\begin{figure}[ht]
\begin{center}
\includegraphics[clip,scale=0.63]{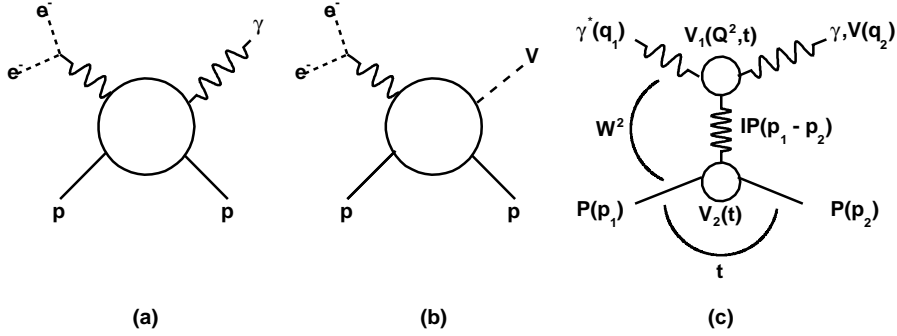}
\end{center}
\vspace{-0.5cm}
\caption{(a) Diagram of DVCS and (b) diagram of VMP
 at HERA; (c) DVCS (VMP) amplitudes in a Regge-factorized
form.} \label{fig:fig1}
\end{figure}

 The forward slope of the differential cross sections for elastic and quasi-elastic reactions, e.g. Deeply Virtual Compton Scattering (DVCS) or
Vector Meson production (VMP),  is known to be related to the masses/virtualities of the interacting particles.
 This phenomenon is clearly seen on Fig.~\ref{fig:slope}, where the forward slope $B(\widetilde Q^2)$
 is plotted against the variable $\widetilde Q^2=Q^2+M_V^2$. Here $Q^2$ and $M_V^2$ are respectively the square of the virtuality and of the mass of the produced particle and the notation is evident from Fig.~\ref{fig:fig1}.

%\begin{figure}[hb]
 % \vspace{-0.2cm}
  %\begin{center}\includegraphics[clip,trim = 0mm 0mm 10mm 14mm,scale=0.6]{diagrams}\end{center}
  %\vspace{-1.0cm}
  %\caption{ \label{fig:diagram1} Diagrams of DVCS (a) and VMP (b); (c) DVCS (VMP) amplitude in a Regge-factorized form.}
%\end{figure}

%More precisely, 
The slope, proportional to the ``interaction radius'' $R(\widetilde Q^2)$, decreases with
 increasing $\widetilde Q^2$, reaching some saturation value determined by the
 finite mass of the nucleons in the lower vertex of Fig. \ref{fig:fig1}(c). Thus, in this ``geometrical" picture,
 the largest slope (radius) is expected for a real Compton scattering, where $\widetilde Q^2=0$.
 %, that may require a separate treatment.

 In the present paper we consider exclusive electroproduction
 of real photons (DVCS) and vector mesons (VMP)  making use of the above geometrical considerations, by writing the scattering amplitude in the form 
 \begin{equation}
 A(s,t)\sim e^{\tilde b(\widetilde Q^2)t},
 \label{e1}
 \end{equation}
 with $\tilde b(\widetilde Q^2)\sim 1/f(\widetilde Q^2).$ This approach was used in Ref.~\cite{Francesco} for the simpler case of photoproduction ($Q^2=0$).

The off-mass shell continuation of the $S$-matrix, or, in
particular, the introduction of any dependence of the Regge-pole
amplitude on the virtuality, $Q^2$ and masses of external particles
remains an open problem for the theory. Empirically, it
was approached in various models, of which the best known is the
introduction of $Q^2$-dependent parameters in the Regge (e.g. the
Pomeron) trajectories, namely, the intercept $\alpha_0\rightarrow\alpha_0(Q^2)$
and the slope $\alpha'\rightarrow\alpha'(Q^2)$. Since, by definition (i.e.
by Regge-factorization), the trajectory should not depend on the
properties (masses and virtualities) of the external particles,
such Regge poles and trajectories are called ``effective" ones.
It is the relevant vertex function that
bears information about the  masses and virtualities of the
particles coupled to them. In Ref.~\cite{Capua} a simple model for
DVCS and VMP, based on the use of a new variable defined as
$z=t-Q^2$ was put forward.

In the view of the limitations of that model as discussed in \cite{PR}, below we scrutinize an alternative
approach \cite{EDS, Canarias} by combining the Regge pole model with geometrical ideas,
resulting in Reggeometry=(Regge+geometry).
In the nearly forward direction, where the
cross sections decrease almost exponentially, $d\sigma/dt\sim e^{Bt}$,
the slope $B$ is related to the interaction radius, which is a
function of the inverse mass (virtuality) of the particles,
\begin{equation}
\label{geometry}
B=R^2\sim 1/f(\widetilde Q^2),
\end{equation}
%where $\widetilde Q^2=M_V^2+Q^2,$ (or $\widetilde Q^2=(M_V^2+Q^2)/4$),
reflecting the
geometrical nature of the slope, proportional to the ``interaction
radius", or to the inverse squared masses. More precisely,
$B=B_1+B_2=R^2_1+R^2_2,$ where the two radii correspond to the
lower and upper vertices of Fig. \ref{fig:fig1}. This dependence Eq.~(\ref{geometry}) will be given explicitly in Sec.~\ref{sec:Model}.

There are however several caveats in this simple and appealing
interpretation of the slope, namely: \begin{itemize}
\item The relation $\widetilde Q^2=M_V^2+Q^2,$ is model-dependent; an
alternative relation, e.g. $\widetilde Q^2=M_V^2+cQ^2,\ \ c$ being a 
parameter to be determined by fits to experimental data, is equally legitimate. Another degree of freedom enters
putting the factor $4$ as denominator in the expression of  $\widetilde Q^2$, i.e. writing
$\widetilde Q^2=(M_V^2+cQ^2)/4,$ (see, Ref.~\cite{Nikolaev}).
\item Empirically, the slope can be given by the relation $B\sim(\widetilde Q^2)^{-n}$. It includes an exponent $n$, with $n>1$,
whose value is not necessarily an integer and may even depend on
$\widetilde Q^2$ (see, for instance,  see Fig. 10.26 and Eqs.~(10.72) - (10.74) of Ref.~\cite{Nikolaev}.
\item Relation (\ref{geometry})
%and its modifications were
was established only within certain classes of
reactions, notably VMP, $\gamma (\gamma^*) p\rightarrow V
p.$ It remains an open question if it is also applicable  to 
DVCS, and, moreover, to purely hadronic reactions.
%Another validity
%test could be the possible symmetry between the upper lines of
%Fig. \ref{fig:diagram1}, i.e. DVCS (production of a real photon
%by a virtual one, with $Q^2=M_V^2$, on the one hand, and, on the
%other hand, photoproduction of a vector meson with mass $M_V.$
\end{itemize}

Studies of all these options involve also $s$ and $t$
dependences of the relevant scattering amplitude. We use
simple models of $s-$ and $t-$dependences to concentrate on the less trivial $\tilde Q^2$ dependence.
Our model is based on two pillars: the ``Reggeometric" form of the amplitude, introduced in the Introduction, and its two-component nature. The latter implies that the unique Pomeron has two components - a ``soft" and a ``hard" one, and that their relative weight varies with $\widetilde Q^2$, thus providing a unify description of both soft and hard collisions.
The model was presented at the 2011 EDS conference in Quy Nhon \cite{EDS} and at the Diffraction 2013 conference at the Canarias \cite{Canarias}. Anticipating a global fit (with a universal set of the free parameters), to appear \cite{FFJS}, here we make the first important step by fitting the HERA data on several particular processes, one by one, to a single Reggeometric term.

This one-component effective model is fitted, to the HERA data on DVCS and VMP. The complete, two-component ("soft" and "hard") model is anticipated in Appendix~\ref{app:Hard and soft}.
%....................................................................

The paper is organized as follows. In Sec.~\ref{sec:Model} Reggeometry is introduced.
In Sec.~\ref{sec:Results} we consider a simplified  version of the model involving a single term only and fitted to DVCS and a number of VMP data. The resulting fits are shown in Sec.~\ref{sec:Conclusions}.

\section{The Reggeometric model} \label{sec:Model}
According to the arguments presented in the Introduction and in \cite{EDS, Canarias}, the DVCS or VMP  amplitudes of the Reggeometric model with a single Pomeron term 
can be written as

\begin{equation}\label{Amplitude0}
A(s,t,M,Q^2)=\frac{\widetilde{A_0}}{\Bigl(1+\frac{\widetilde{Q^2}}{{Q_0^2}}\Bigr)^{n}}\xi(t)\beta(t,M,Q^2)(s/s_0)^{\alpha(t)},
\end{equation}
where $A_0$ is a normalization factor,  $Q_0^2$ is a scale for the virtuality, $n$ is a free parameter with $n>1$, $\xi(t)=e^{-i\pi\alpha(t)}$ is the signature factor, $s_0$ is a scale for the squared energy and $\alpha(t)$ is the Pomeron trajectory.
$\beta(t,M,Q^2)$ is the residue factor to be specified as
\begin{equation}
\label{residue0}
\beta(t,M,Q^2)=\exp\Bigl[-2\Bigl(\frac{a}{\widetilde Q^2}+\frac{b}{2m_N^2}\Bigr)|t|\Bigr],
\end{equation}
where $a$ and $b$ are free parameters and $m_N$ is the nucleon mass.
%Substituting Eq.~(\ref{residue0}) of the residue factor to Eq.~(\ref{Amplitude0}), the amplitude reads
Thus:
\begin{equation}
\label{Amplitude1}
A(s,t,M,Q^2)=\frac{\widetilde{A_0}}{\Bigl(1+\frac{\widetilde{Q^2}}{{Q_0^2}}\Bigr)^{n_s}}\xi(t)(s/s_0)^{\alpha(t)} e^{-2\Bigl(\frac{a}{\widetilde Q^2}+\frac{b}{2m_N^2}\Bigr)|t|}.
\end{equation}
The differential cross section is
\begin{equation}
\frac{d\sigma}{d|t|}=\frac{\pi}{s^2}|A(s,t,M,Q^2)|^2.
\label{eq3}
\end{equation}
%so that the forward slope $B$ is equal to $2\tilde b$ due to Eqs.~(\ref{e1}) and (\ref{3}).
Using Eq.~(\ref{Amplitude1})
we get
\begin{equation}
\label{eq:dcsdt}
\frac{d\sigma}{d|t|}=\frac{|A_0|^2}{\Bigl(1+\frac{\widetilde{Q^2}}{{Q_0^2}}\Bigr)^{2n}}(s/s_0)^{2(\alpha(t)-1)}e^{-4\Bigl(\frac{a}{\widetilde Q^2}+\frac{b}{2m_N^2}\Bigr)|t|},
\end{equation}
where $A_0=\frac{\sqrt{\pi}}{s_0}\widetilde{A_0}$.

Assuming a linear Regge trajectory for the Pomeron 
$\alpha(t) = {\alpha}_0 - {\alpha'}|t|,$
the differential cross section takes the form
\begin{equation}
\frac{d\sigma}{d|t|}=\frac{|A_0|^2}{\Bigl(1+\frac{\widetilde{Q^2}}{{Q_0^2}}\Bigr)^{2n}}(s/s_0)^{2(\alpha_0-1)}e^{-\Bigl[2\alpha'\ln(s/s_0)+4\Bigl(\frac{a}{\widetilde Q^2}+\frac{b}{2m_N^2}\Bigr)\Bigr]|t|}=Ce^{-B|t|}.
\end{equation}
where the function $C$ is independent of $t$.
From this expression we obtain the forward slope:
\begin{equation}
\label{eq:Bslope}
B(s,\widetilde Q^2)=2\alpha'\ln(s/s_0)+4\Bigl(\frac{a}{\widetilde Q^2}+\frac{b}{2m_N^2}\Bigr).
\end{equation}
Then the elastic cross section is obtained by integrating the differential cross section. The result, in the case of linear Regge trajectory, is
\begin{equation}
\label{integrated}
\sigma=\frac{1}{B}\frac{d\sigma}{dt}\Bigr|_{t=0},
\end{equation}
and due to Eqs.~(\ref{eq:dcsdt}) and (\ref{eq:Bslope}) it reads
\begin{equation}
\label{eq:cs}
\sigma=
\frac{|A_0|^2}{\Bigl(1+\frac{\widetilde{Q^2}}{{Q_0^2}}\Bigr)^{2n}}
\frac{\left(\frac{s}{s_0}\right)^{2(\alpha_0-1)}}
{2\alpha'\ln(s/s_0)+4\Bigl(\frac{a}{\widetilde Q^2}+\frac{b}{2m_N^2}\Bigr)}.
\end{equation}

%\newpage
\section{Fitting strategy and results} \label{sec:Results}
As already mentioned in Introduction, in this paper we consider a simple model with a single term for the Pomeron,
``soft" or ``hard" - depending on the kinematic region and class of reaction\footnote{In the Appendix~\ref{app:Alternative} we present a possible alternative, empirical model with $\widetilde Q^2$-dependent effective" trajectories, mimicking the transition between soft and hard dynamics. This model has been presented to the workshop ``Diffraction 2012''~\cite{Canarias}.}. 
The expressionsfor the cross-sections and the slope are given by Eqs.~(\ref{eq:Bslope}), (\ref{eq:dcsdt}) and (\ref{eq:cs}).

\subsection{Strategy}
There are eight free parameters in our model namely: $A_0$, $Q_0^2$, $n$, $\alpha_0$, $\alpha'$, $a$, $b$ and $s_0$. For simplicity we set $s_0 = 1.0$ GeV$^2$, so seven free parameters  remain: the normalization factor $A_0$; the scale $Q_0^2$ for the virtuality; the exponent $n$ for $Q-$dependent factor in the amplitude; the intercept $\alpha_0$ and the slope $\alpha'$ of the linear Pomeron trajectory; the two coefficients $a$ and $b$, which set the $Q-$dependence for the forward slope. These seven free parameters have been found from the fit to the HERA data on the DVCS and the electroproduction of $\phi$ and $J/\psi$ mesons. Each process was treated separately.

Each class of reaction contains various measurables i.e. the data for the forward slope $B$, for the differential cross section $d\sigma/d|t|$ and for the integrated cross section $\sigma$ as function of virtuality $Q^2$ at fixed energy $W$, or for the integrated cross section $\sigma$ as function of the energy $W$ at fixed virtuality $Q^2$.Thus, for each separate reaction we fitted $d\sigma/d|t|$, $\sigma$ and $B$ simultaneously.

The units of the fitted parameters next
$$[A_0]=\frac{\sqrt{nb}}{GeV},\; [n]=[\alpha_0]=1,\; [\alpha']=GeV^{-2},\; [a]=[b]=[Q_0^2]=GeV^2.$$

\subsection{DVCS}
\begin{figure}[h,t,p,d,!]
\begin{center}
\includegraphics[clip,scale=0.475]{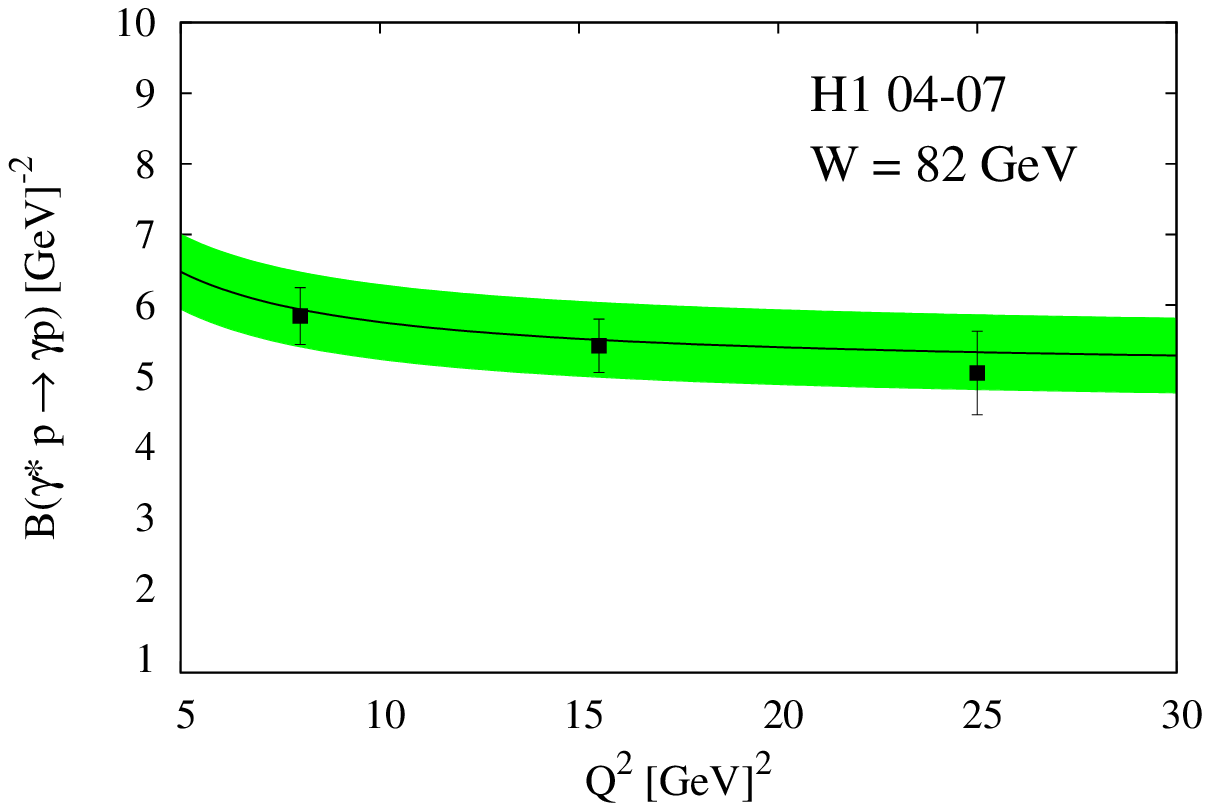}
\includegraphics[clip,scale=0.475]{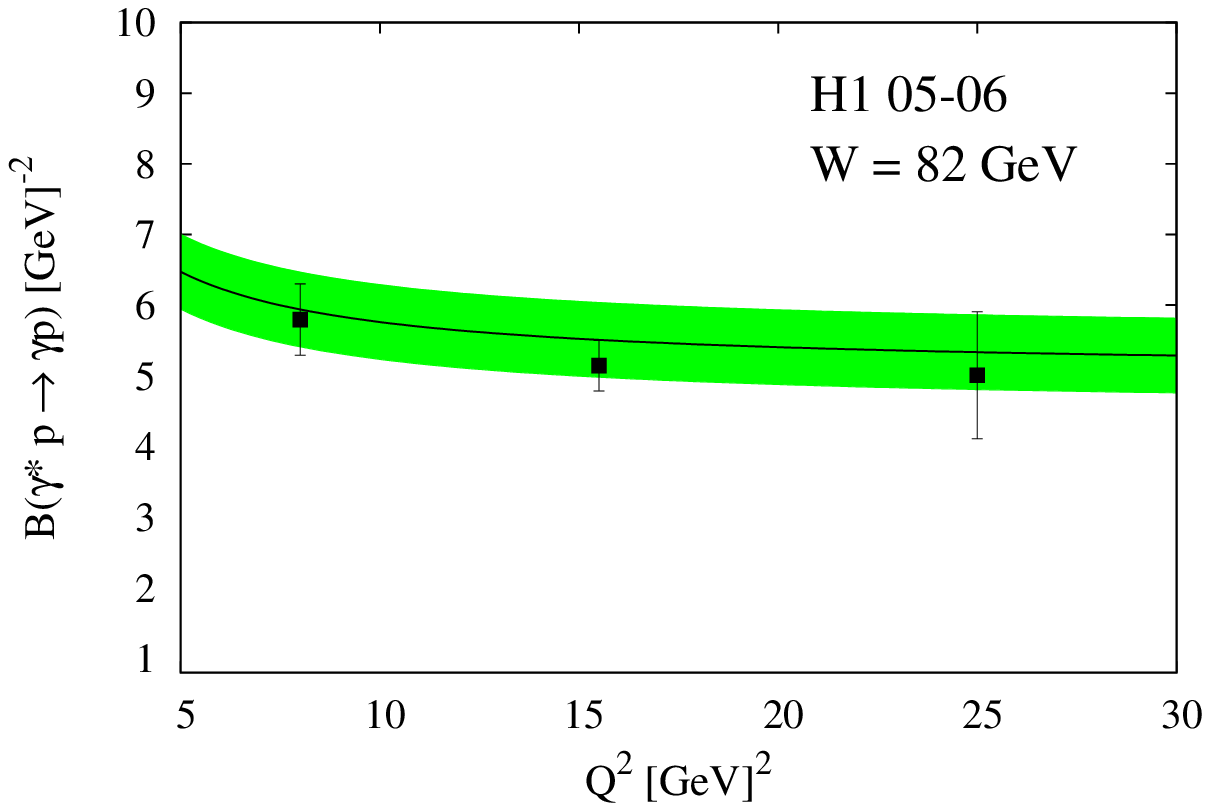}
\includegraphics[clip,scale=0.475]{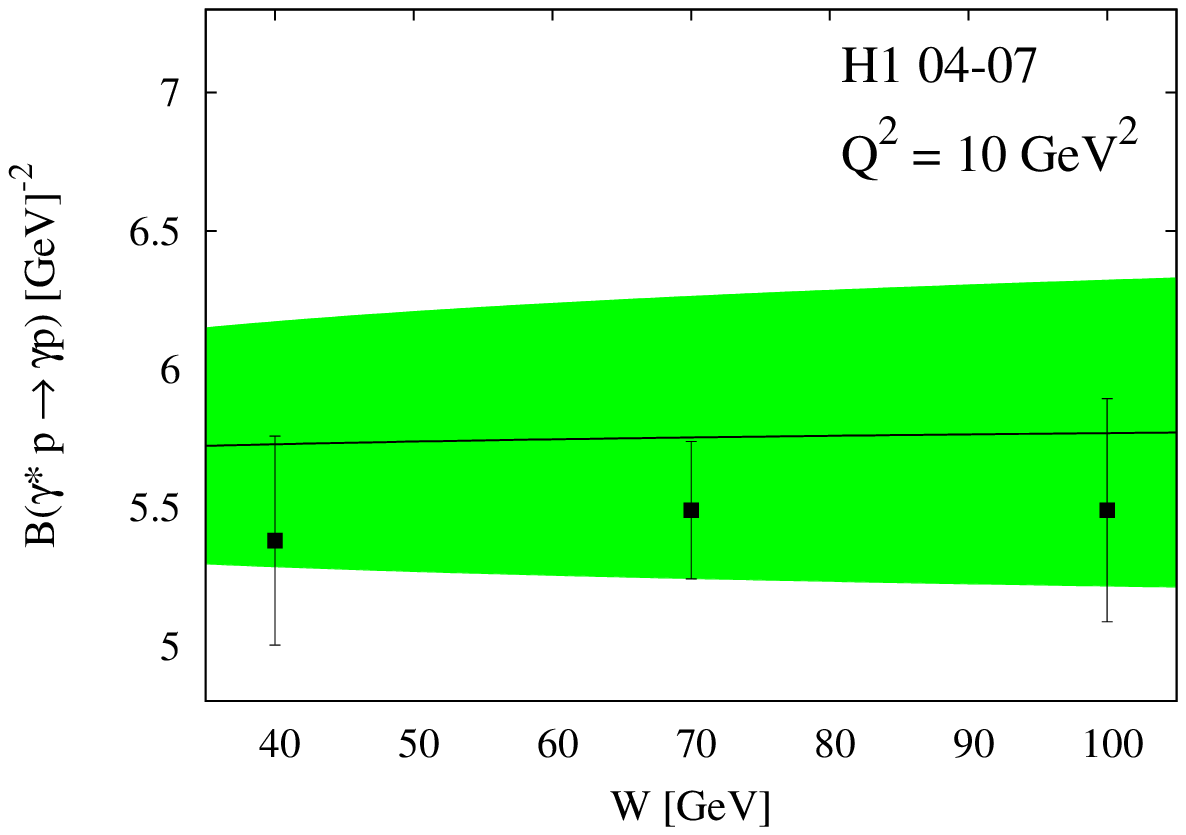}
\includegraphics[clip,scale=0.475]{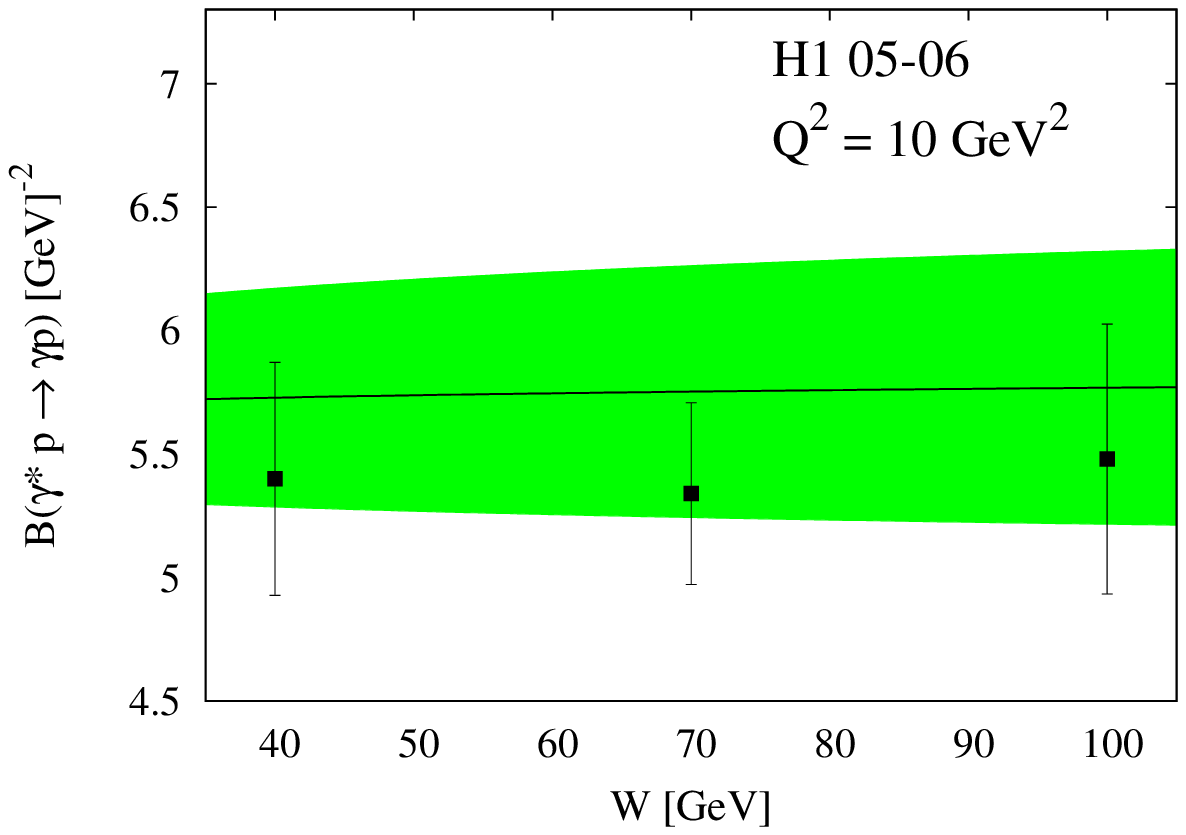}
\end{center}
\vspace{-0.5cm}
 \caption{Fit of Eq.~(\ref{eq:Bslope}) to the H1 data on the forward slopes % as functions of $W$ and(or) $Q^2$
  for $\gamma^*p\rightarrow\gamma p$.}\label{fig:B_DVCS}
\end{figure}

\begin{figure}[h,t,p,d,!]
\begin{center}
\includegraphics[clip,scale=0.475]{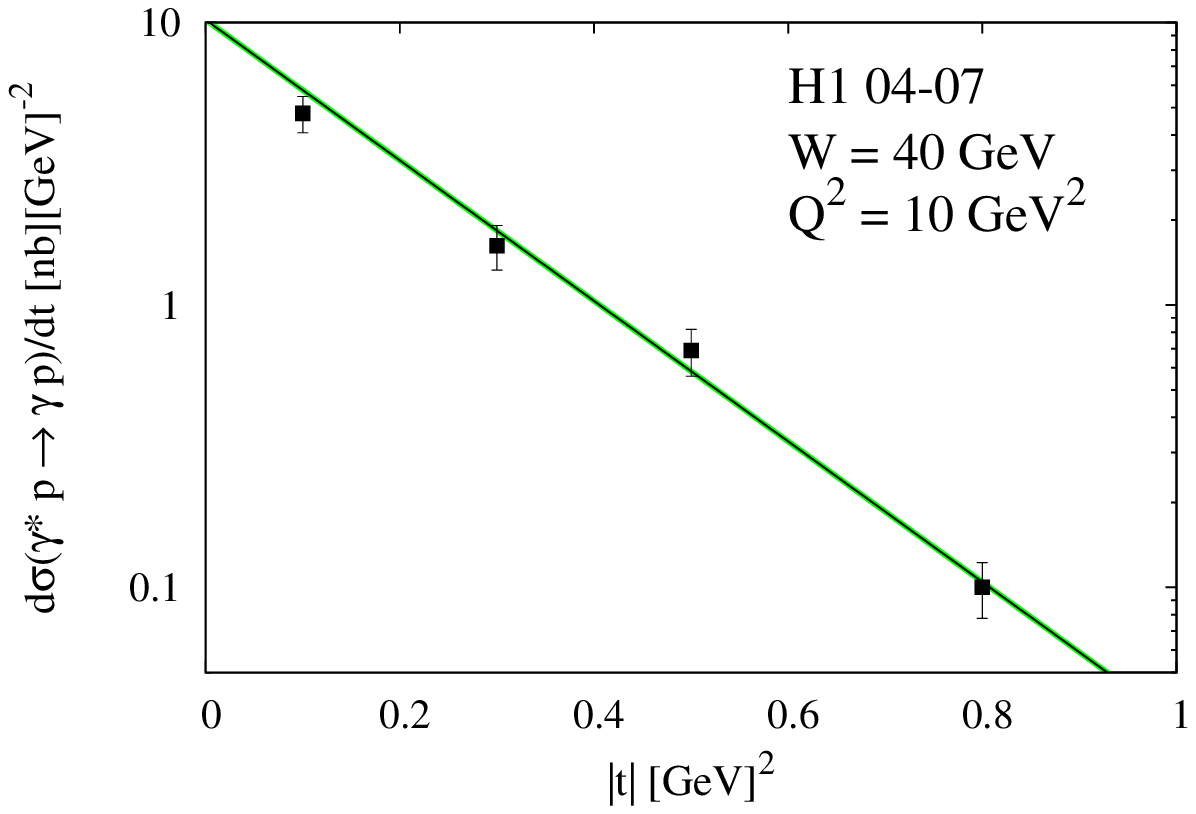}
\includegraphics[clip,scale=0.475]{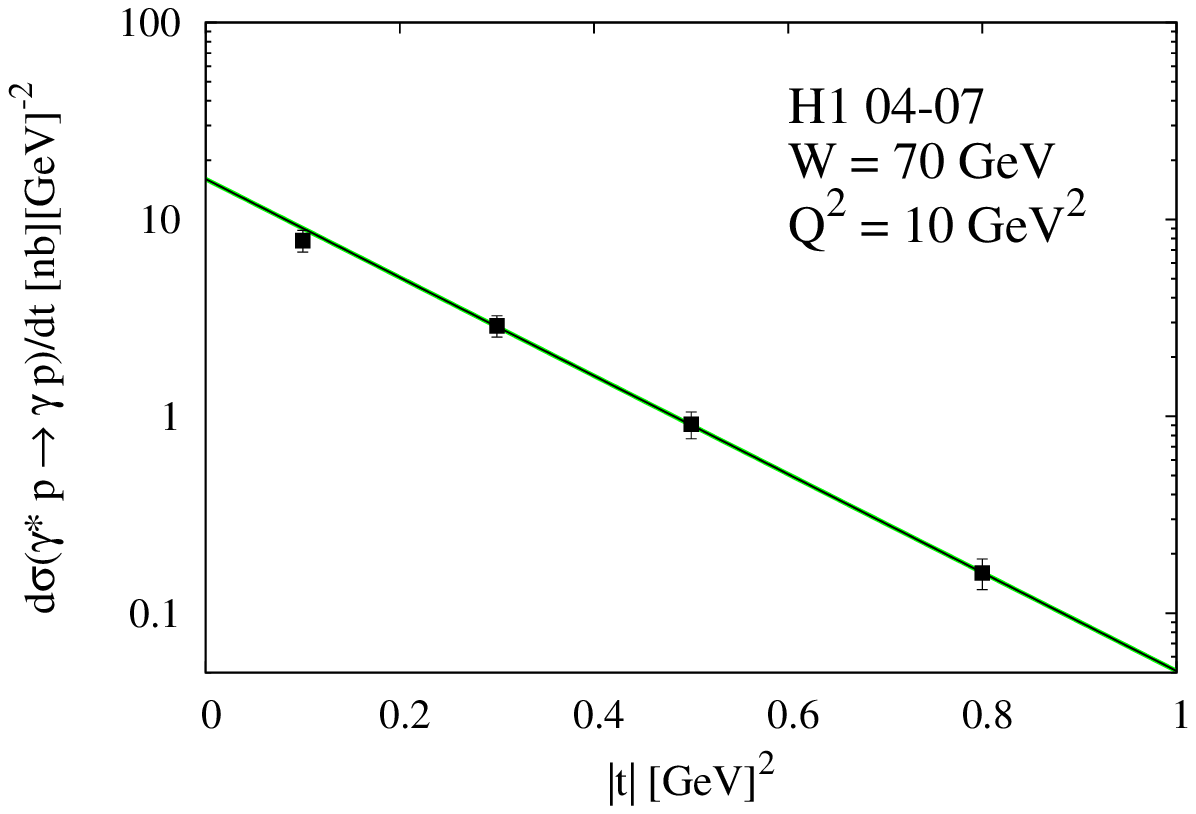}
\includegraphics[clip,scale=0.475]{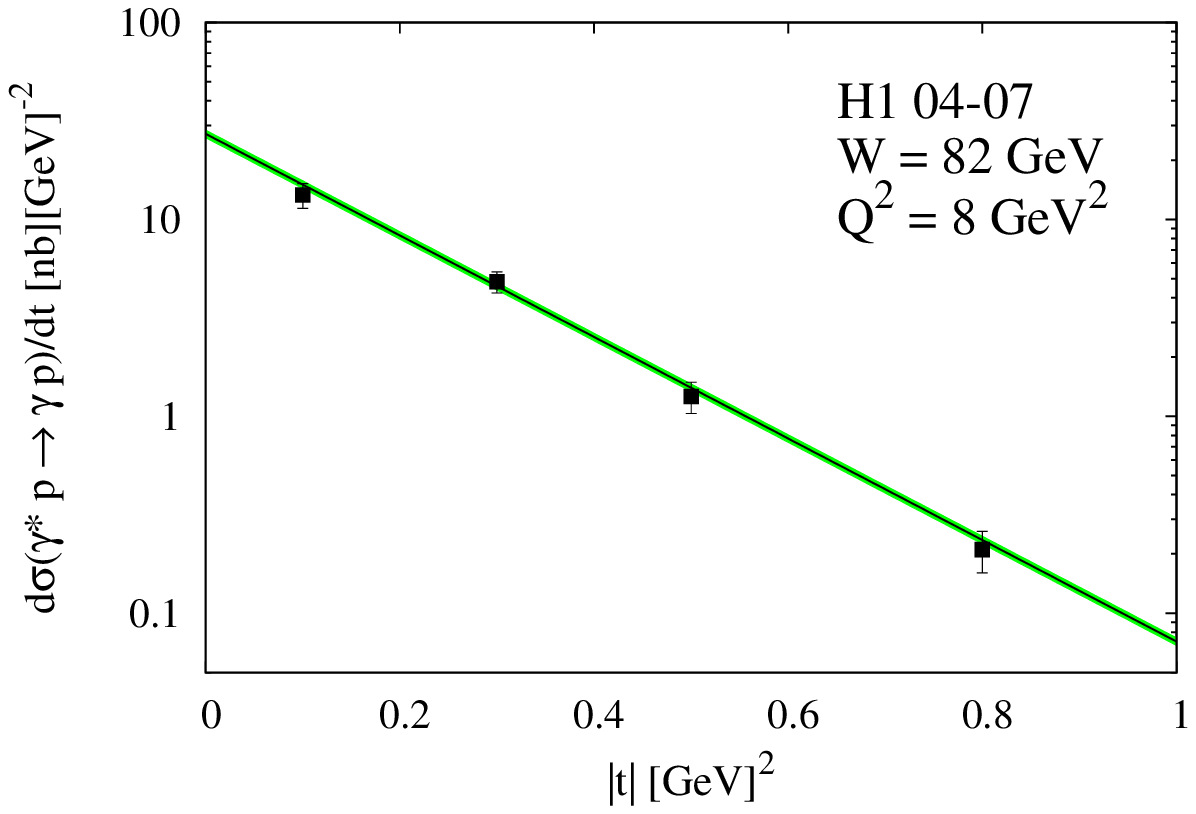}
\includegraphics[clip,scale=0.475]{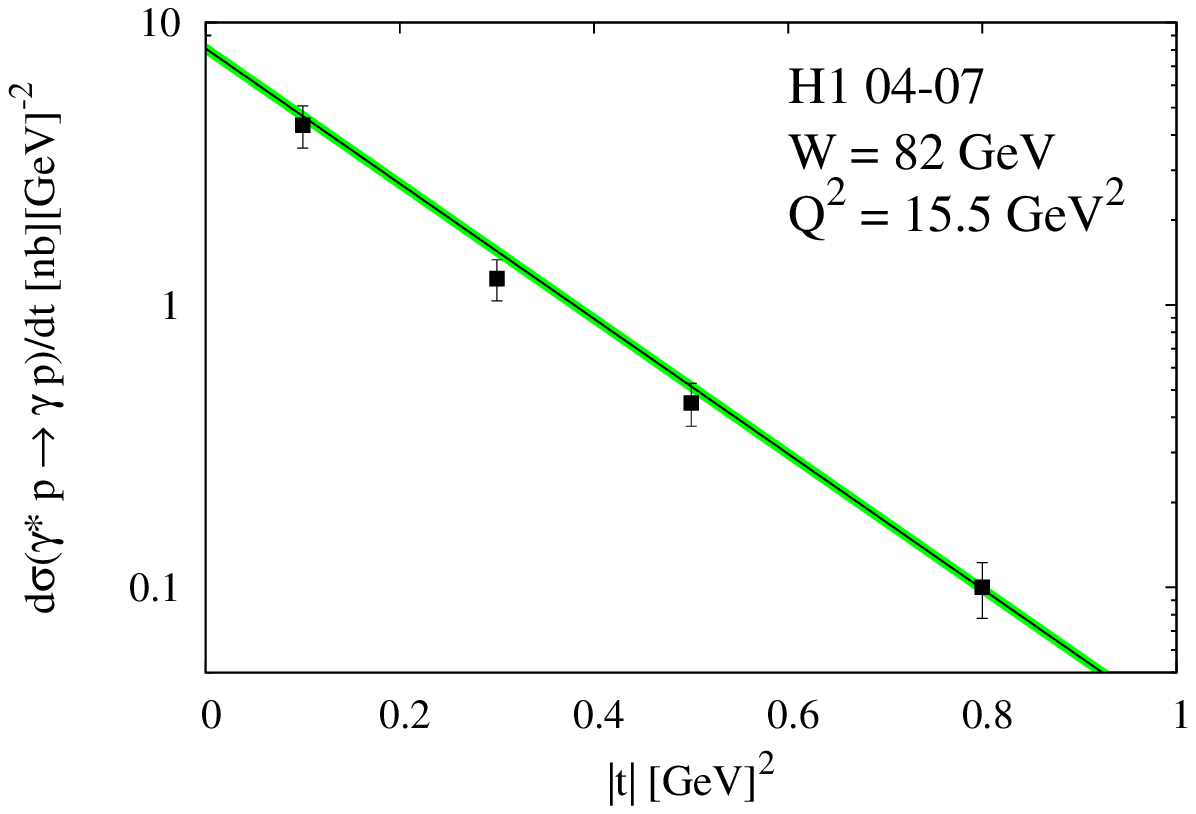}
\includegraphics[clip,scale=0.475]{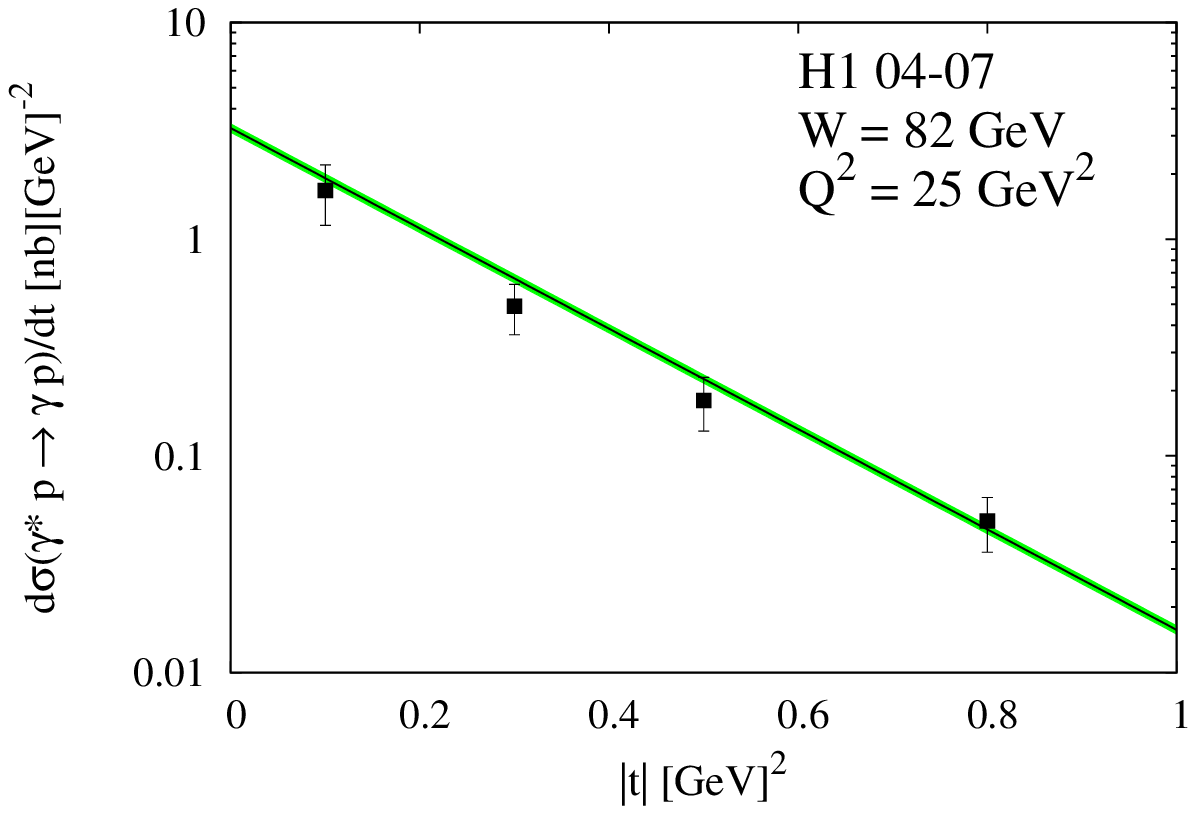}
\includegraphics[clip,scale=0.475]{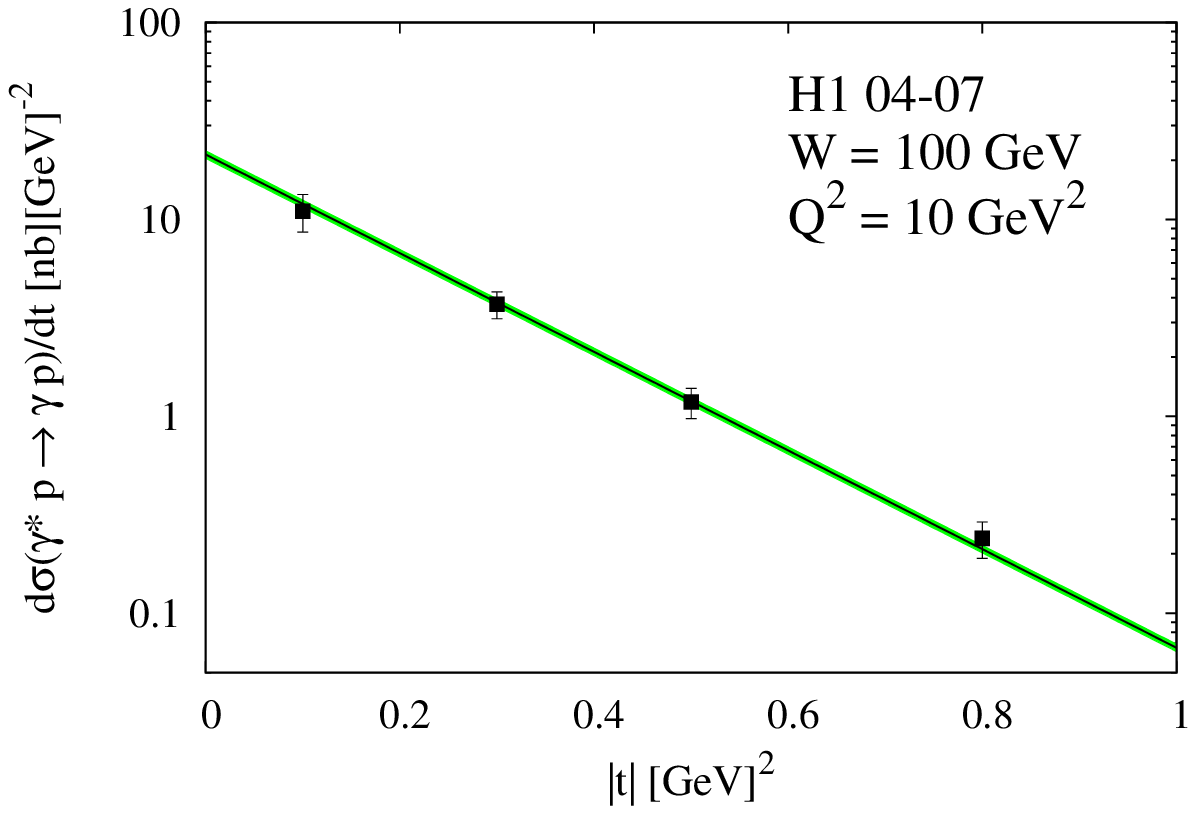}
\includegraphics[clip,scale=0.475]{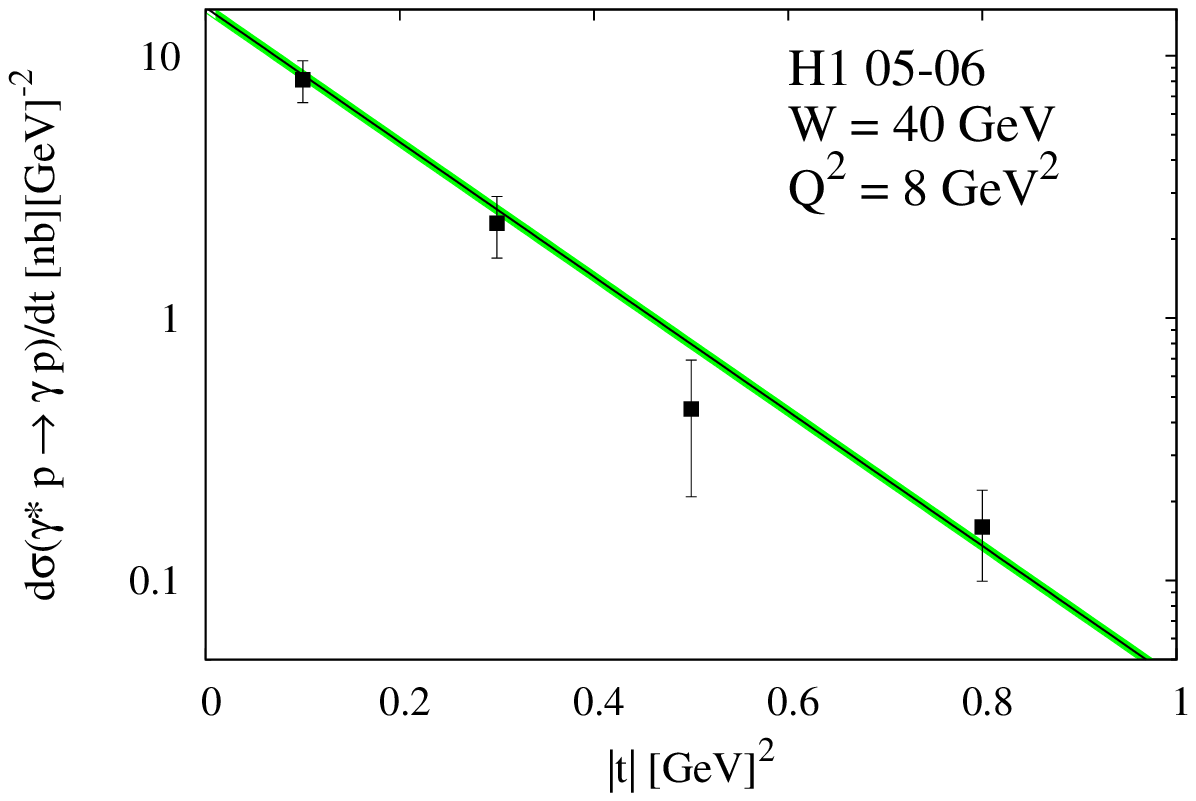}
\end{center}
\vspace{-0.5cm}
 \caption{Fit of Eq.~(\ref{eq:dcsdt}) to the H1 data on the differential cross sections %as functions of $|t|$ 
 for $\gamma^*p\rightarrow\gamma p$.}\label{fig:dcsdt1_DVCS}
\end{figure}

\begin{figure}[h,t,p,d,!]
 \begin{center}
\includegraphics[clip,scale=0.475]{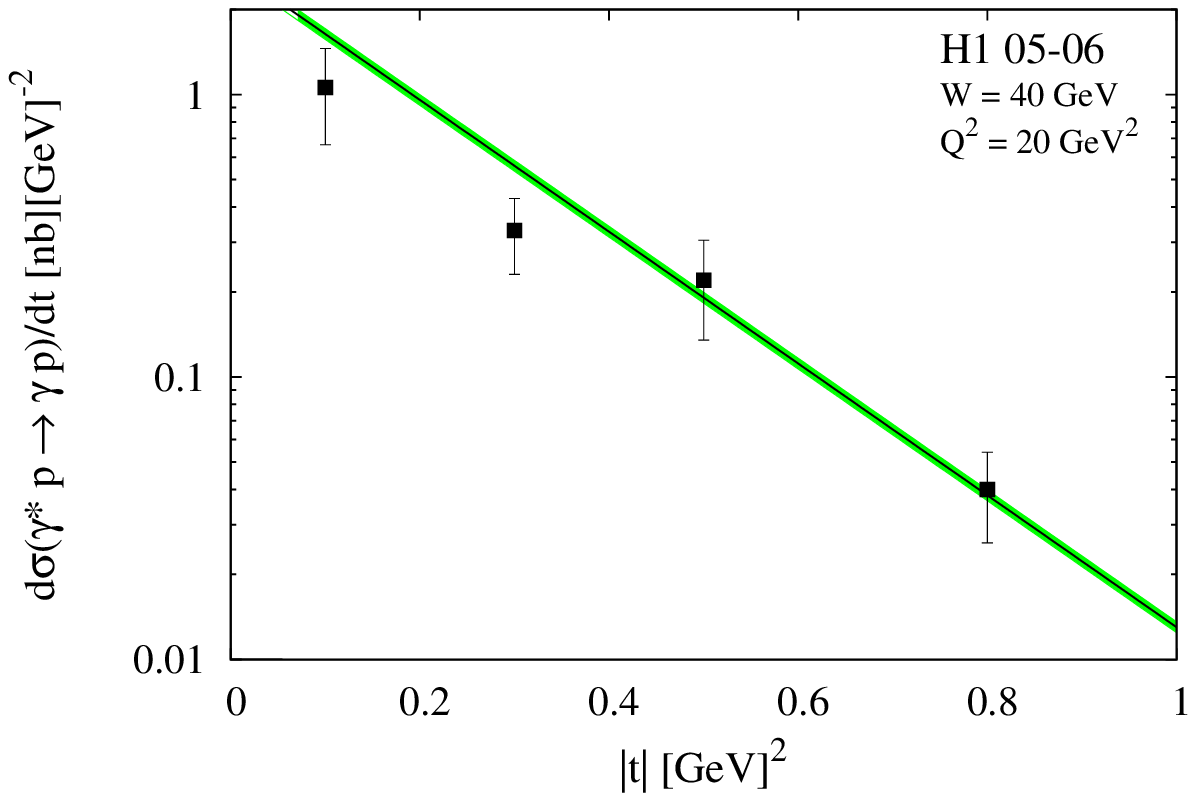}
\includegraphics[clip,scale=0.475]{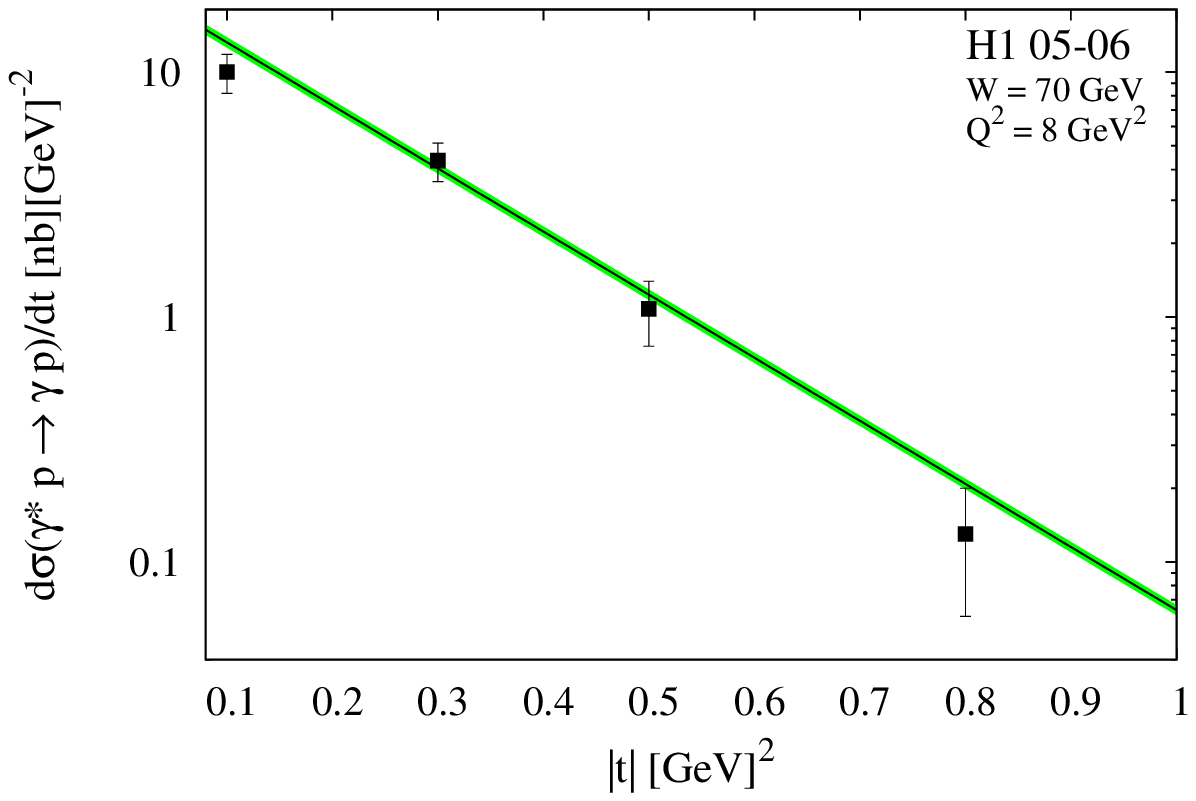}
\includegraphics[clip,scale=0.475]{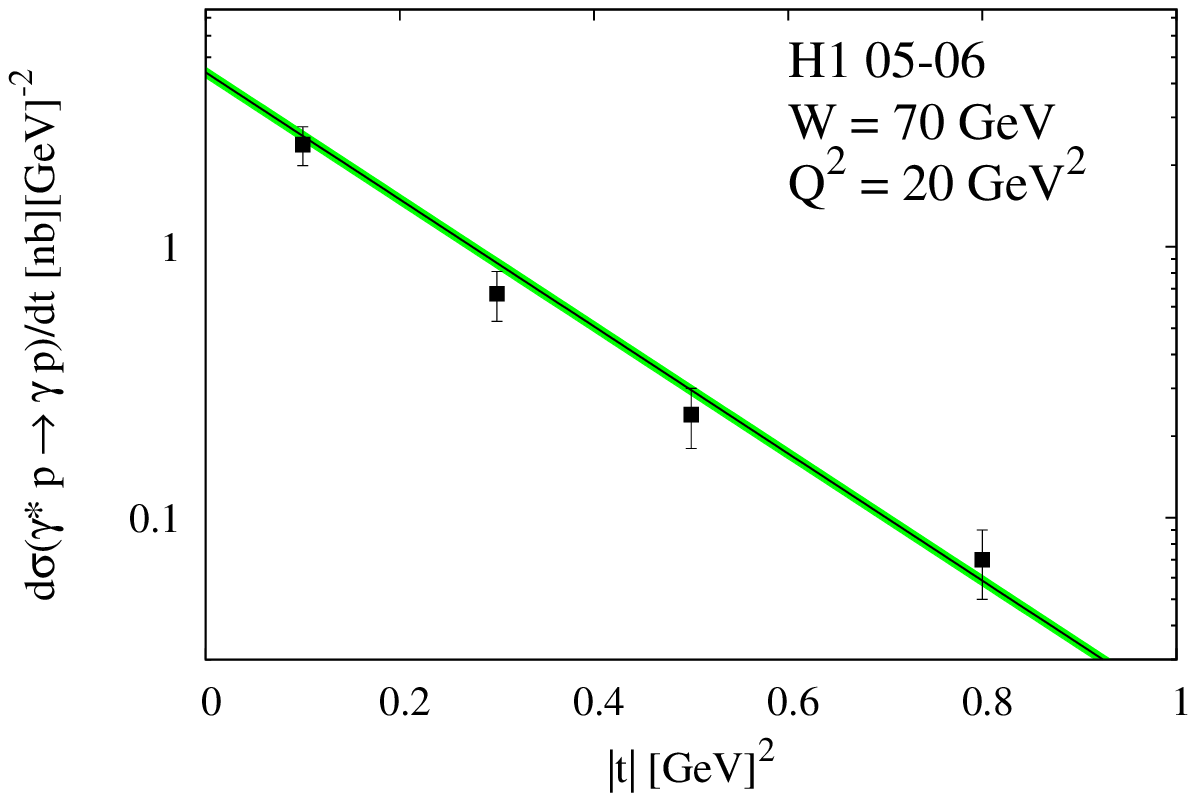}
\includegraphics[clip,scale=0.475]{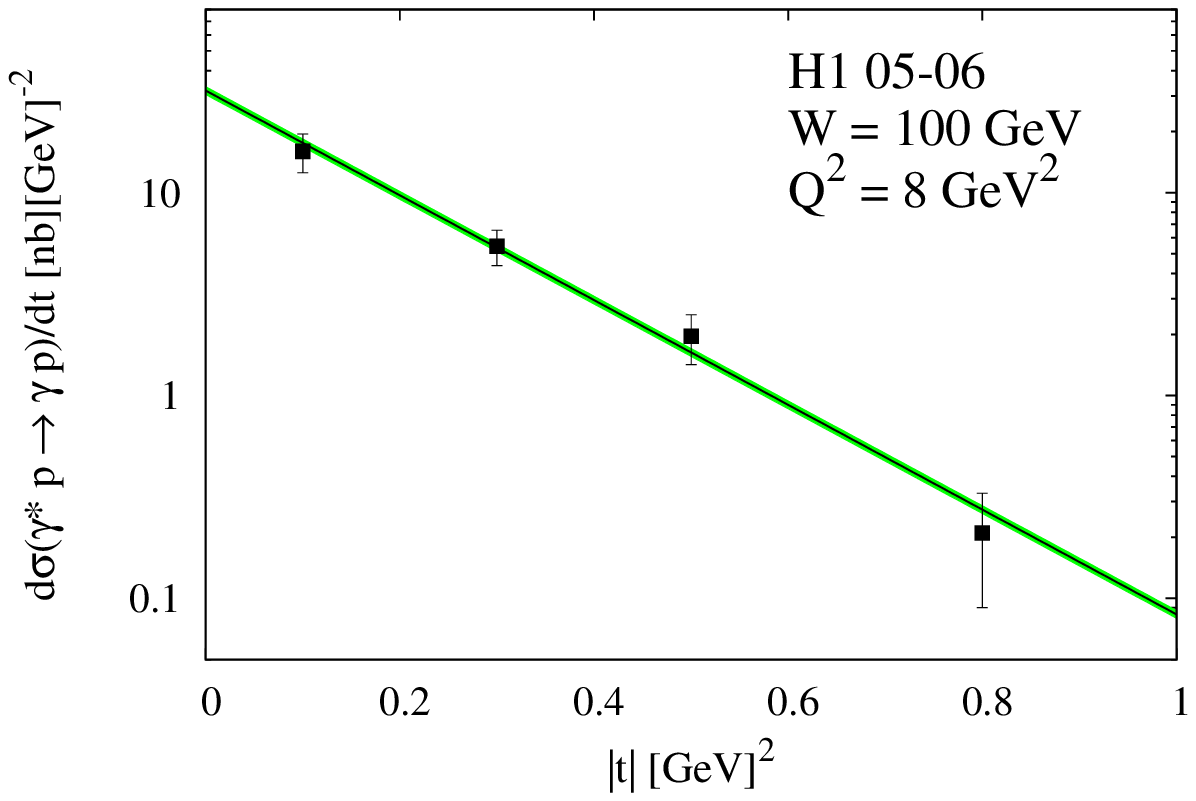}
\includegraphics[clip,scale=0.475]{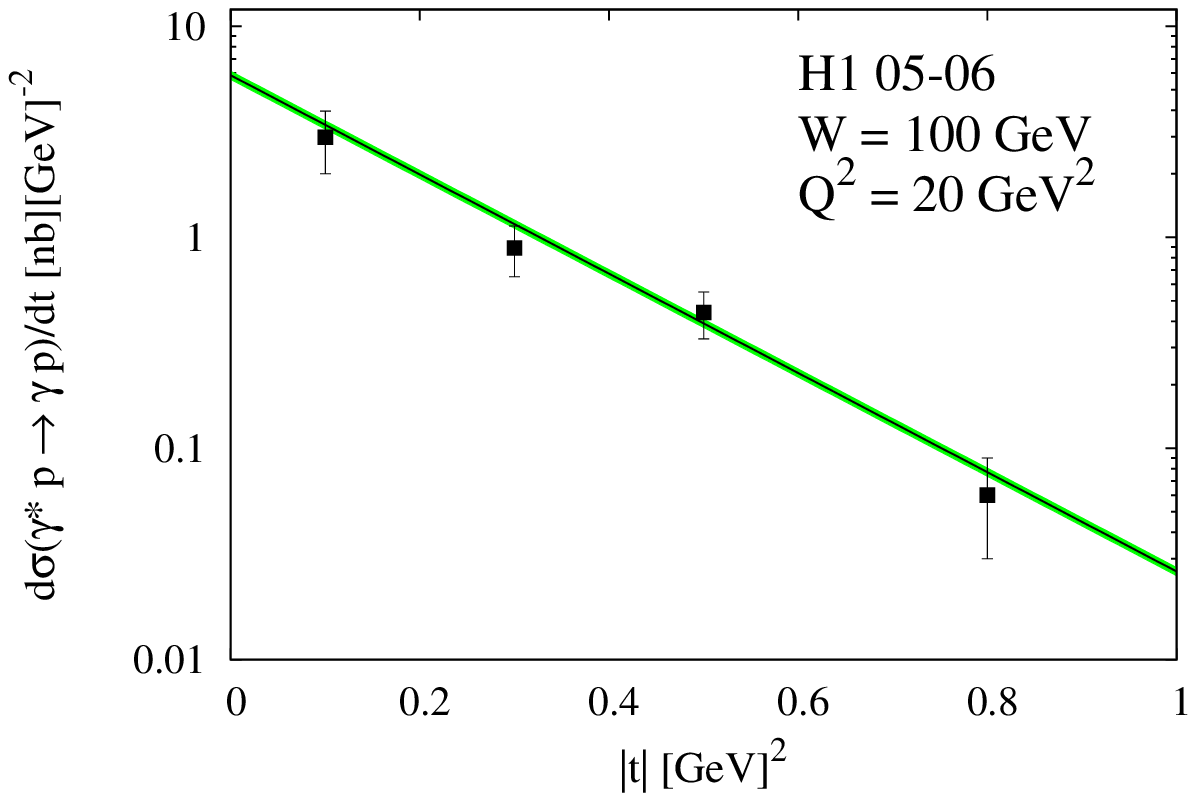}
\includegraphics[clip,scale=0.475]{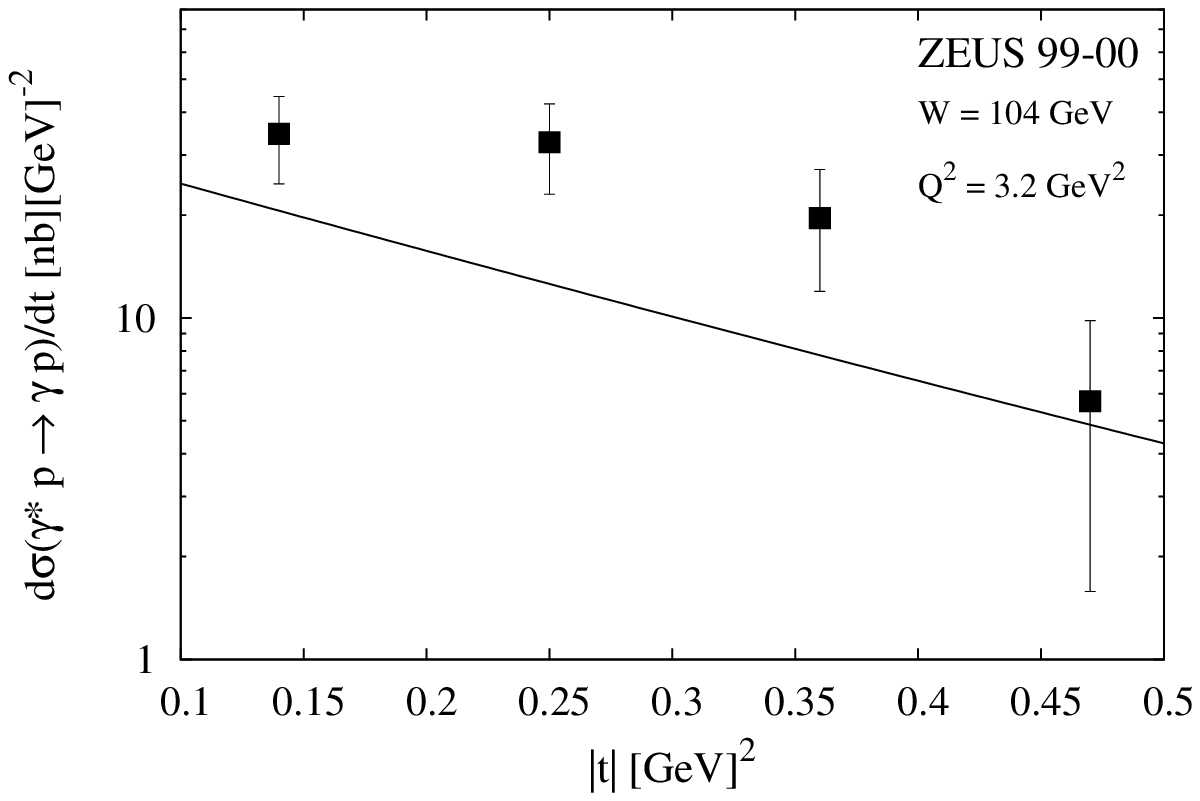}
\end{center}
\vspace{-0.5cm}
 \caption{Fit of Eq.~(\ref{eq:dcsdt}) to the H1 an ZEUS data on the differential cross sections %as functions of $|t|$ 
 for $\gamma^*p\rightarrow\gamma p$.} \label{fig:dcsdt2_DVCS}
\end{figure}

\begin{figure}[h,t,p,d,!]
\begin{center}
\includegraphics[clip,scale=0.475]{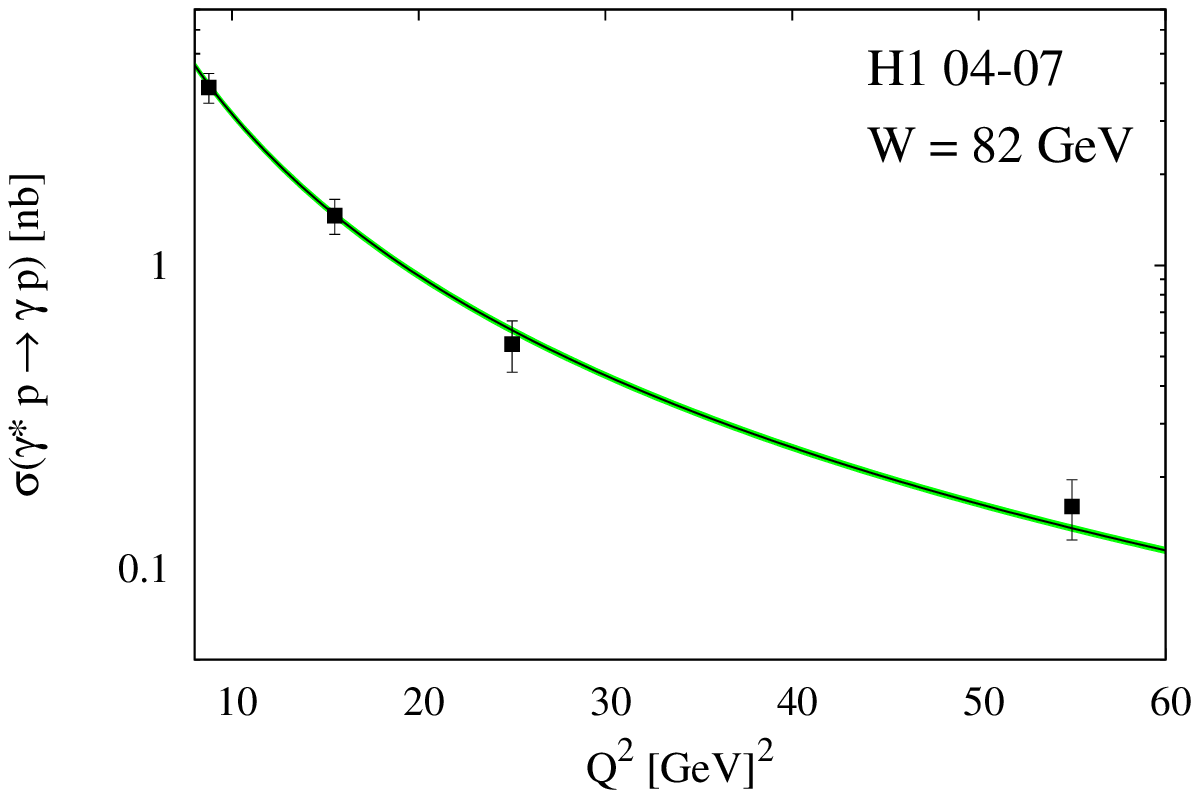}
\includegraphics[clip,scale=0.475]{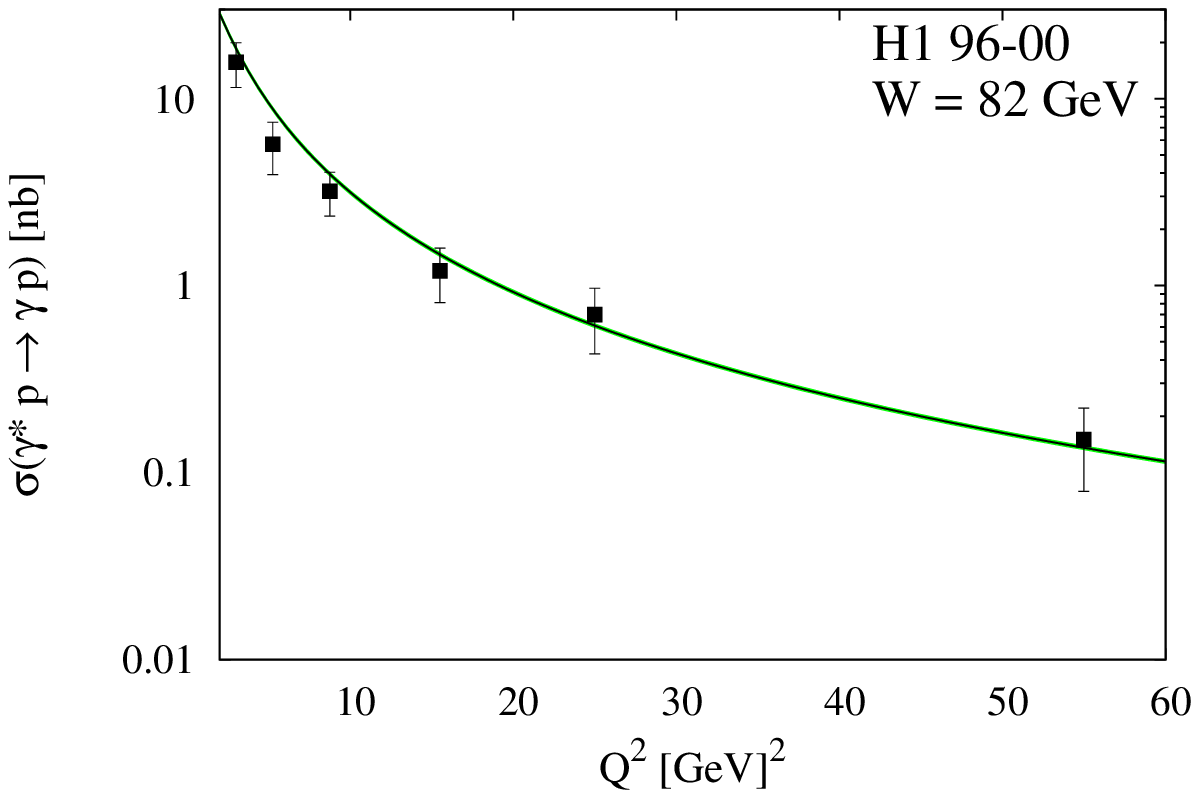}
\includegraphics[clip,scale=0.475]{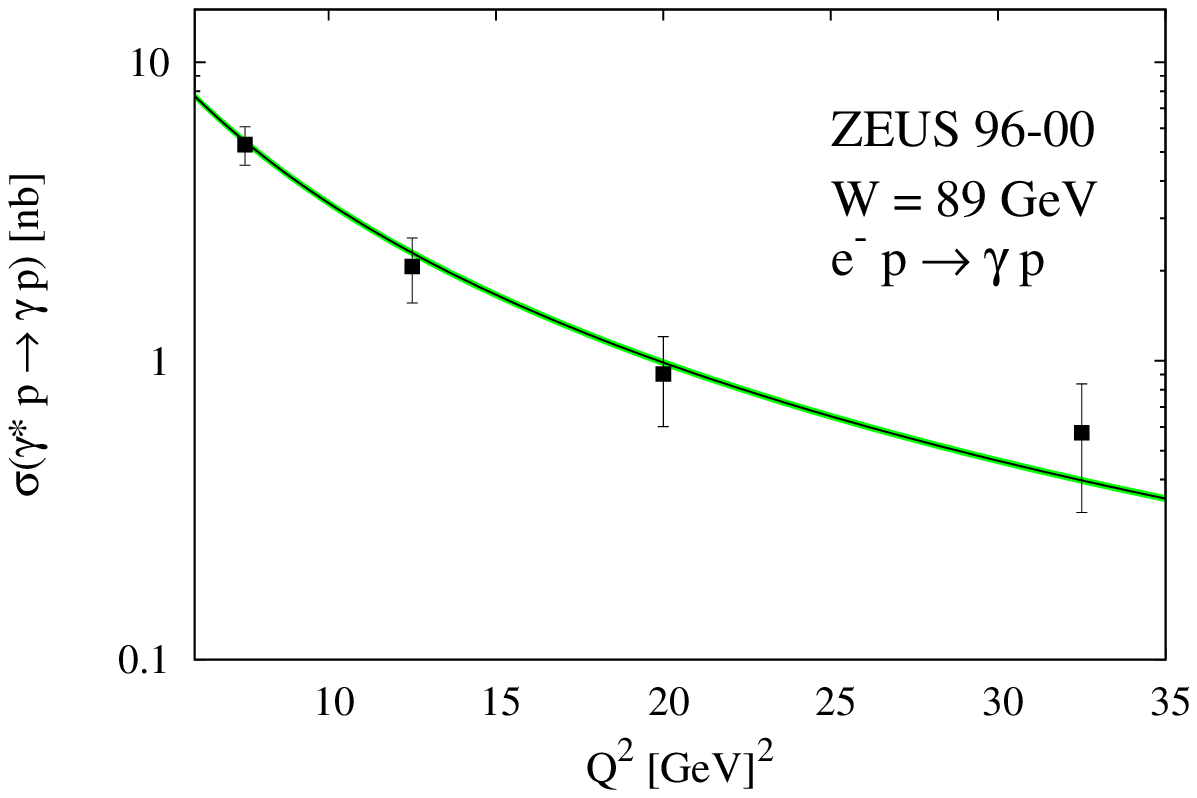}
\includegraphics[clip,scale=0.475]{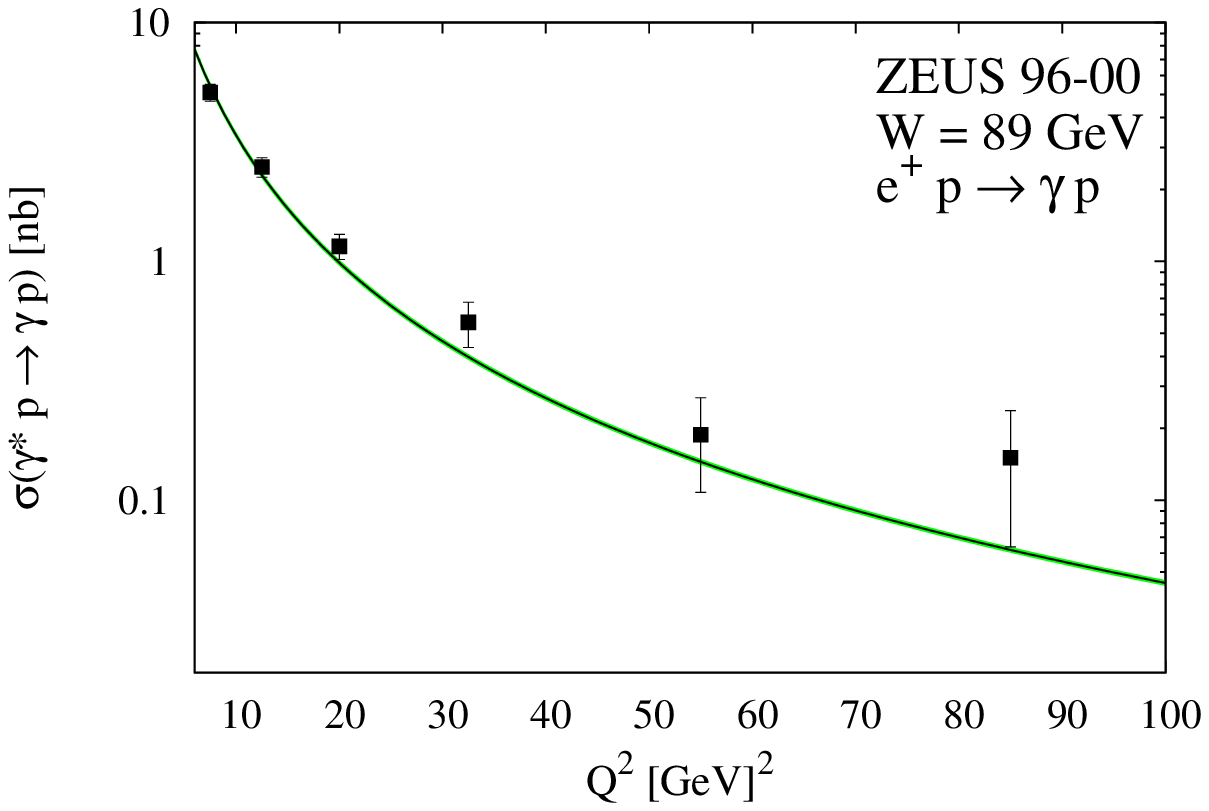}
\includegraphics[clip,scale=0.475]{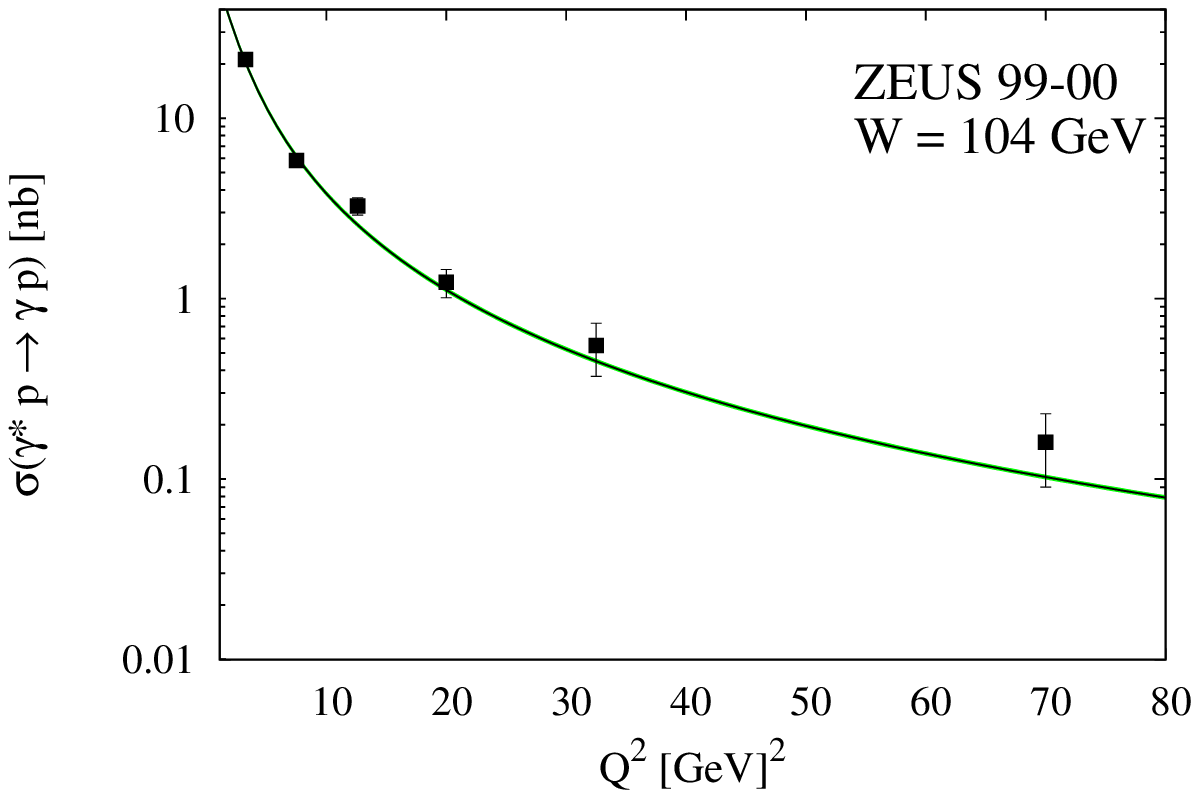}
\end{center}
\vspace{-0.5cm}
 \caption{Fit of Eq.~(\ref{eq:cs}) to the H1 and ZEUS data on the integrated cross sections % as functions of $Q^2$ 
 for $\gamma^*p\rightarrow\gamma p$.}\label{fig:csQ2_DVCS}
\end{figure}

\begin{figure}[h,t,p,d,!]
\begin{center}
\includegraphics[clip,scale=0.475]{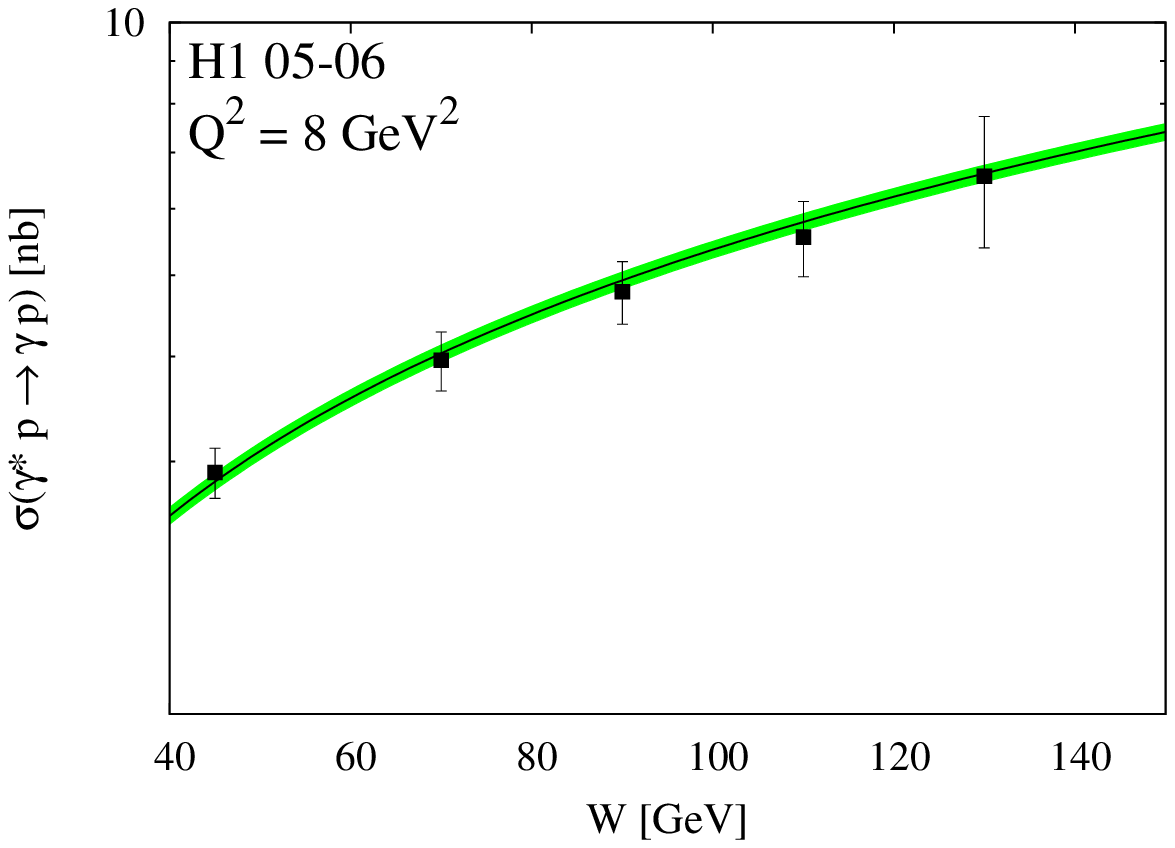}
\includegraphics[clip,scale=0.475]{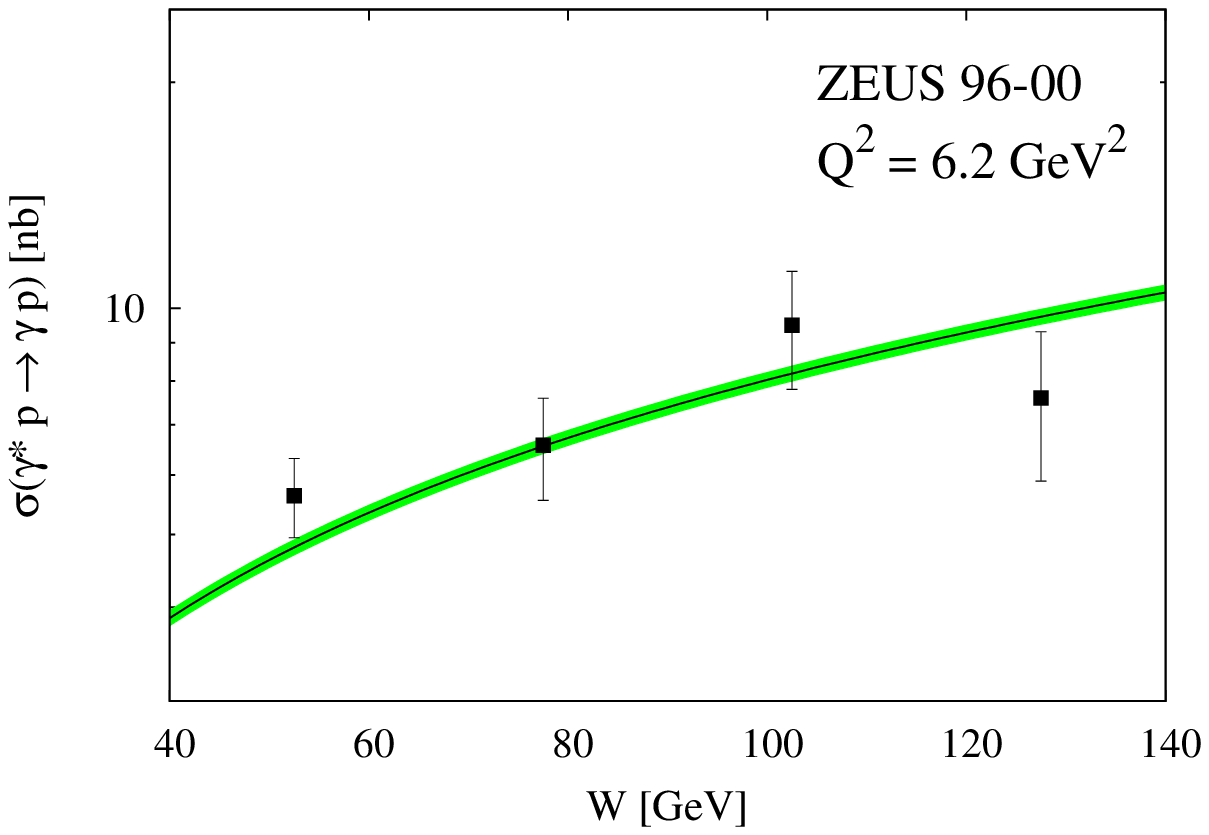}
\includegraphics[clip,scale=0.475]{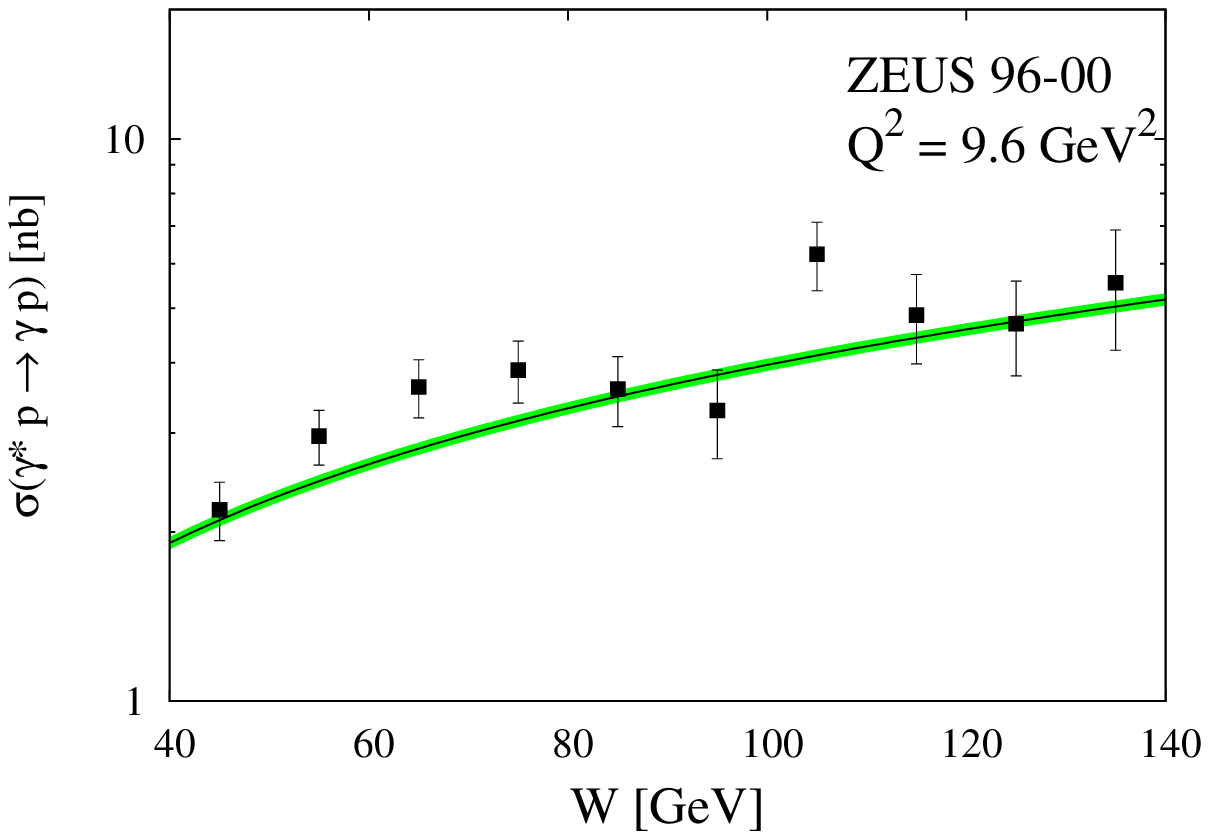}
\includegraphics[clip,scale=0.475]{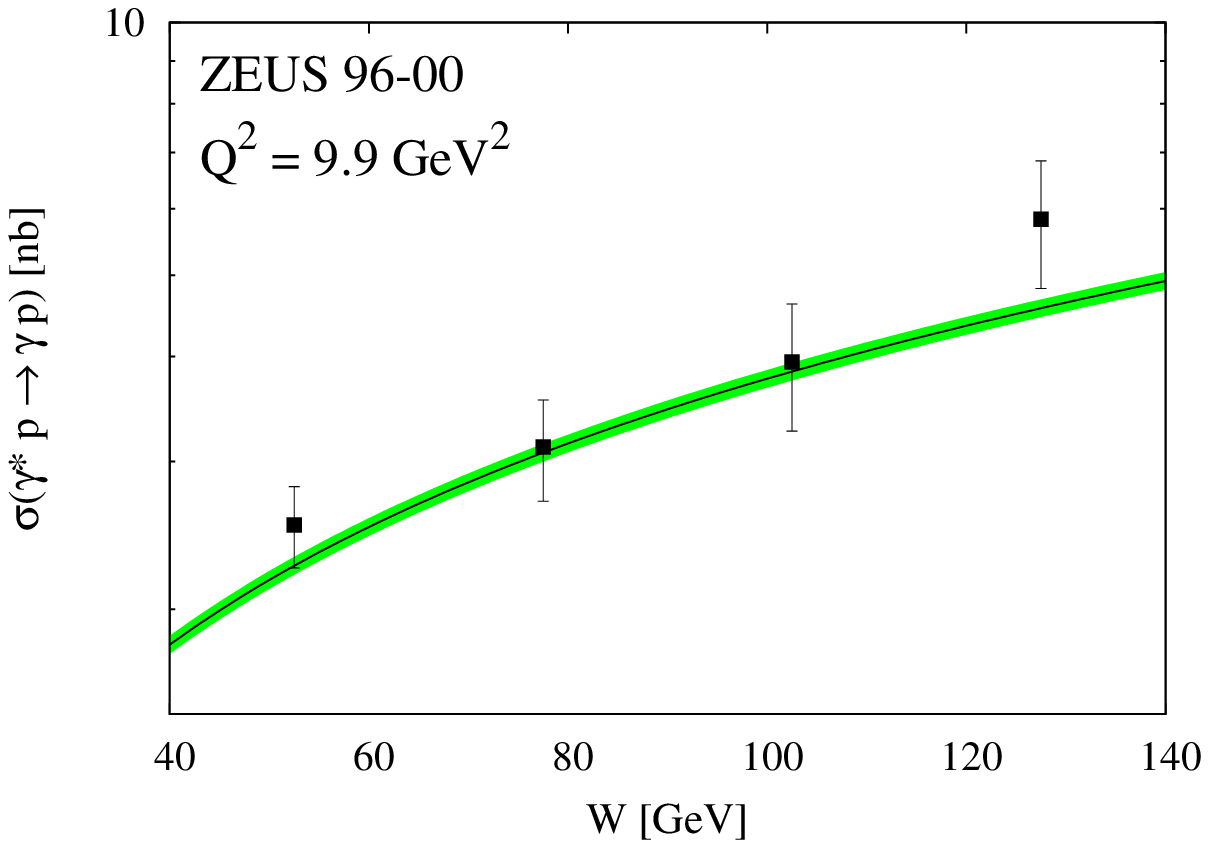}
\includegraphics[clip,scale=0.475]{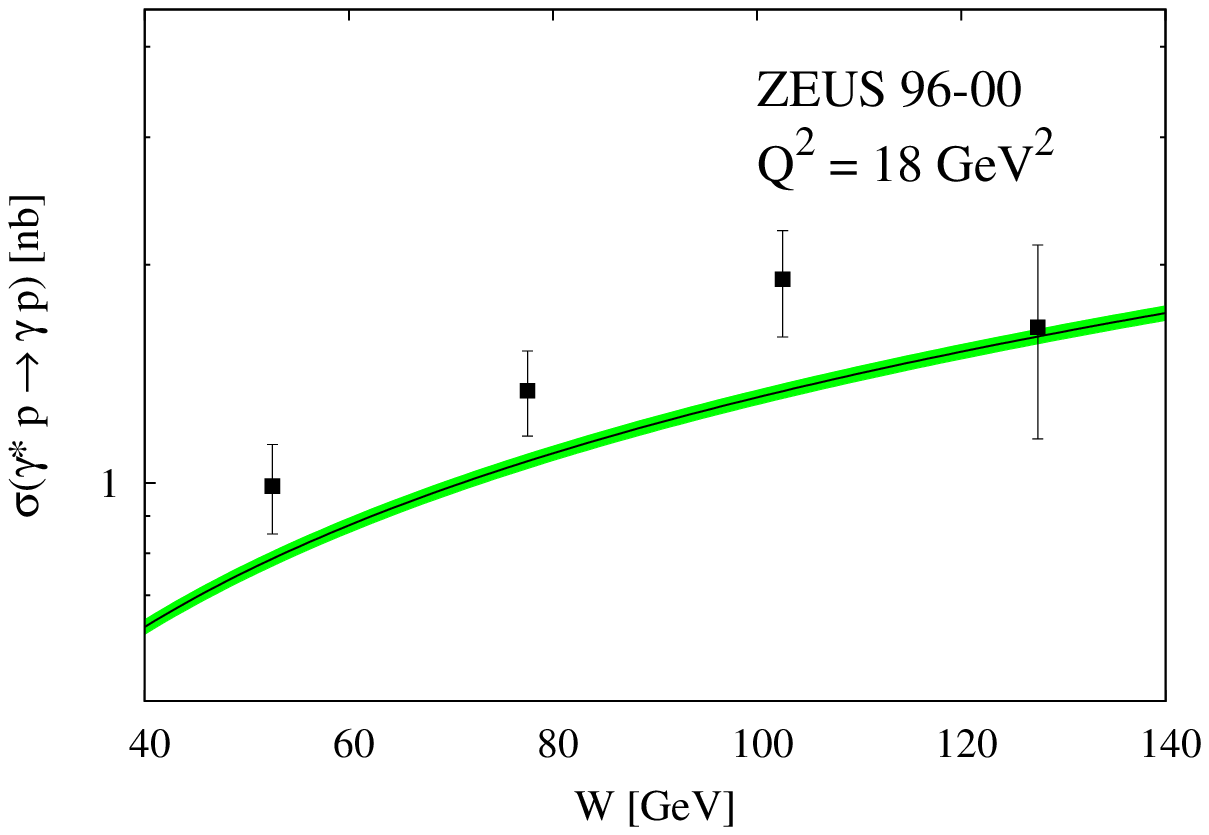}
\includegraphics[clip,scale=0.475]{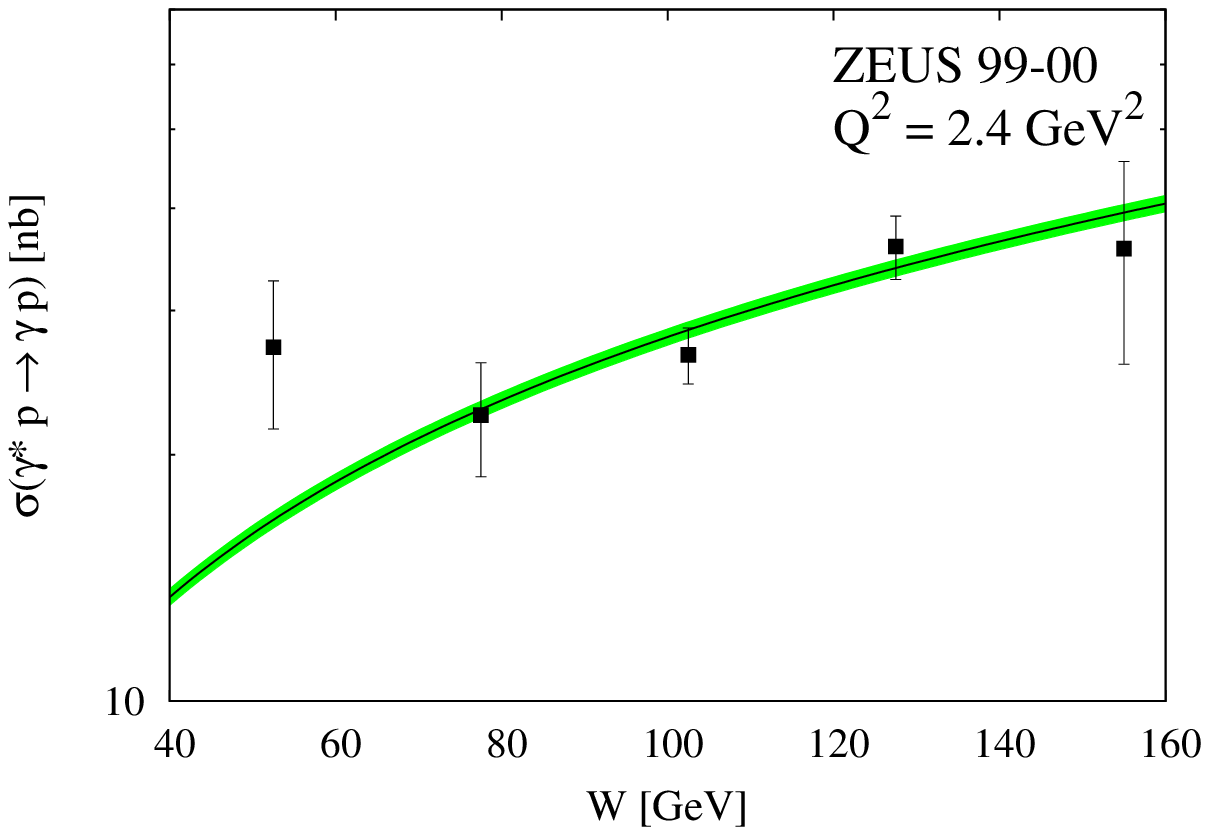}
\includegraphics[clip,scale=0.475]{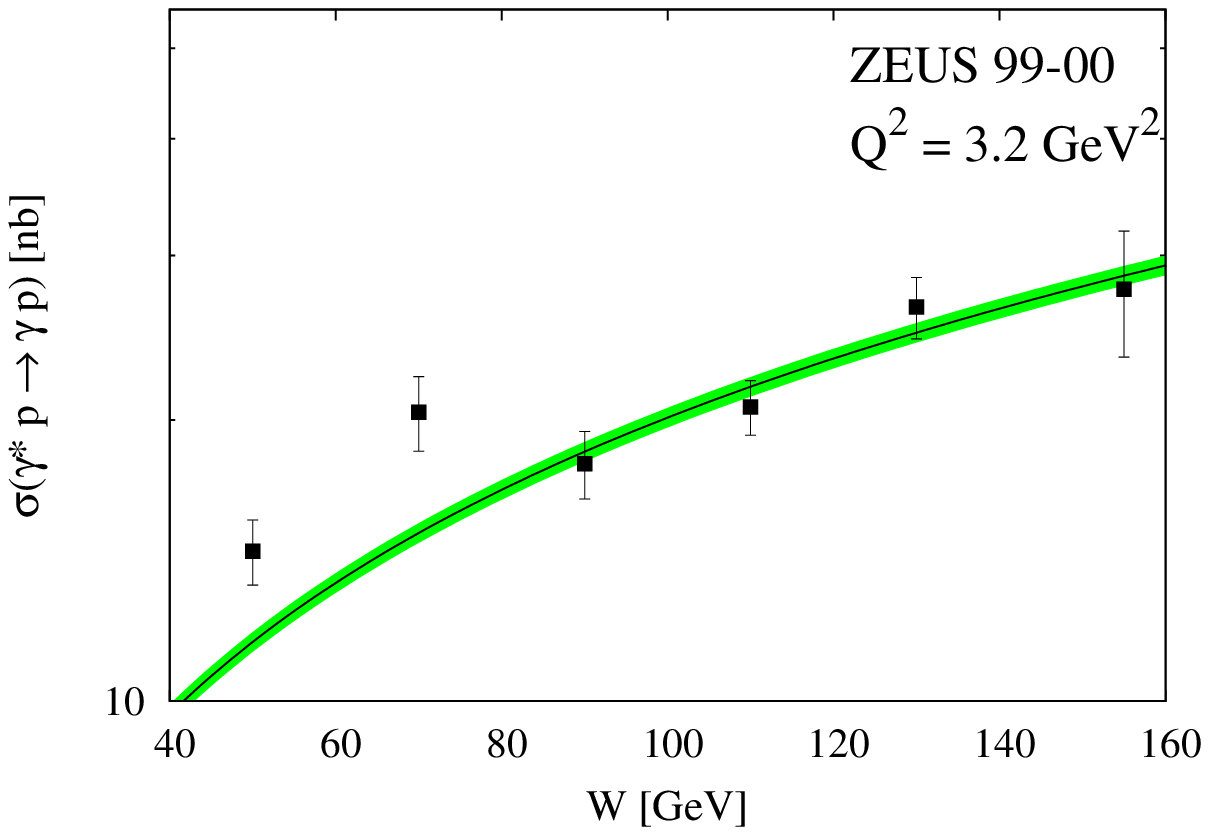}
\end{center}
\vspace{-0.5cm}
 \caption{Fit of Eq.~(\ref{eq:cs}) to the H1 and ZEUS data on the integrated cross sections% as functions of $W$ 
 for $\gamma^*p\rightarrow\gamma p$.} \label{fig:csW_DVCS}
\end{figure}

Here we fit our model to the data on DVCS published in \cite{d1,d3,d4}. Notice that for DVCS we have $\widetilde Q^2 = Q^2$, since $M_\gamma =0$ GeV.
The resulting fit is shown on Figs.~\ref{fig:B_DVCS}~-~\ref{fig:csW_DVCS}, with the values of the fitted parameters and the relevant $\chi^2/d.o.f.$, given in Table \ref{tab:DVCS}.
\begin{table}[h,t,p,d,!]
 \caption{Fitted parameters for DVCS}\label{tab:DVCS}
 \centering
  \begin{tabular}{c|c|c|c}\hline \hline
    $A_0$  &  $Q^2_{0}$  & n  & $\alpha_0$  \\\hline
    7.98 $\pm$ 0.09& 1.00 $\pm$ 0.34 & 1.00 $\pm$ 0.04 & 1.20$\pm$ 0.02 \\\hline \hline
    $\alpha'$  & a & b  & $\chi^2/d.o.f.$  \\\hline
    0.01 $\pm$ 0.03& 1.78 $\pm$ 0.19 & 2.14 $\pm$ 0.30 & 0.85\\
  \end{tabular}
\end{table}
%%
%%############### Figures ###################
%%

\newpage

\newpage

\newpage

\newpage

%%########### Jpsi electroproduction ###################################
%\newpage
\subsection {$J/\psi$ meson electroproduction}
\begin{figure}[p,h,t,p,d,!]
\begin{center}
\includegraphics[clip,scale=0.475]{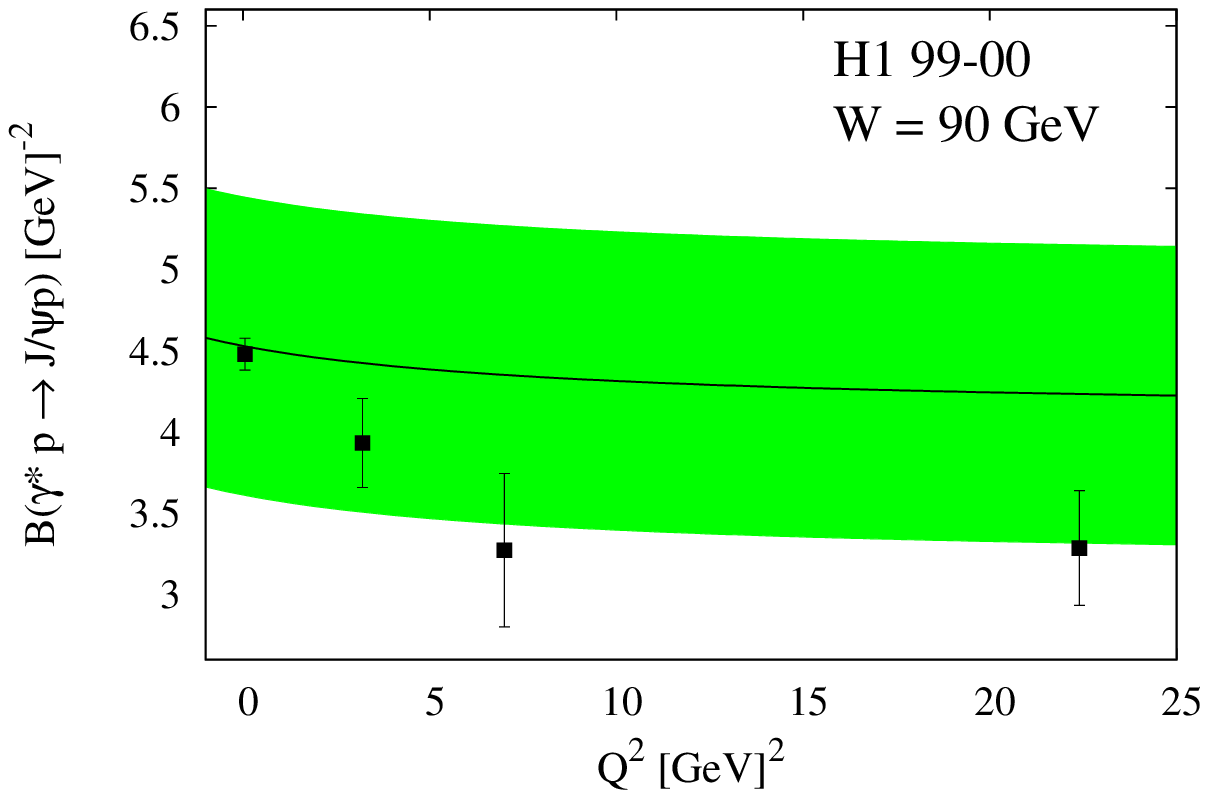}
\includegraphics[clip,scale=0.475]{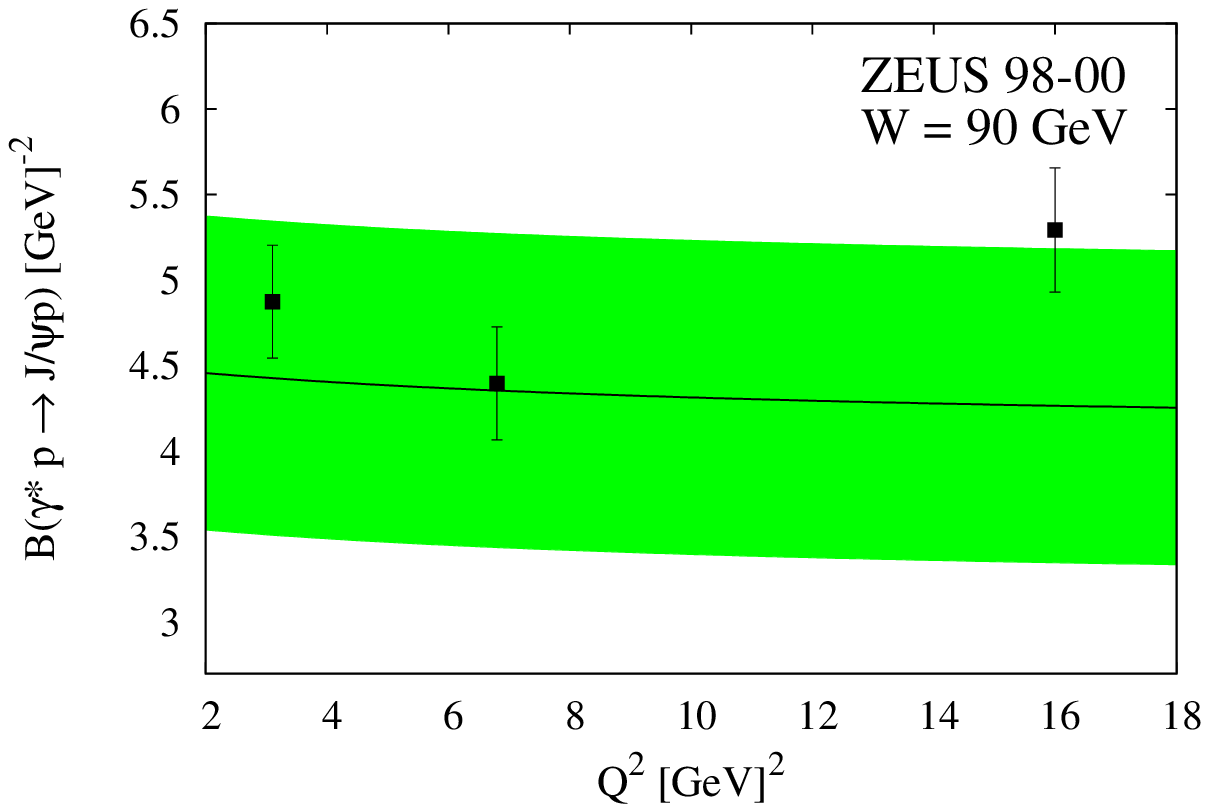}
\includegraphics[clip,scale=0.475]{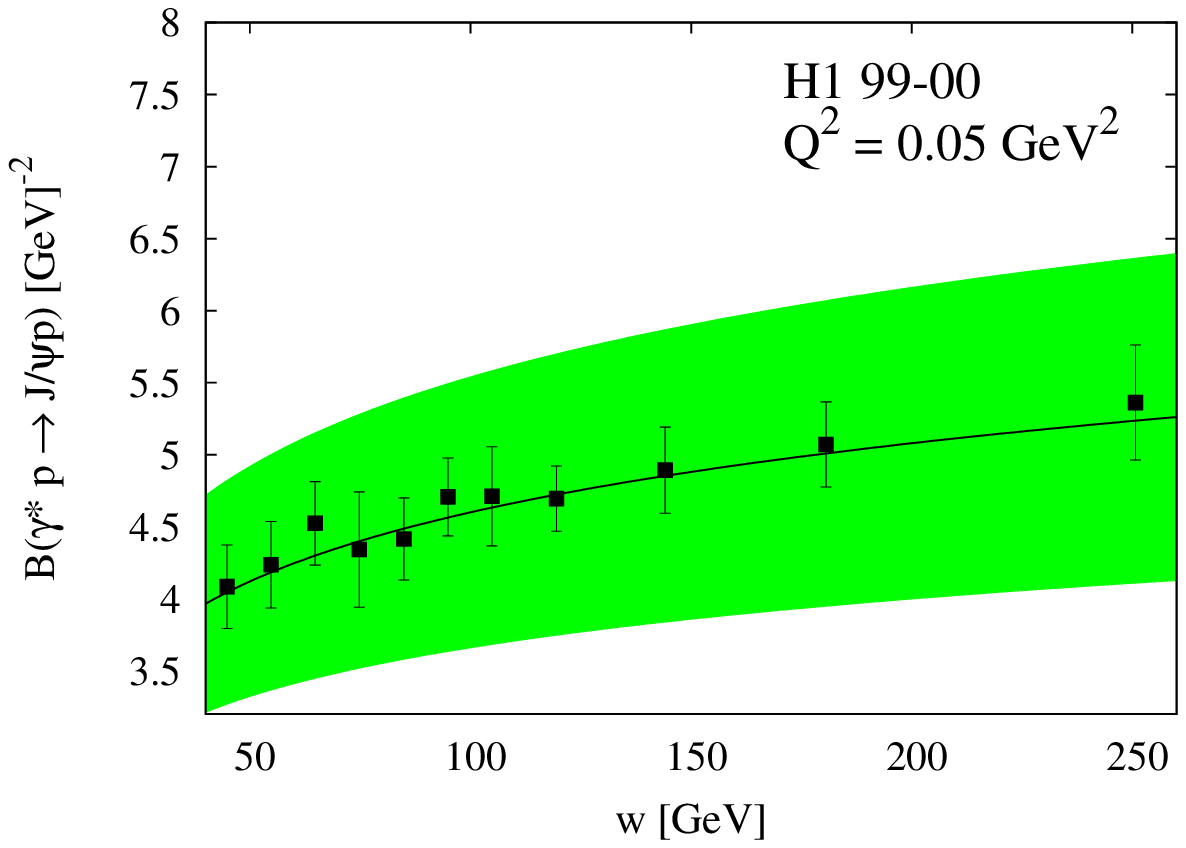}
\includegraphics[clip,scale=0.475]{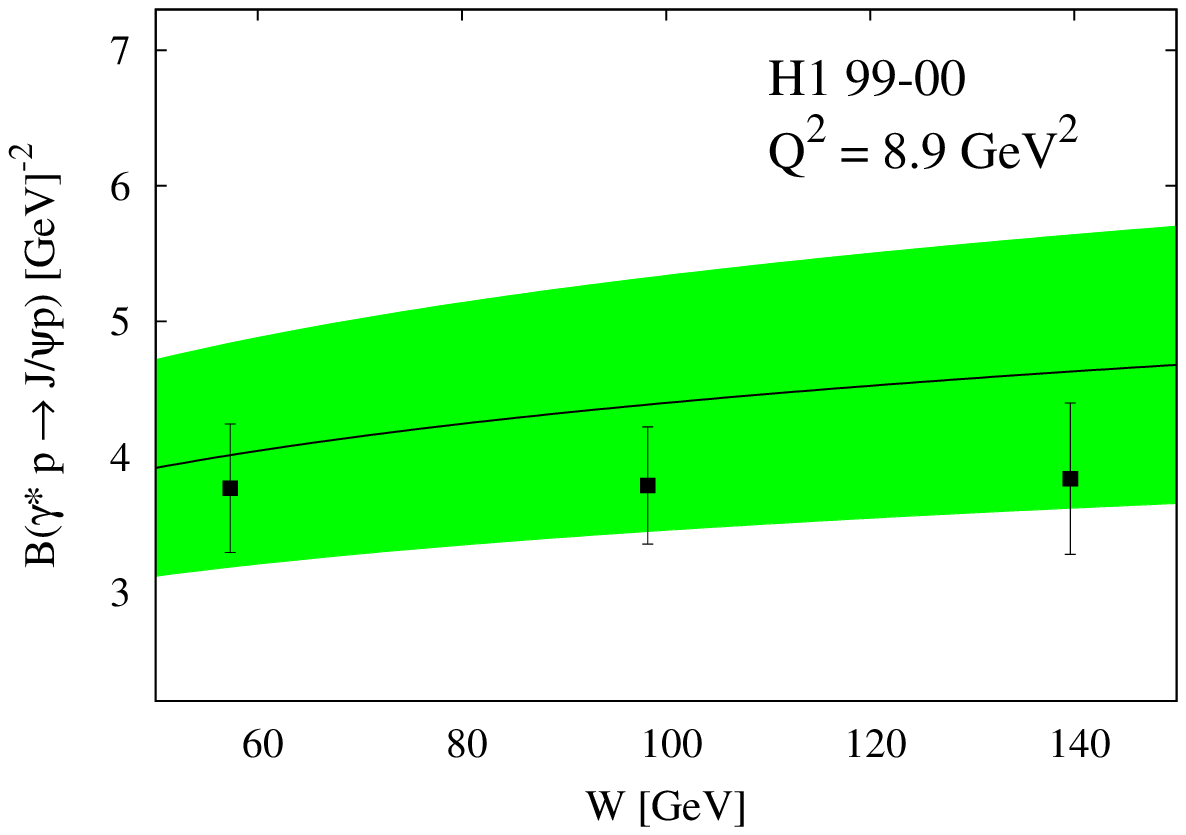}
\end{center}
\vspace{-0.5cm}
\caption{Fit of Eq.~(\ref{eq:Bslope}) to the H1 and ZEUS data on the forward slopes % as functions of $W$ and(or) $Q^2$ 
for $\gamma^*p\rightarrow J/\psi p$.}\label{fig:B_Jpsi}
\end{figure}

\begin{figure}[p,h,t,p,d,!]
\begin{center}
\includegraphics[clip,scale=0.475]{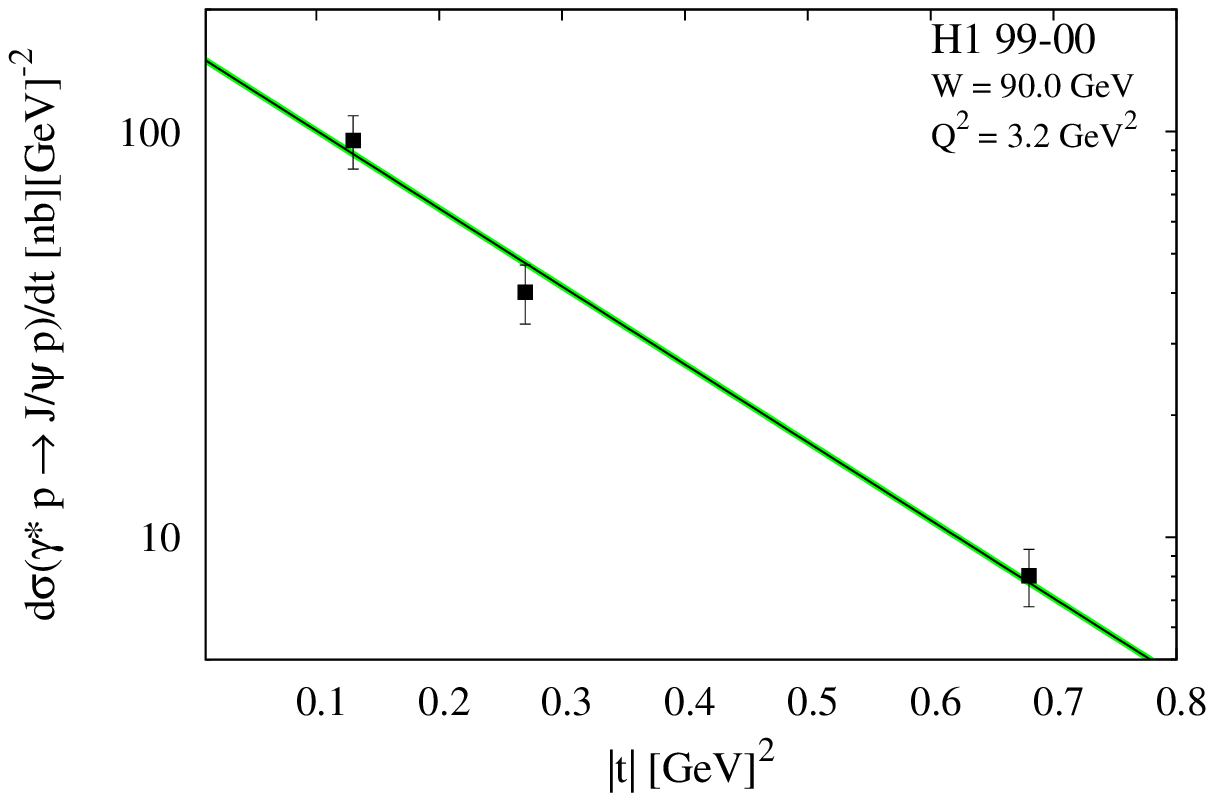}
\includegraphics[clip,scale=0.475]{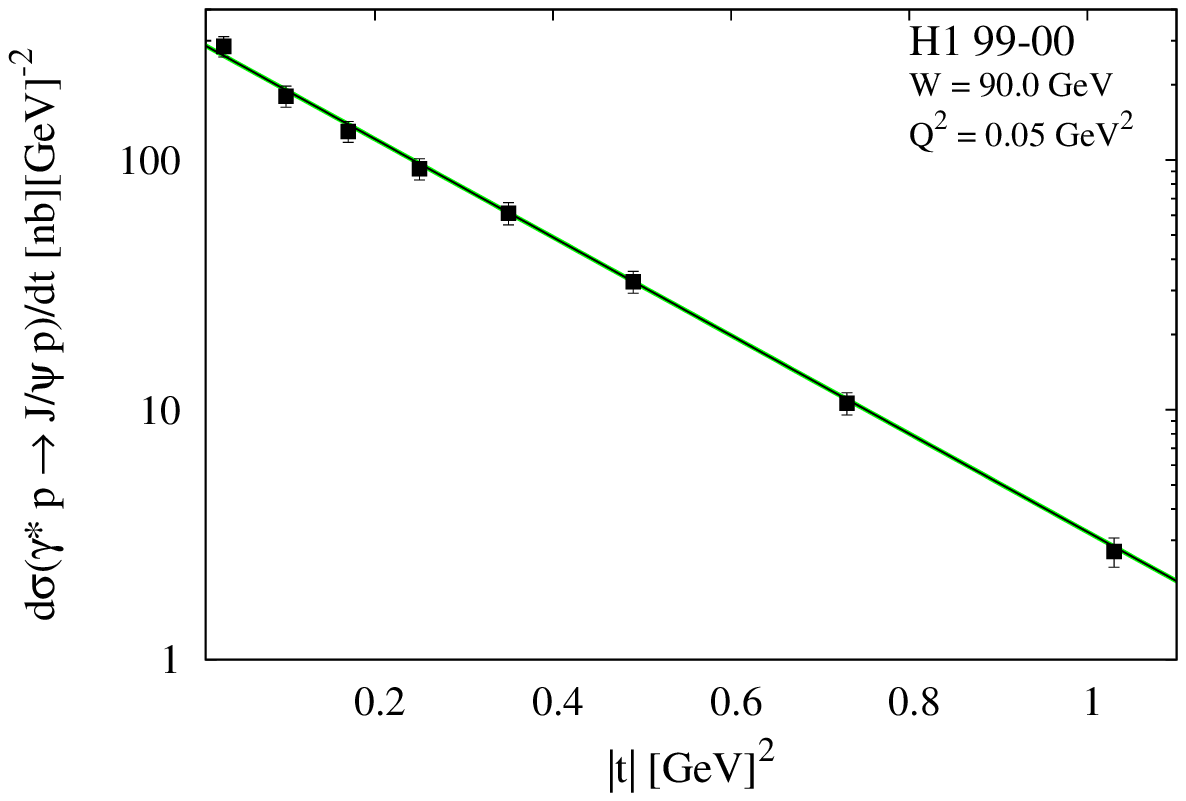}
\includegraphics[clip,scale=0.475]{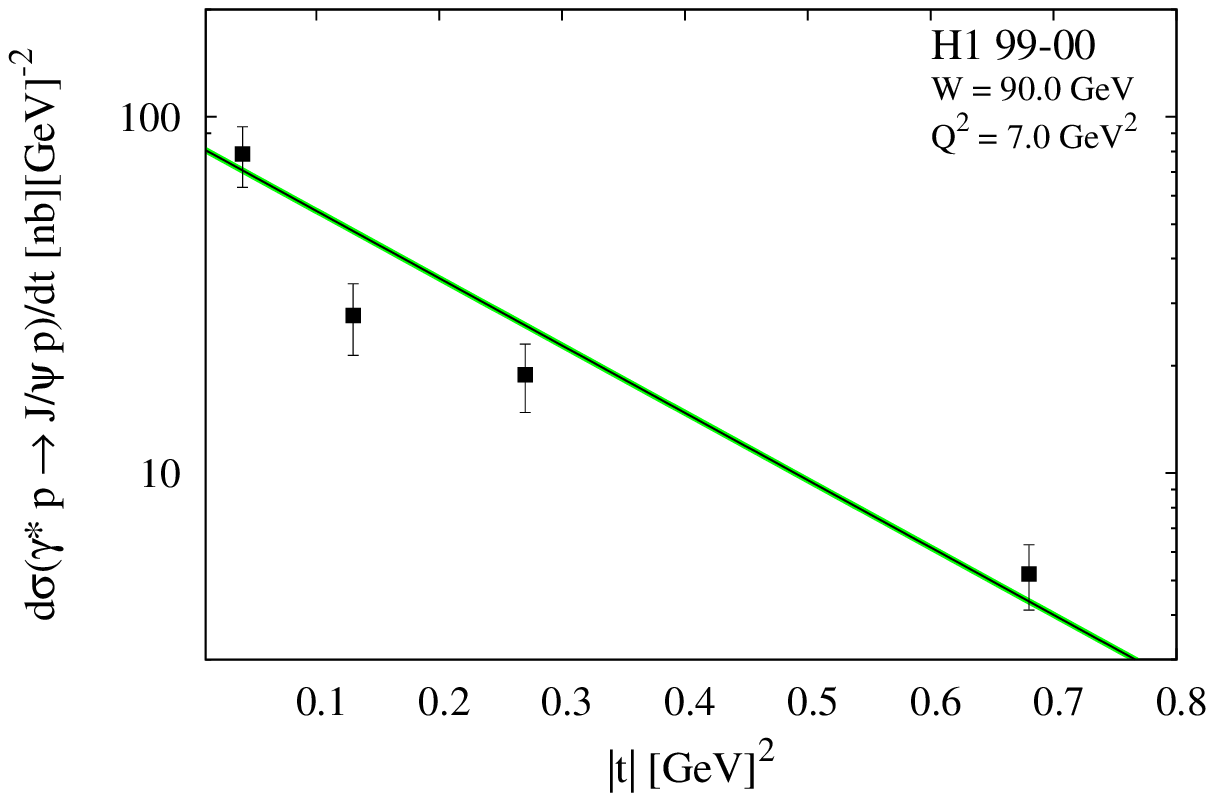}
\includegraphics[clip,scale=0.475]{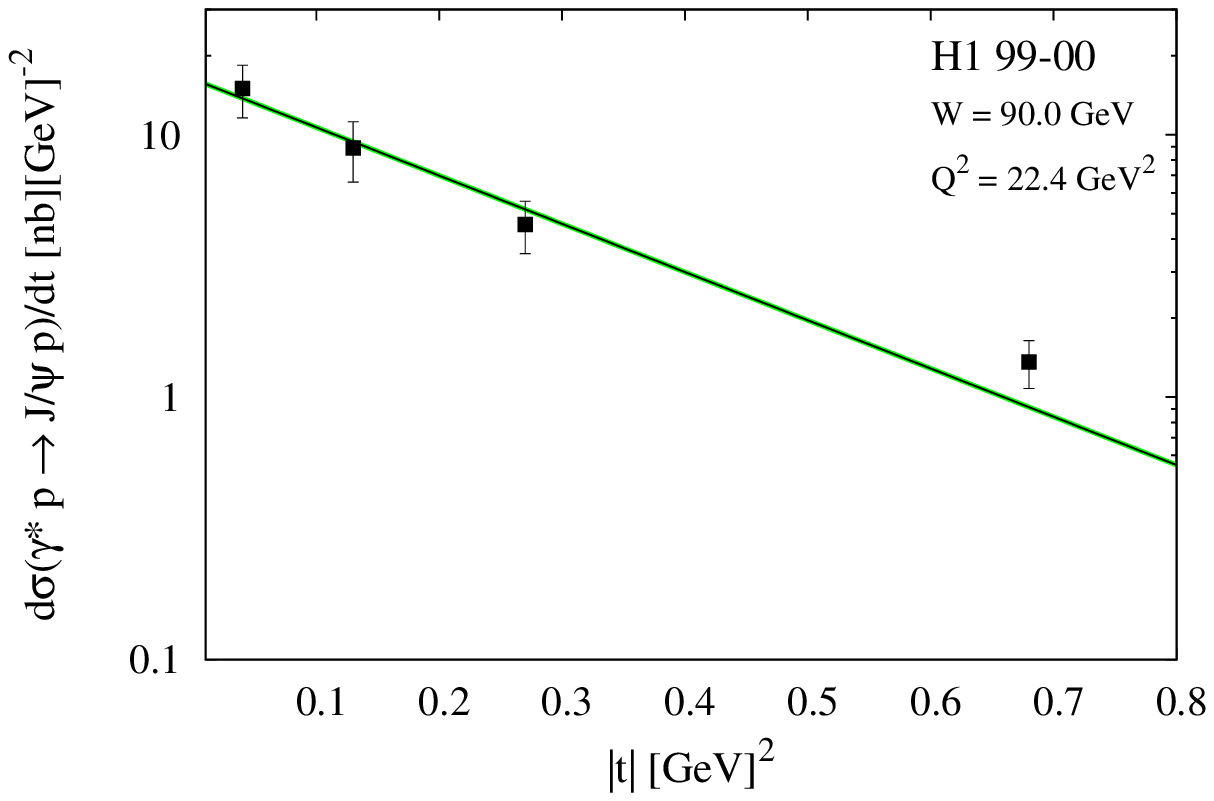}
\includegraphics[clip,scale=0.475]{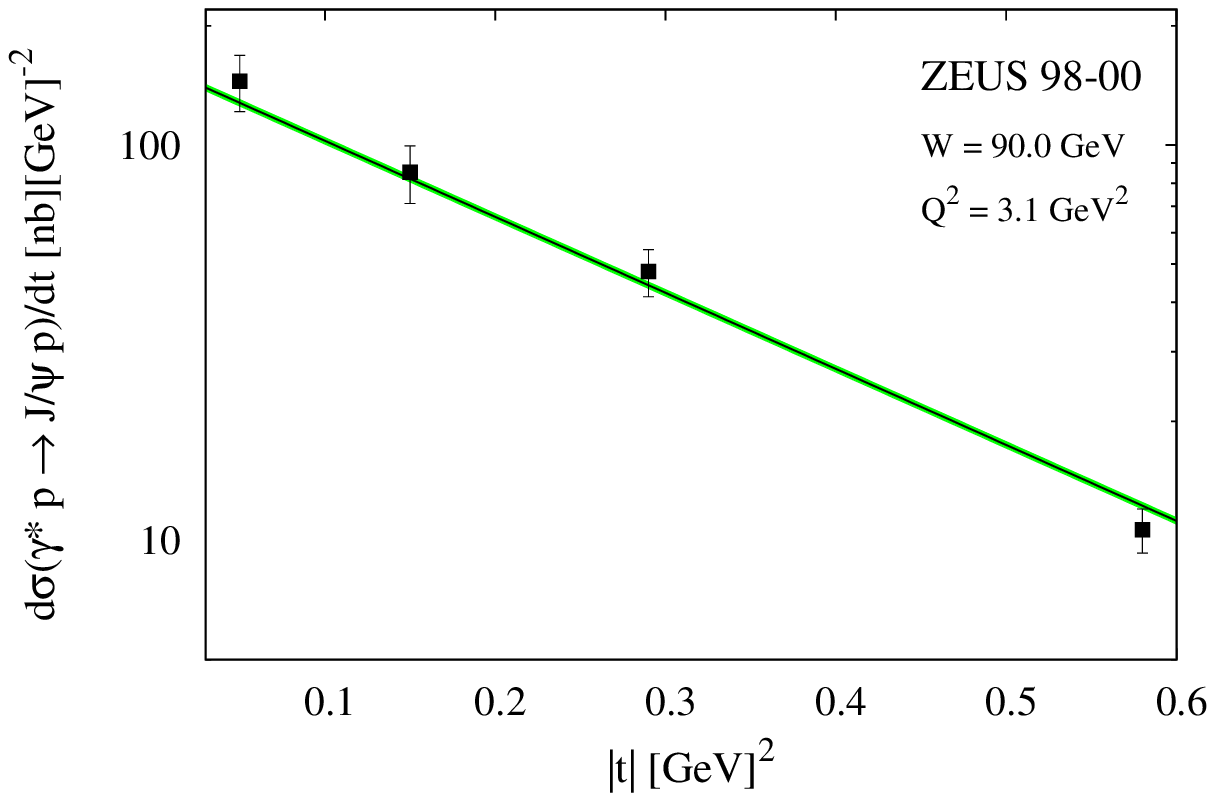}
\includegraphics[clip,scale=0.475]{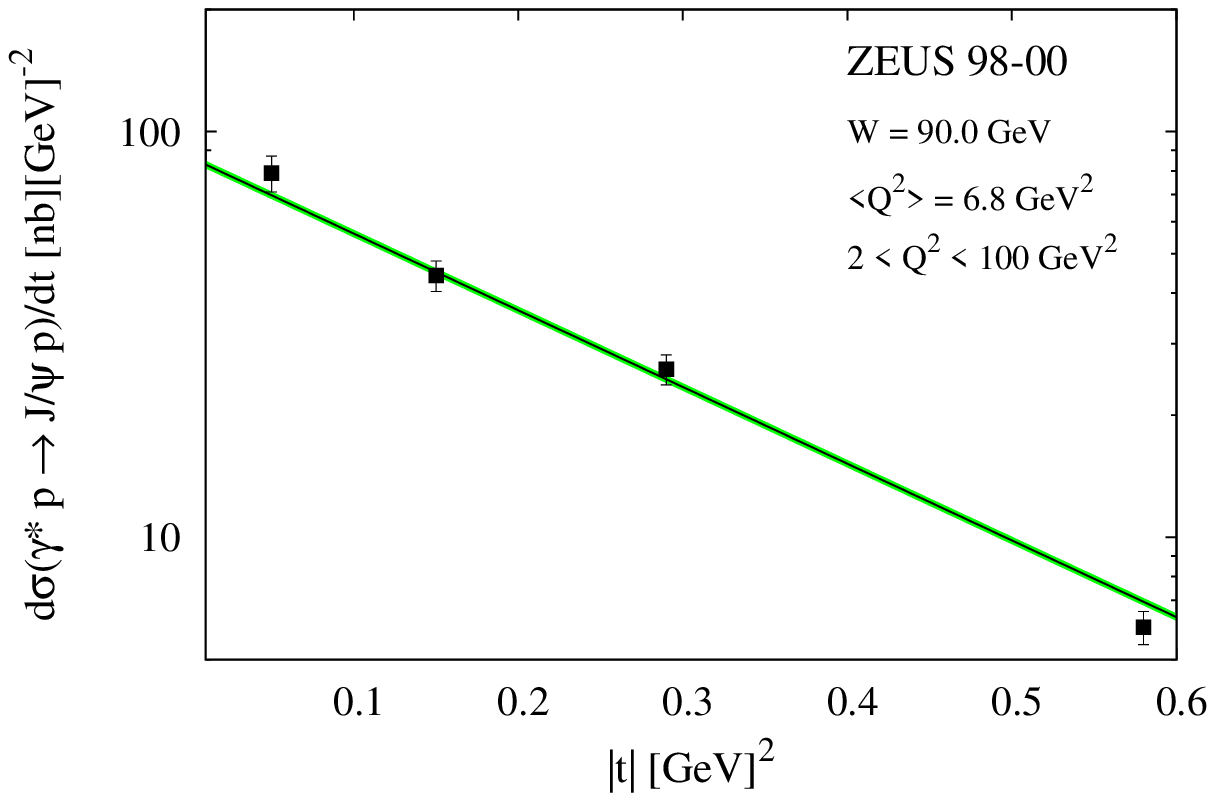}
\includegraphics[clip,scale=0.475]{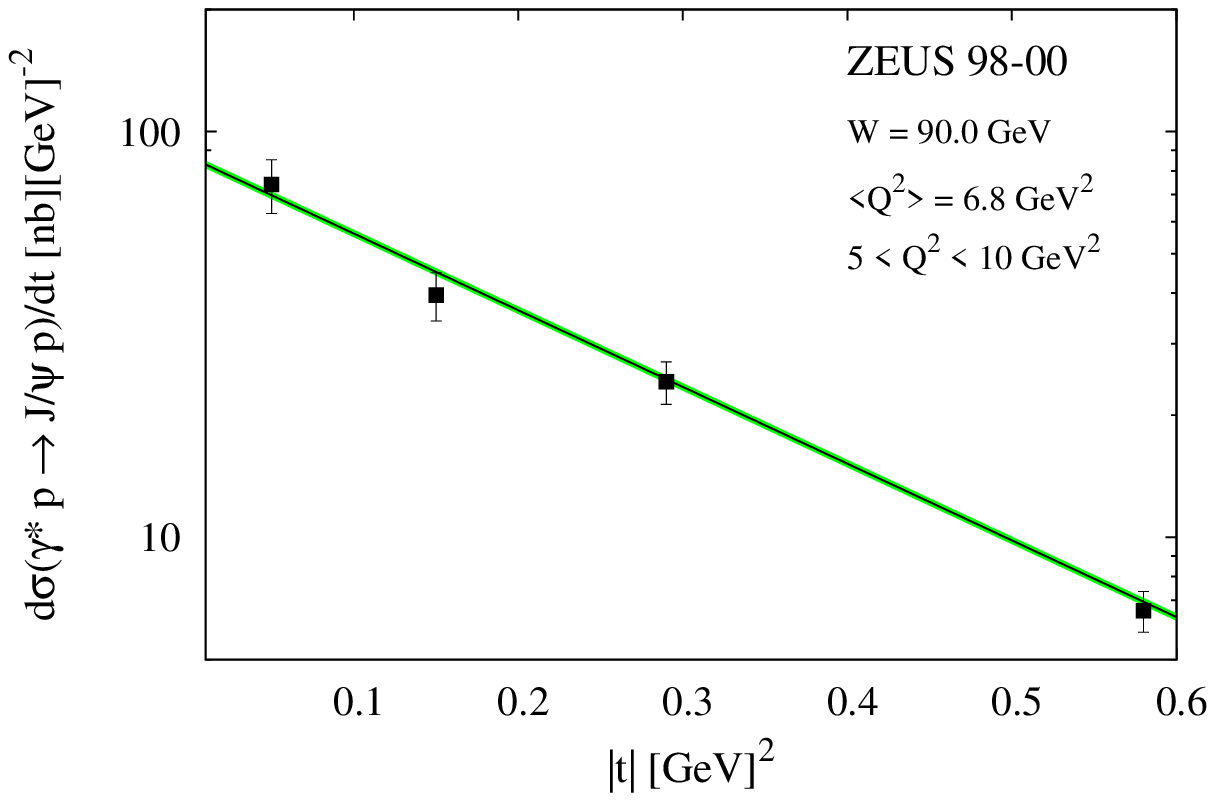}
\includegraphics[clip,scale=0.475]{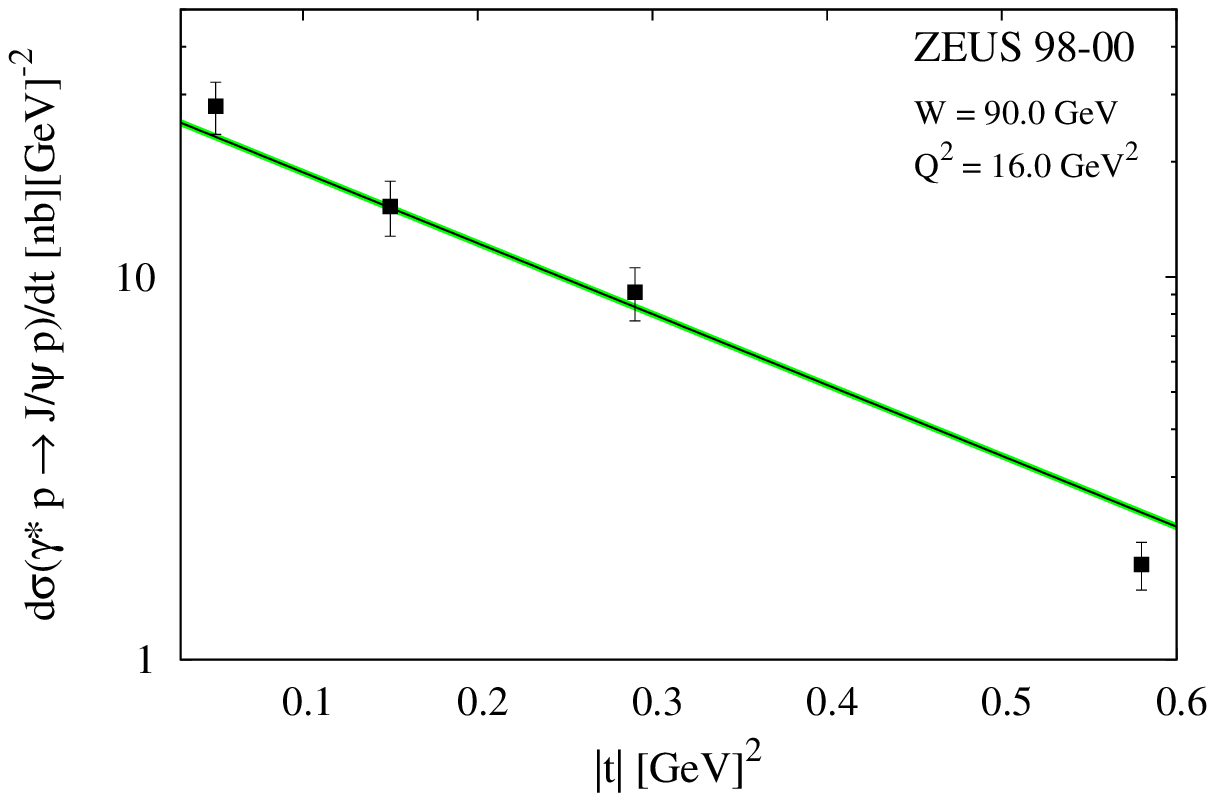}
\end{center}
\vspace{-0.5cm}
\caption{Fit of Eq.~(\ref{eq:dcsdt}) to the H1 and ZEUS data on the differential cross sections %as functions of $|t|$ 
for $\gamma^*p\rightarrow J/\psi p$.}\label{fig:dcsdt_Jpsi}
\end{figure}

\begin{figure}[p,h,t,p,d,!]
\begin{center}
\includegraphics[clip,scale=0.475]{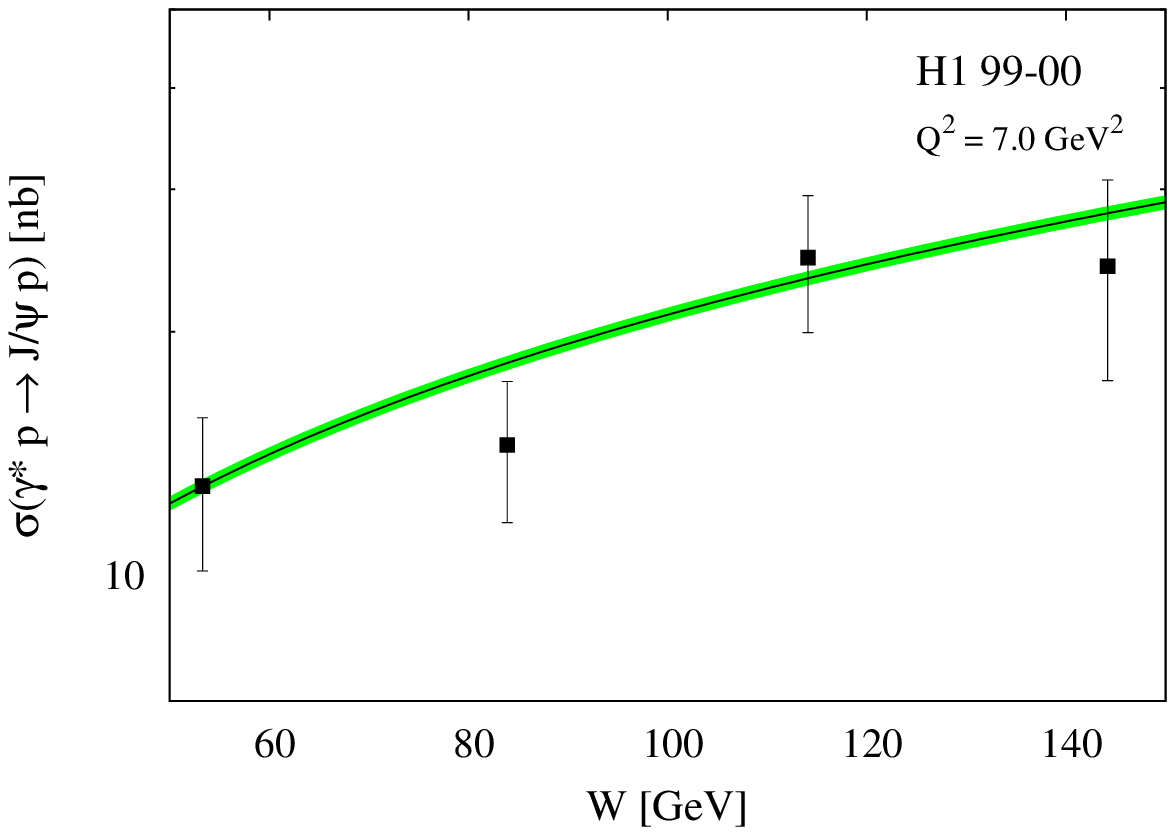}
\includegraphics[clip,scale=0.475]{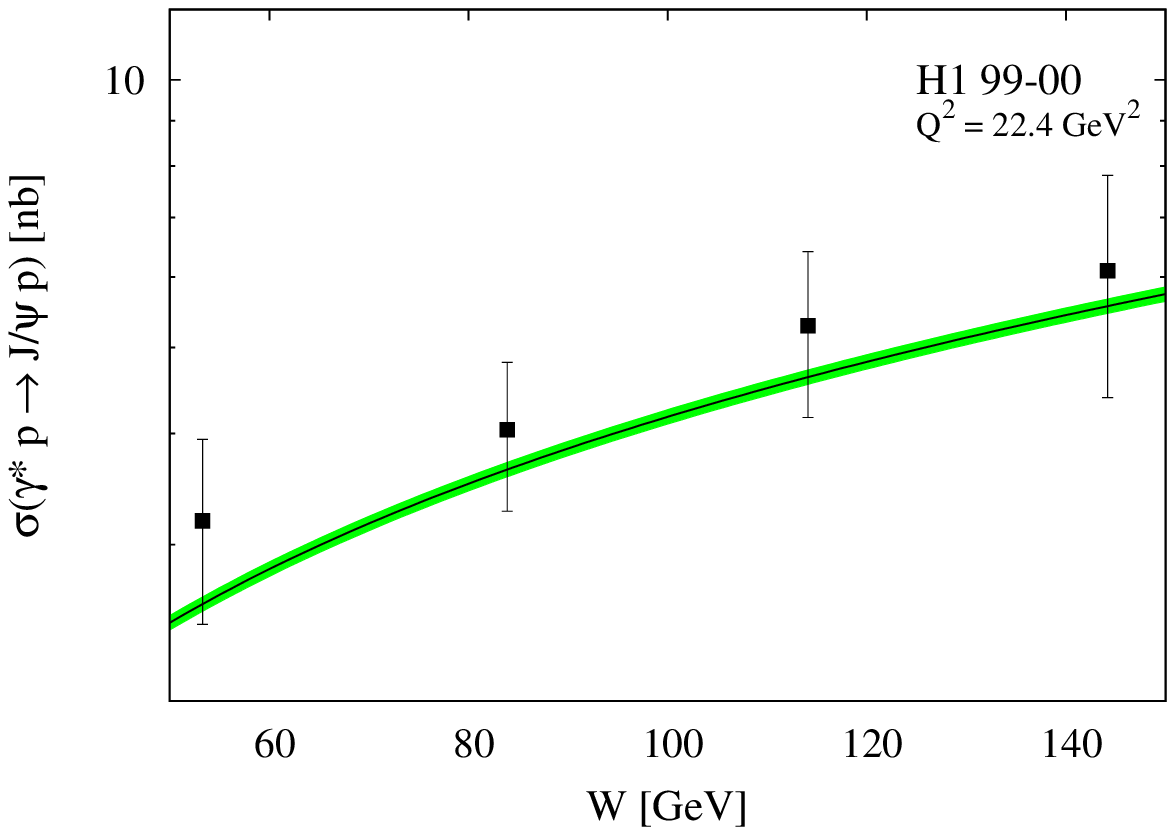}
\includegraphics[clip,scale=0.475]{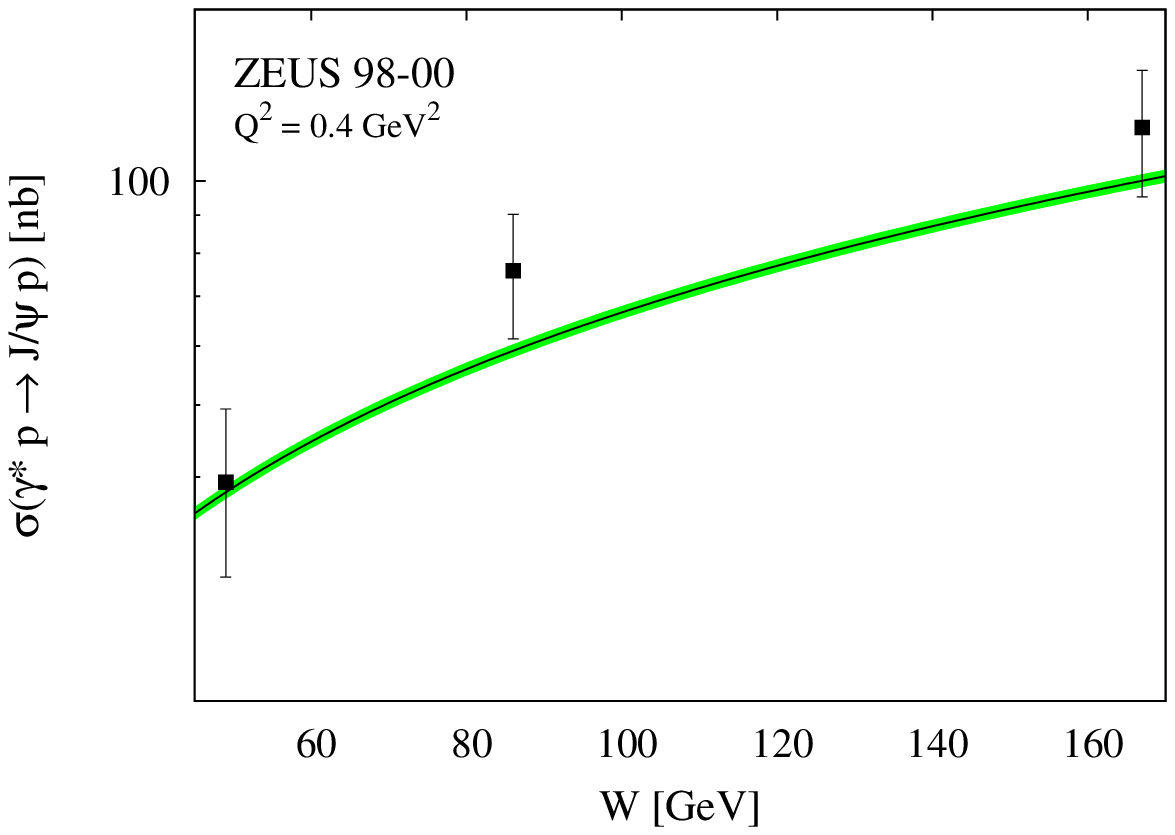}
\includegraphics[clip,scale=0.475]{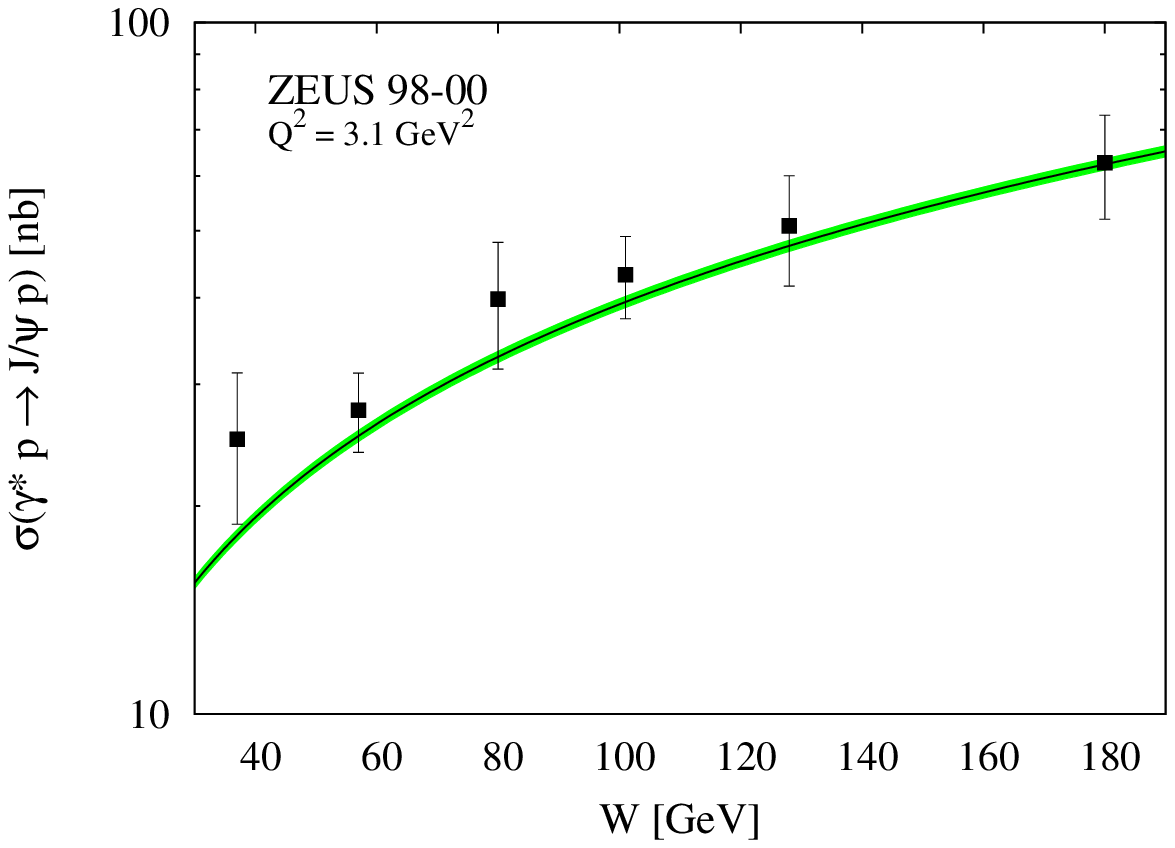}
\includegraphics[clip,scale=0.475]{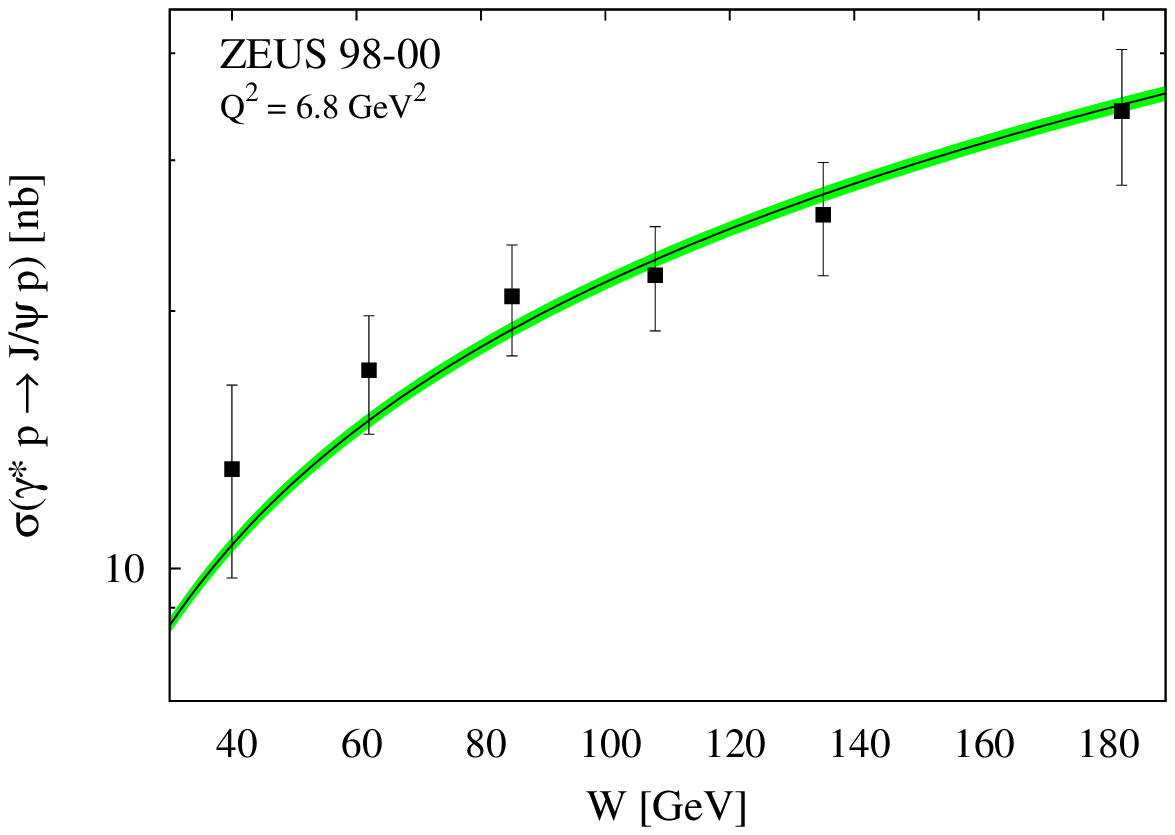}
\includegraphics[clip,scale=0.475]{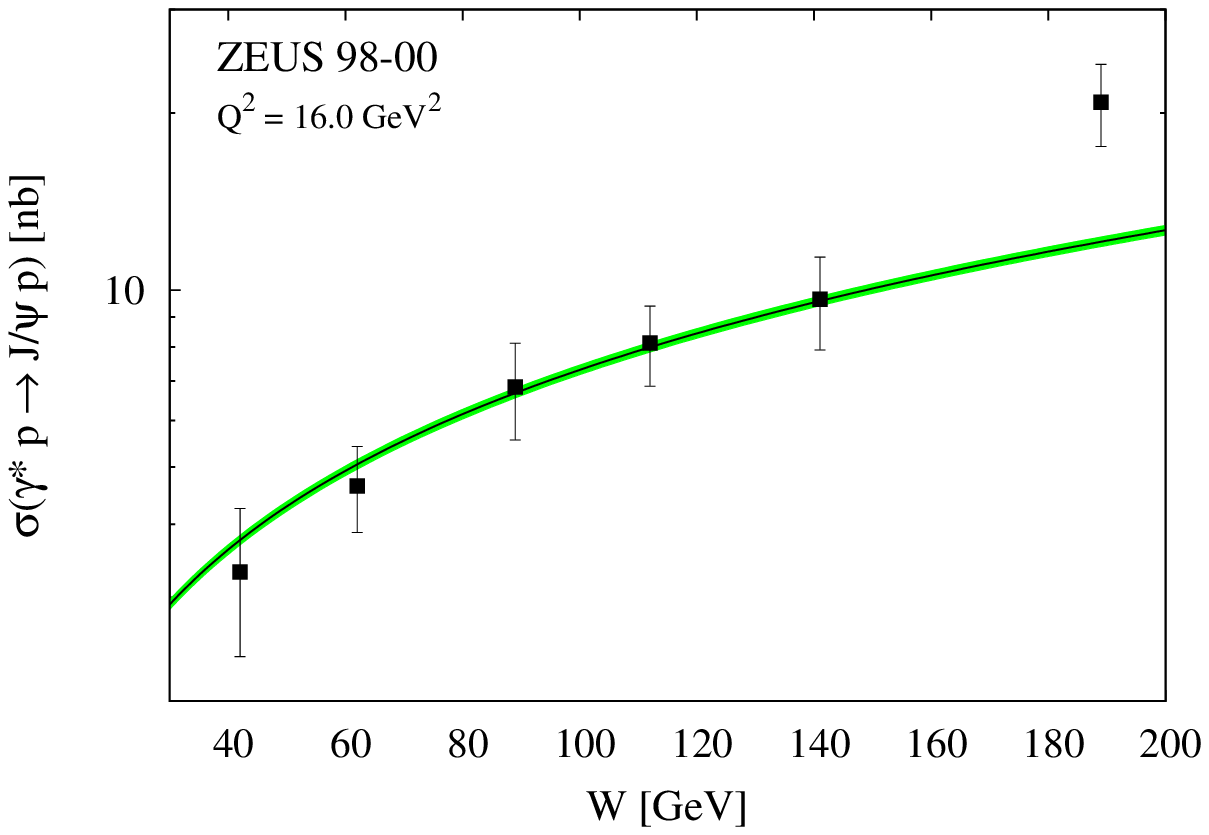}
\includegraphics[clip,scale=0.475]{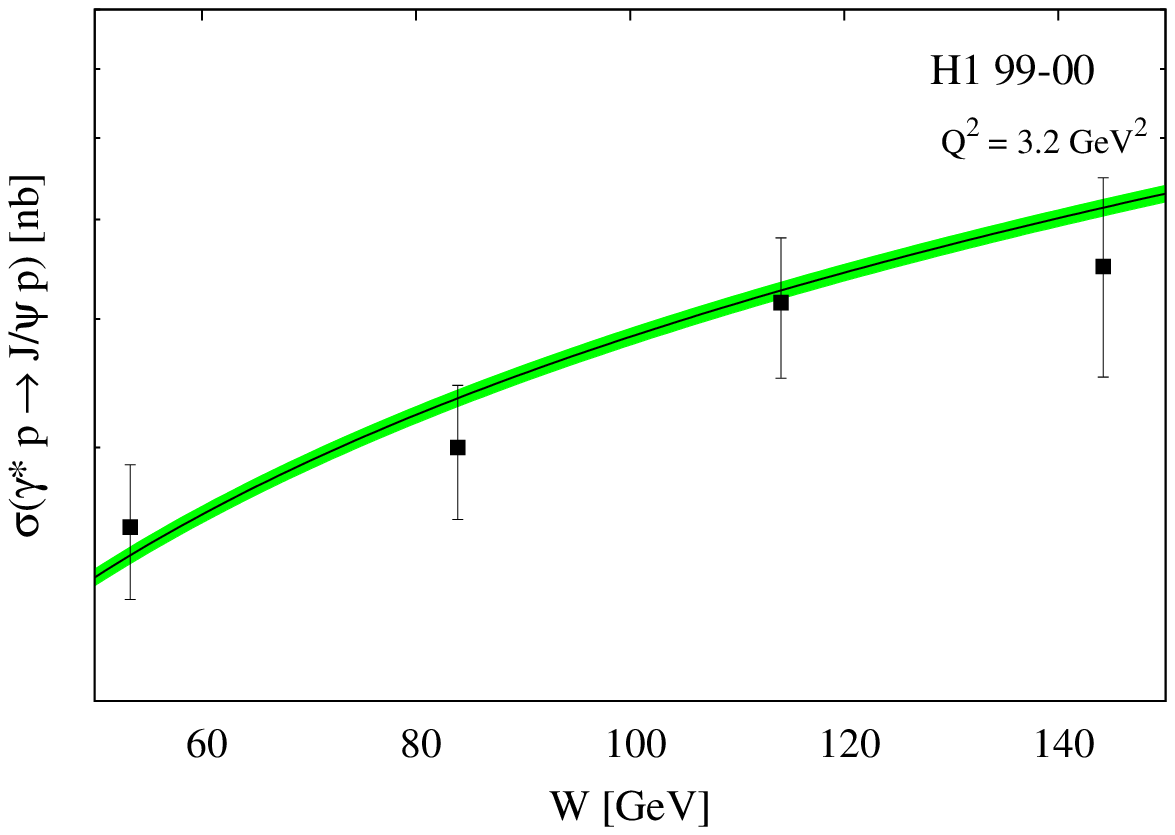}
\end{center}
\vspace{-0.5cm}
\caption{Fit of Eq.~(\ref{eq:cs}) to the H1 and ZEUS data on the integrated cross sections %as functions of $W$ 
for $\gamma^*p\rightarrow J/\psi p$.}\label{fig:csW_Jpsi}
\end{figure}

\begin{figure}[p,h,t,p,d,!]
\begin{center}
\includegraphics[clip,scale=0.475]{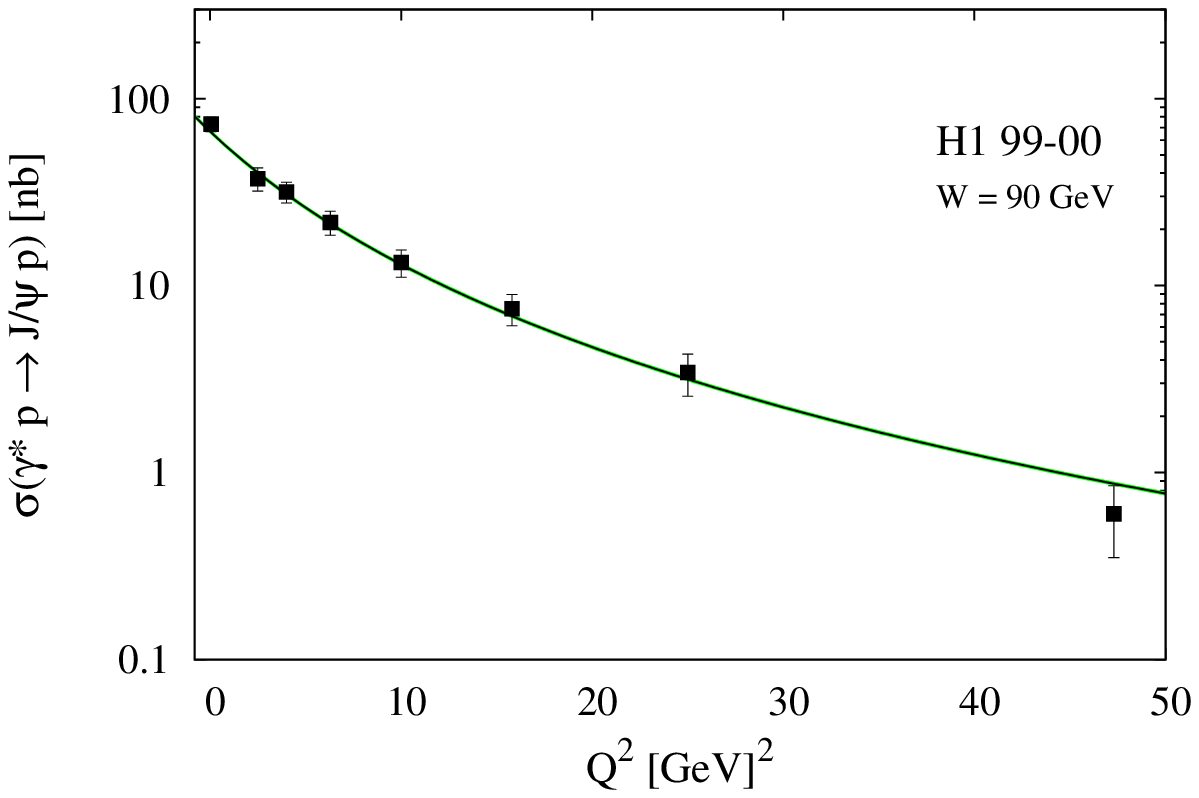}
\includegraphics[clip,scale=0.475]{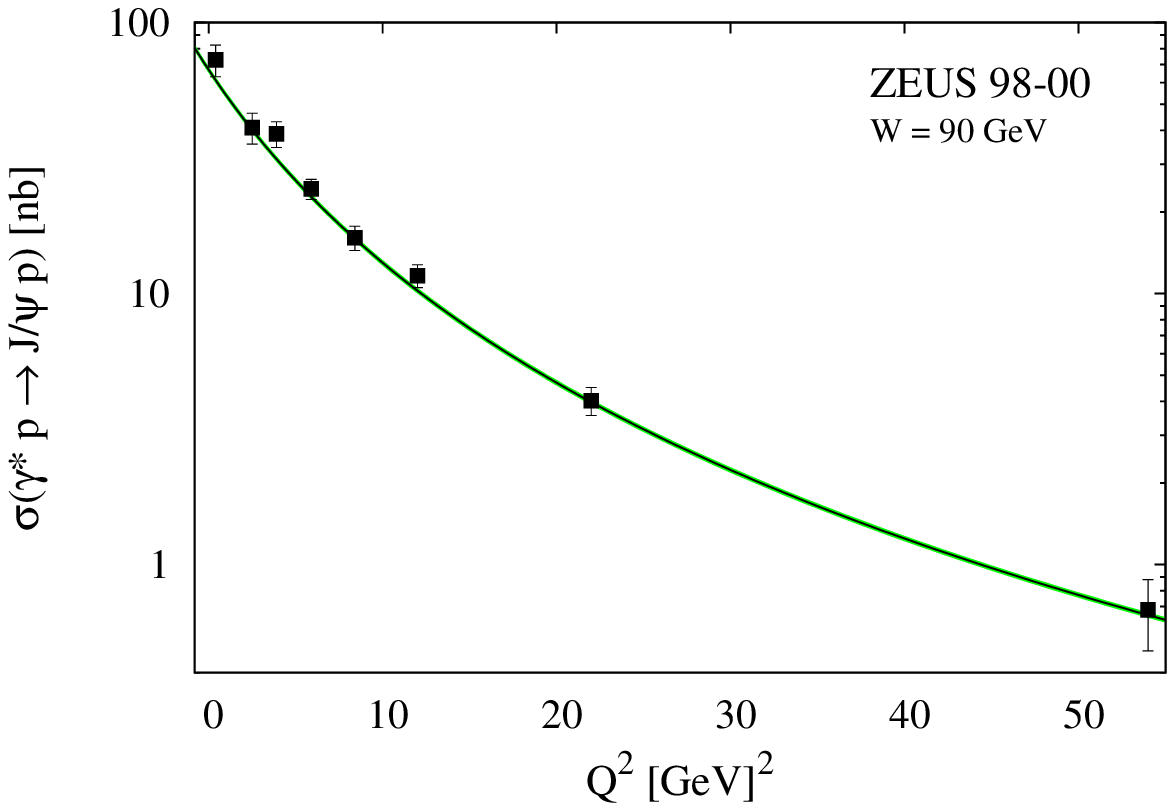}
\end{center}
\vspace{-0.5cm}
\caption{Fit of Eq.~(\ref{eq:cs}) to the H1 and ZEUS data on the integrated cross sections %as functions of $Q^2$ 
for $\gamma^*p\rightarrow J/\psi p$.}\label{fig:csQ2_Jpsi}
\end{figure}

Here we show a fit of the model to the HERA data on $J/\psi$ electroproduction \cite{j1,j2}.
The resulting fit is shown on Figs.~\ref{fig:B_Jpsi}~-~\ref{fig:csQ2_Jpsi}, with the values of the fitted parameters and the relevant $\chi^2/d.o.f.$, given in Table \ref{tab:Jpsi}.
\begin{table}[h,t,p,d,!]
 \caption{Fitted parameters for $J/\psi$ electroproduction.} \label{tab:Jpsi}
 \centering
  \begin{tabular}{c|c|c|c}\hline \hline
    $A_0$ & $Q^2_{0}$ &  n & $\alpha_{0}$ \\\hline 
    29.8 $\pm$ 2.8&2.1 $\pm$ 0.4&1.37 $\pm$ 0.14&1.20 $\pm$ 0.02\\\hline\hline
    $\alpha'$  & a & b  & $\chi^2$/d.o.f.  \\\hline 
    0.17 $\pm$ 0.05 & 1.01 $\pm$ 0.11 & 0.44 $\pm$ 0.08 & 1.12 \\
  \end{tabular}
\end{table}
%%
%%############### Figures ###################
%%

\newpage
%\newpage

\newpage

%%########### phi electroproduction ###################################
%\pagebreak
%\newpage
\subsection {$\phi$ meson electroproduction}
\begin{figure}[p,h,t,p,d,!]
\begin{center}
\includegraphics[clip,scale=0.5]{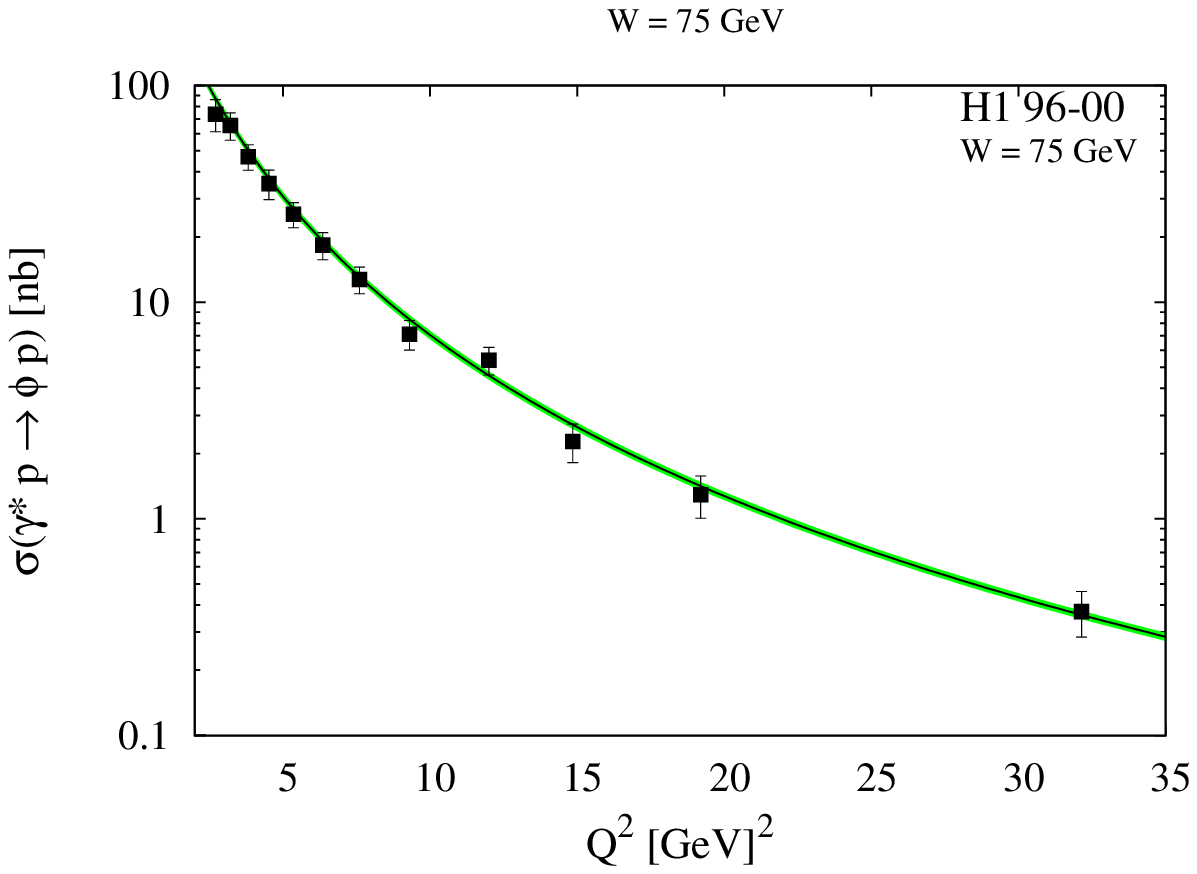}
\end{center}
\vspace{-0.5cm}
\caption{Fit of Eq.~(\ref{eq:cs}) to the H1 data on the integrated cross sections %as a function of $Q^2$
for $\gamma^*p\rightarrow\phi p$.} \label{fig:csQ2_phi}
\end{figure}

\begin{figure}[p,h,t,p,d,!]
\begin{center}
\includegraphics[clip,scale=0.475]{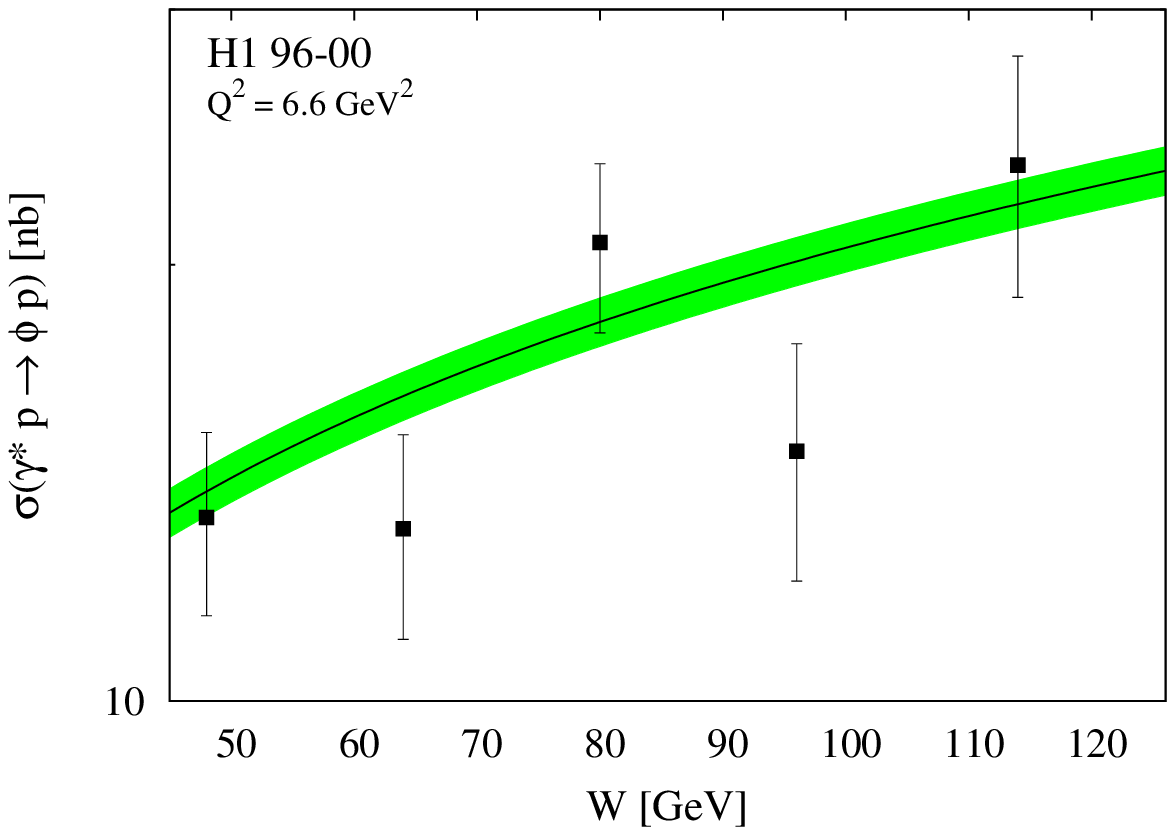}
\includegraphics[clip,scale=0.475]{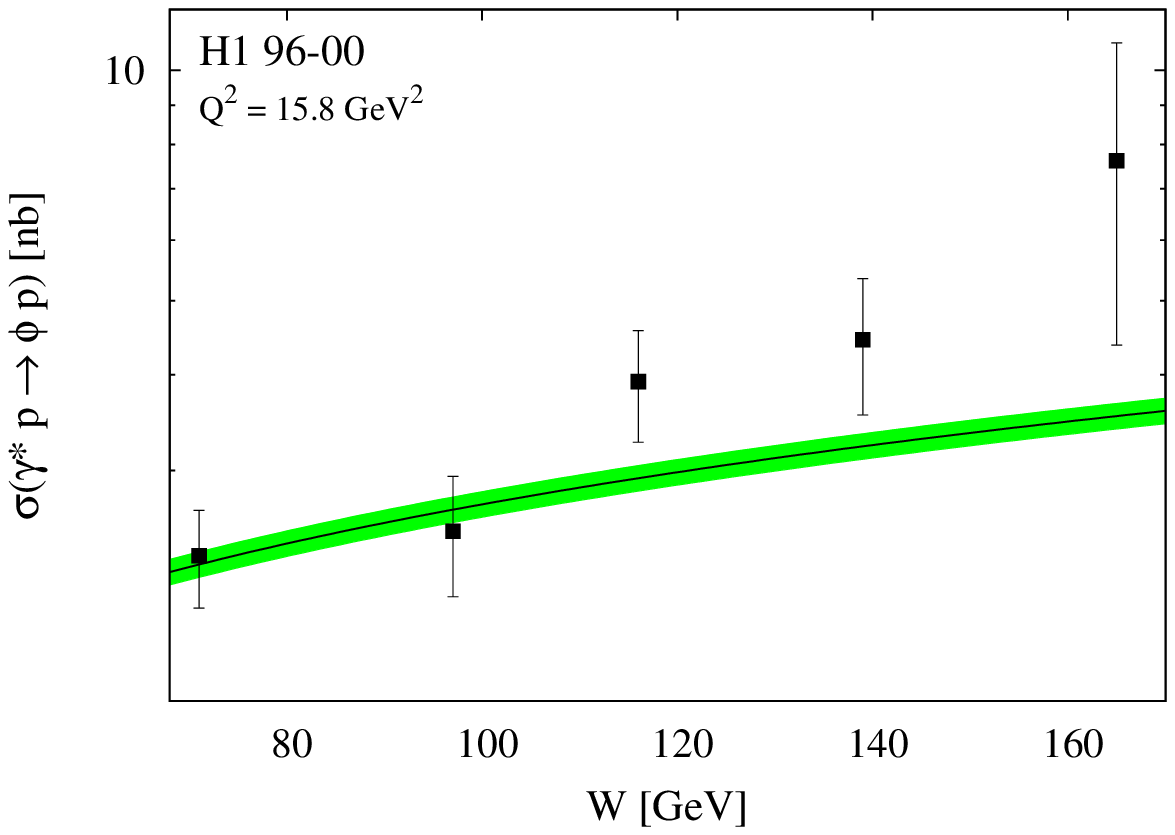}
\includegraphics[clip,scale=0.475]{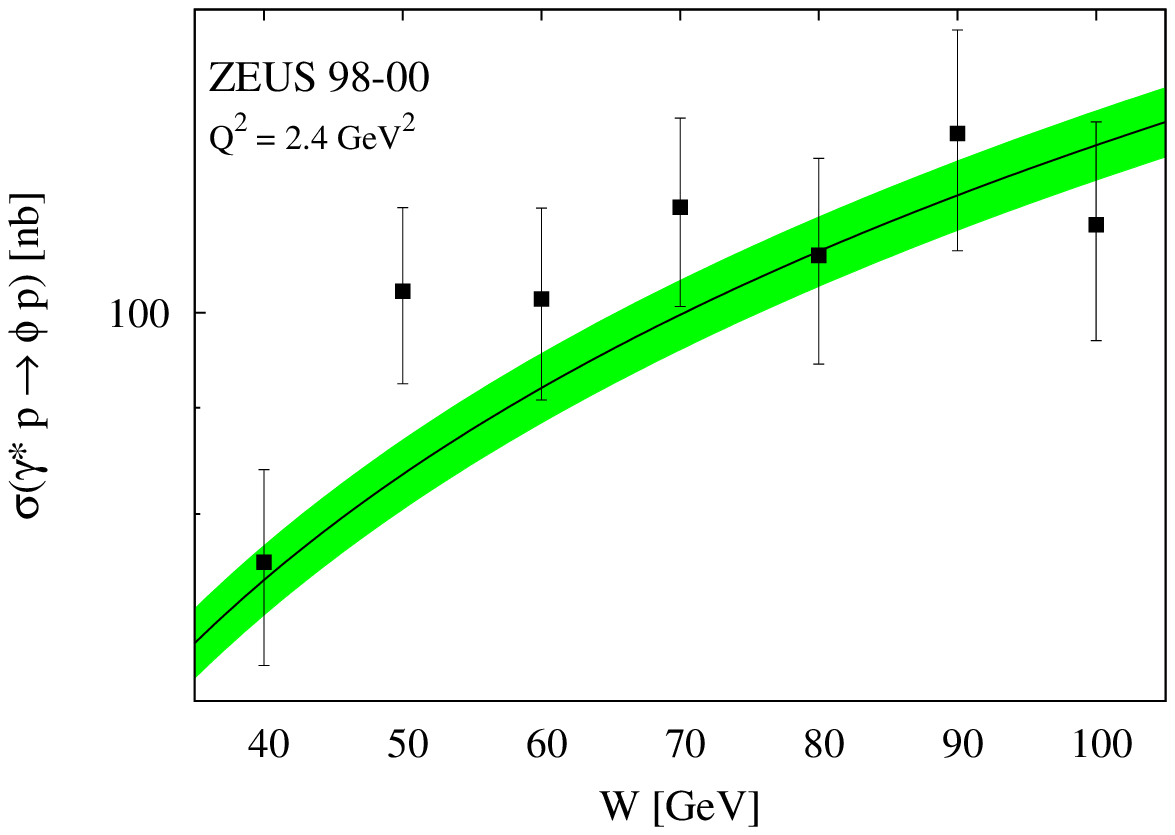}
\includegraphics[clip,scale=0.475]{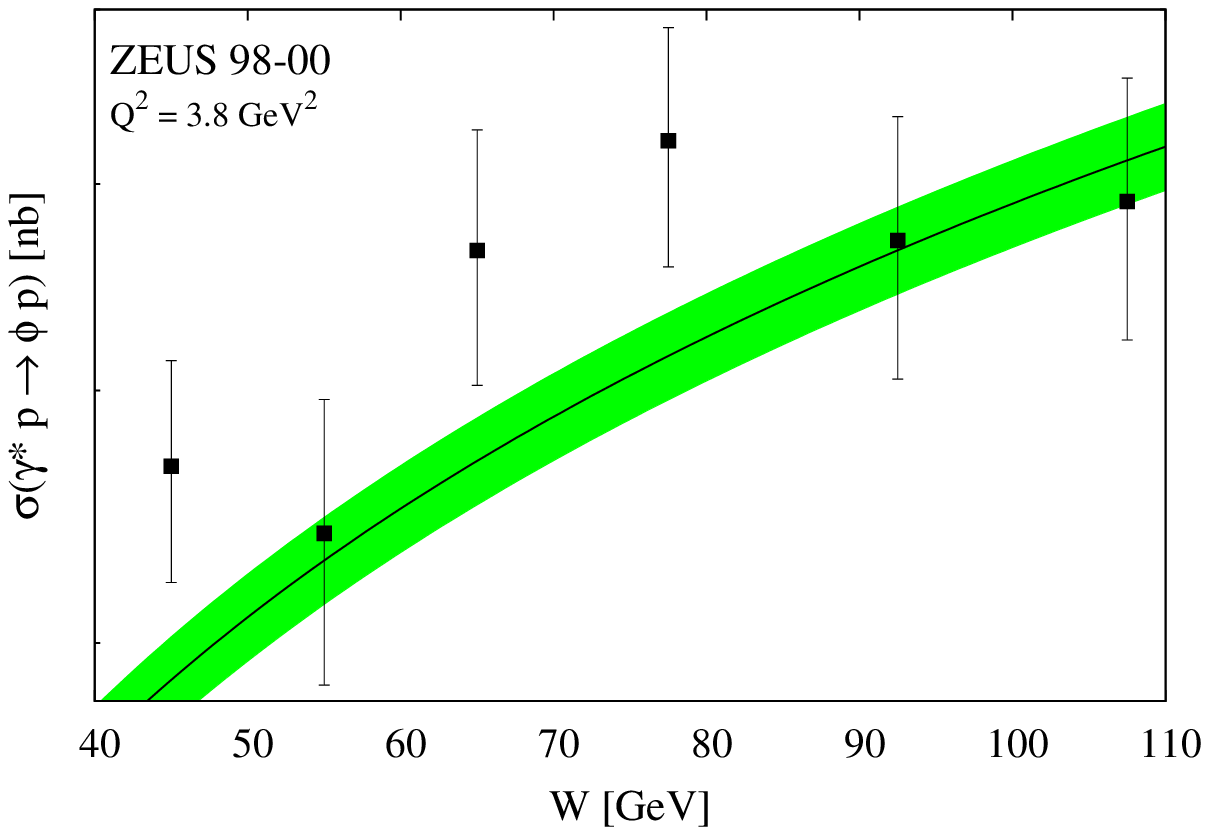}
\includegraphics[clip,scale=0.475]{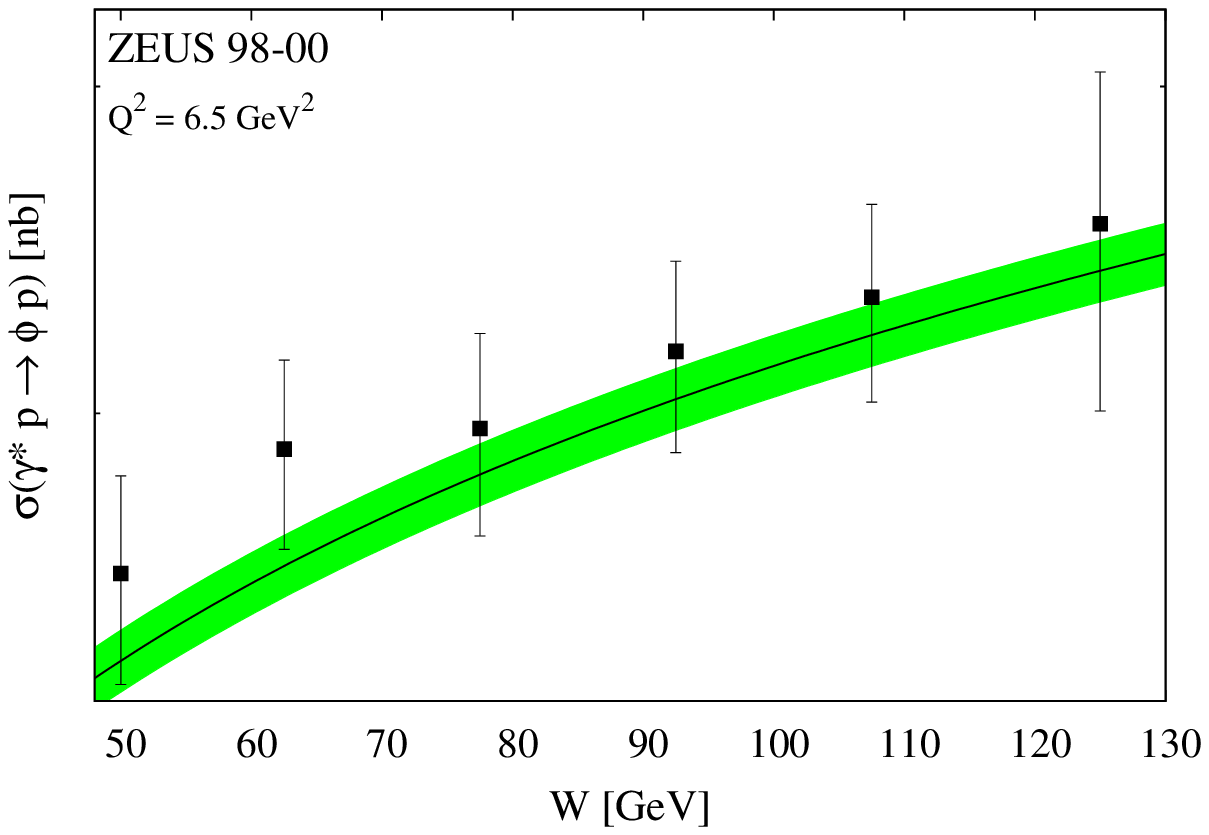}
\includegraphics[clip,scale=0.475]{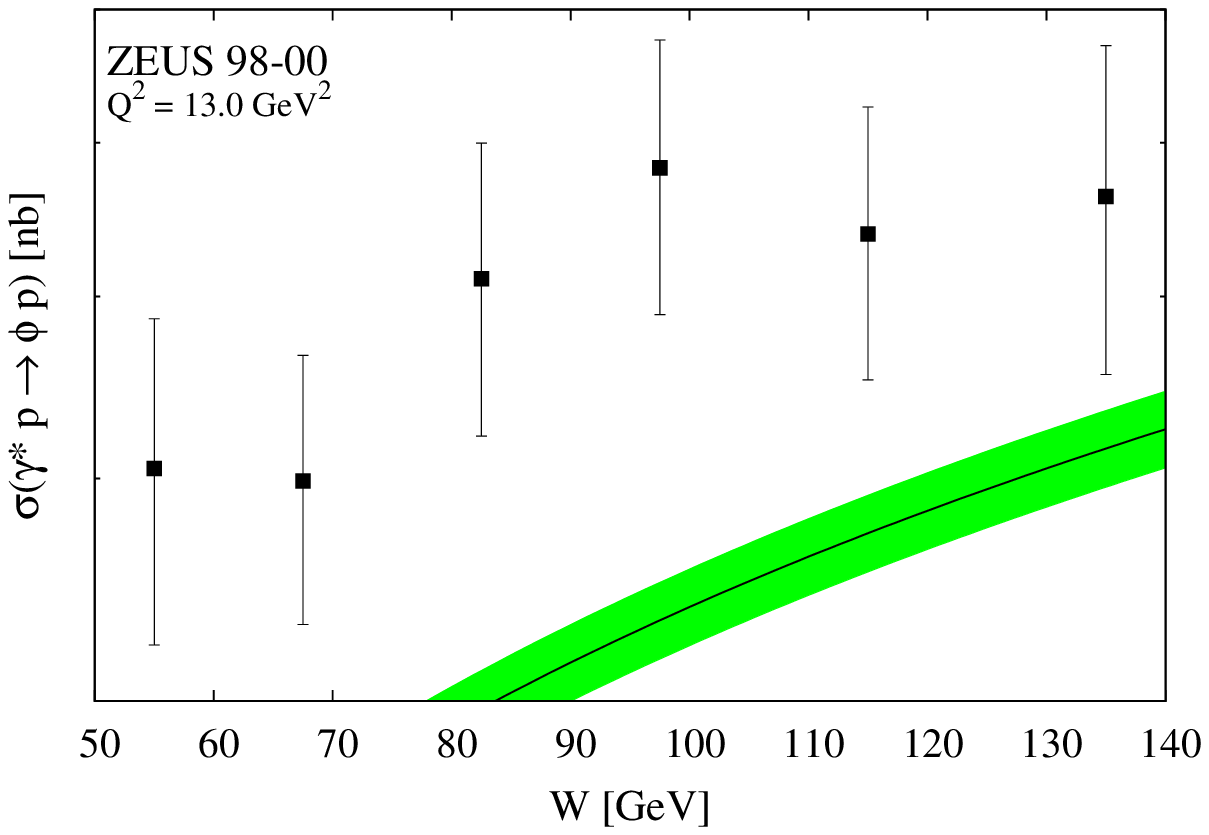}
\includegraphics[clip,scale=0.475]{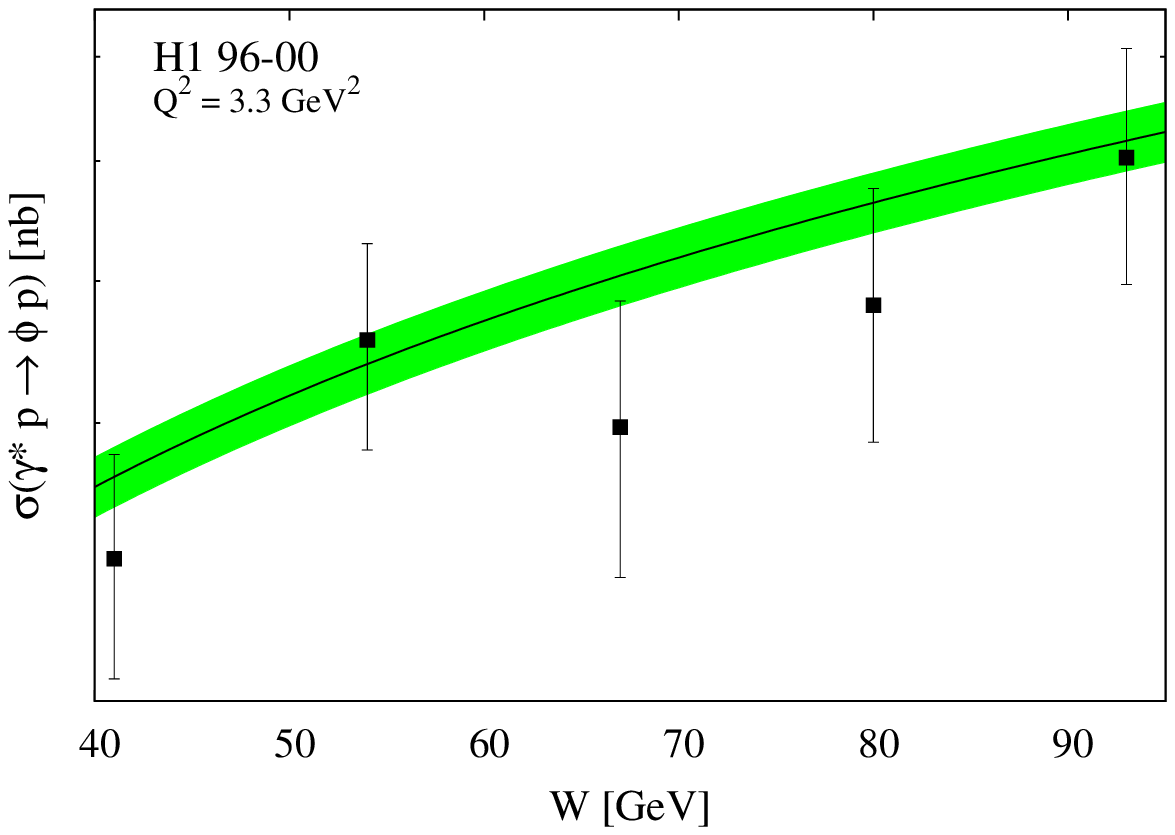}
\end{center}
\vspace{-0.5cm}
\caption{Fit of Eq.~(\ref{eq:cs}) to the H1 and ZEUS data on the integrated cross sections %as functions of $W$
for $\gamma^*p\rightarrow \phi p$.}\label{fig:csW_phi}
\end{figure}

\begin{figure}[p,h,t,p,d,!]
\begin{center}
\includegraphics[clip,scale=0.475]{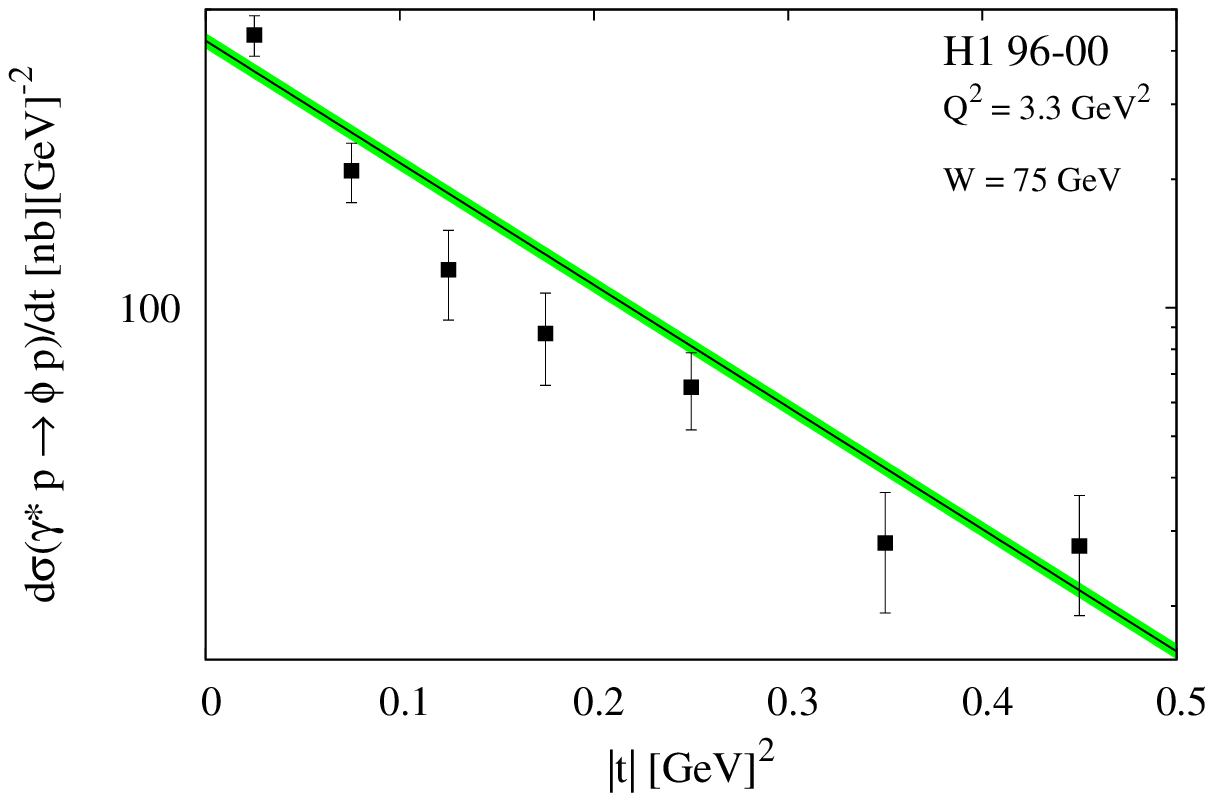}
\includegraphics[clip,scale=0.475]{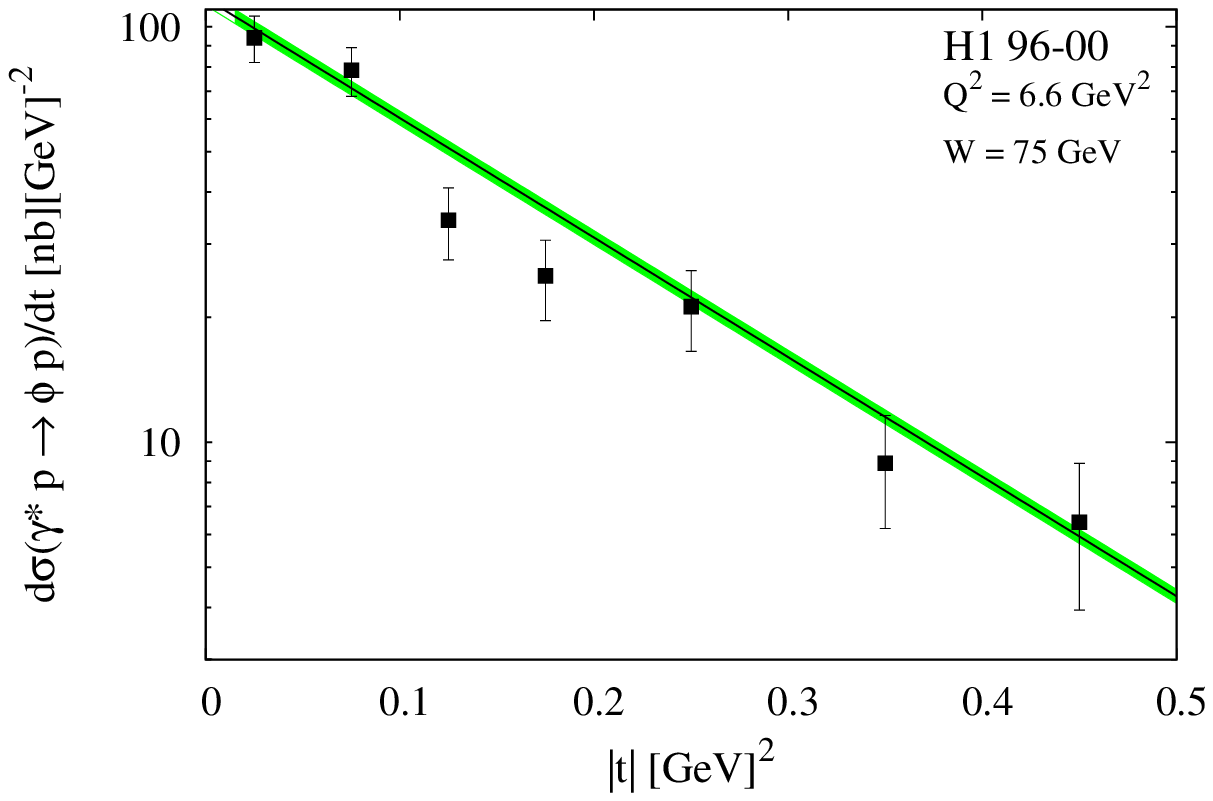}
\includegraphics[clip,scale=0.475]{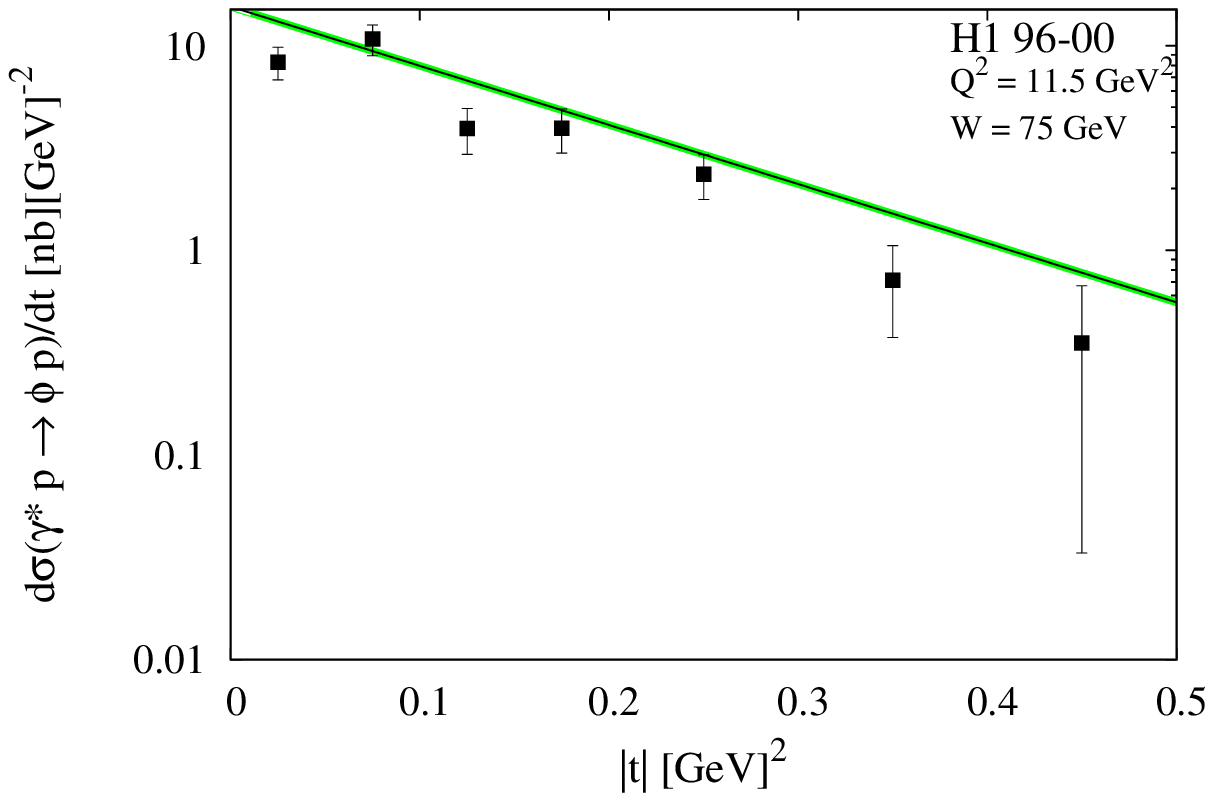}
\includegraphics[clip,scale=0.475]{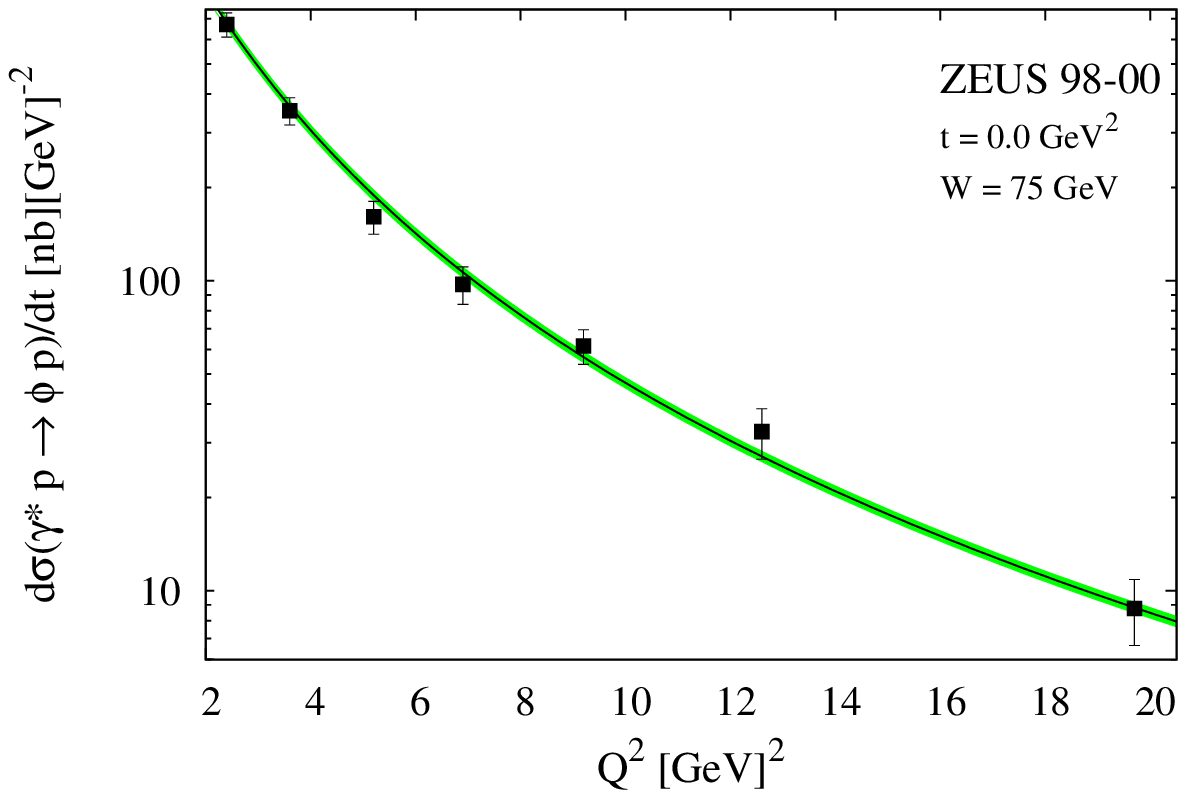}
\end{center}
\vspace{-0.5cm}
\caption{Fit of Eq.~(\ref{eq:dcsdt}) to the H1 and ZEUS data on the differential cross sections %as functions of $|t|$ (and $Q^2$) 
for $\gamma^*p\rightarrow\phi p$.}\label{fig:dcsdt_phi}
\end{figure}

Here we present a fit of the model to the HERA data on $\phi$ electroproduction, see~\cite{r1,phi_zeus}.
%Take note that for $\phi$ we have $M_V = 1.0956$ GeV. 
%
The resulting fit is shown on Figs.~\ref{fig:csQ2_phi}~-~\ref{fig:dcsdt_phi}, with the values of the fitted parameters and the relevant $\chi^2/d.o.f.$, given in Table \ref{tab:phi}.
\begin{table}[h,t,p,d,!]
 \caption{Fitted parameters for $\phi$ electroproduction.} \label{tab:phi}
 \centering
   \begin{tabular}{c|c|c|c}     \hline \hline
     $A_0$ & $Q^2_{0}$ &  n & $\alpha_{0}$ \\ \hline
     37.2 $\pm$ 0.7 & 1.7 $\pm$ 0.2 & 1.48 $\pm$ 0.13&1.15 $\pm$ 0.27\\\hline\hline
          $\alpha'$  & a & b  & $\chi^2$/d.o.f.  \\\hline
     0.10 $\pm$ 0.05 & -0.09 $\pm$ 0.18 & 2.18 $\pm$ 0.34 & 0.05 \\
   \end{tabular}
\end{table}
%{\tiny * this parameter does not affects significantly our fitting procedure, so it has been fixed as shown above.}
%%
%%############### Figures ###################
%%
%\newpage

\newpage

\newpage

%%########### Conclusion ###################################
\section{Conclusions and discussion} \label{sec:Conclusions}
Our study shows that DVCS and each VM electroproduction reactions can be  separately fitted within the present model with a single Pomeron trajectory.

Some problems arise for $\phi$ production at high virtualities $Q^2\gtrsim13$GeV$^2$ (see~Fig.~\ref{fig:csW_phi}). Also there is problems with fitting photoproduction and electroproduction data simultaneously for $\rho$ (and also $\phi$) production. Light mesons seem to be sensitive to the transition from the soft to the hard regime. So, these may serve as hints that it is not enough to use the current model with one-component Pomeron to describe the whole spectrum of VMP and DVCS data together. For this reason we may need a model with two components Pomeron (i.e. the amplitudes with soft and hard terms), see (see~Appendix~\ref{app:Hard and soft}).

The present fits can serve as a basis (input parameters) for a global fit of the model with both soft and hard terms (see~Appendix~\ref{app:Hard and soft}) with a unique set of parameters for all reactions. Such work is in progress \cite{FFJS}.

Among open problems we mention the need for theoretical arguments to define and constrain the form of the $\widetilde Q^2-$dependent factor in amplitude Eq.~(\ref{Amplitude0}) (or the $H$  factors in front of the soft and hard terms, Eqs.~(\ref{eq:Amplitude_hs}),~(\ref{eq:cs_hs}),~(\ref{eq:dcsdt_hs}) and (\ref{eq:H-factors}) in Appendix~\ref{app:Hard and soft}). These constrains may come from QCD evolution and/or from the unitarity condition.

%%########### Acknowledgements ###################################
%\newpage
\vskip 1cm \centerline{\bf Acknowledgements}
L.J. thanks the Department of Physics of the University of Calabria and the Istituto Nazionale di Fisica
Nucleare - Gruppo Collegato di Cosenza, where part of this work was done, for their hospitality and support.
He was supported also by the Project ``Matter under extreme conditions'' of the Ukrainian National Academy of Sciences.

%%########### Appendix ###################################
\appendix 
\section{A two-component Pomeron} \label{app:Hard and soft}

 Following Refs.~\cite{L, DL}, we comprise soft and hard dynamics within a single Pomeron, which however has two components -
 a soft and a hard one, with relative $\widetilde Q^2-$dependent weights.
 %$H_s(\widetilde Q^2)$ and $H_h(\widetilde Q^2)$.
 These weights are constructed such as to provide the right balance between the soft and hard components, i.e. as $\widetilde Q^2$ increases,
 the weight of the hard term increases, and v.v. These weights
 %$H_s(\widetilde Q^2)-$depebdent factors
 do not affect the Reggeometric form of the model.

% The parameters will be fitted to the data. The spirit of the multi-component Pomeron,
% in a sense, is the spirit of QCD, although there is no direct connection between the present "multicomponent" model and the "Lipatov Pomeron".

The relevant scattering amplitude is 
\begin{eqnarray}
 %$$ 
 \label{eq:Amplitude_hs}
 A(s,t,Q^2,{M_v}^2)=\frac{\widetilde{A_s}}{\Bigl(1+\frac{\widetilde{Q^2}}{{Q_s^2}}
 \Bigr)^{n_s}}
 e^{-i\frac{\pi}{2}\alpha_s(t)}\Bigl(\frac{s}{s_{0s}}\Bigr)^{\alpha_s(t)}  e^{2\Bigl(\frac{a_s}{\widetilde{Q^2}}+\frac{b_s}{2m_N^2}\Bigr)t}\\
 %$$
 %\begin{equation}
 +\frac{\widetilde{{A_h}}\Bigl({\widetilde{Q^2}\over{Q_h^2}}\Bigr)}{{\Bigl(1+\frac{\widetilde{Q^2}}{{Q_h^2}}\Bigr)}^{n_h+1}}e^{-i\frac{\pi}{2}\alpha_h(t)}\Bigl(\frac{s}{s_{0h}}\Bigr)^{\alpha_h(t)} e^{2\Bigl(\frac{a_h}{\widetilde{Q^2}}+\frac{b_h}{2m_N^2}\Bigr)t}.\nonumber
 %\end{equation}  
\end{eqnarray}
Here $\widetilde{A_s}$ and $\widetilde{A_h}$ are normalization factors, $Q_s^2$ and $Q_h^2$ are soft and hard scales for virtuality, $n_s$ and $n_h$ are free parameters to be fitted, as well as four parameters $a_s$, $b_s$, $a_h$ and $b_h$, $s_{0s}$ and $s_{h0}$ are soft and hard scales for the squared energy, $\alpha_{s}(t)$ and $\alpha_{h}(t)$ are Regge trajectories for the soft and hard components of the Pomeron.

 Substituting the amplitude (\ref{eq:Amplitude_hs}) to Eqs.~(\ref{eq3}) and (\ref{integrated}),  we obtain the  differential and  integrated cross sections which are, respectively,

 \begin{eqnarray}
\label{eq:dcsdt_hs}
%$$
    &{d\sigma \over {d|t|}}&=H_s^2e^{2\{L_s(\alpha_s(t)-1)+\mathfrak{g_s}t\}}+H_h^2e^{2\{L_h(\alpha_h(t)-1)+\mathfrak{g_h}t\}}\\
% $$
%  \begin{equation}
&&+2H_sH_he^{\{L_s(\alpha_s(t)-1)+L_h(\alpha_h(t)-1)+(\mathfrak{g_s}+\mathfrak{g_h})t\}}\cos\Bigl({\pi\over2}(\alpha_s(t)-\alpha_h(t))\Bigr),\nonumber
%\end{equation}
%  
\end{eqnarray}

and
\begin{eqnarray}
\label{eq:cs_hs}
     &\sigma=\frac{H_s^2e^{2\{L_s(\alpha_{0s}-1)\}}}{2(\alpha_s'L_s+\mathfrak{g_s})}&
                +\frac{H_h^2e^{2\{L_h(\alpha_{0h}-1)\}}}{2(\alpha_h'L_h+\mathfrak{g_h})}\\
     & &         +2H_sH_he^{L_s(\alpha_{0s}-1)+L_h(\alpha_{0h}-1)}
                 \frac{\mathfrak{B}\cos\phi_0+\mathfrak{L}\sin\phi_0} {\mathfrak{B}^2 + \mathfrak{L}^2}.\nonumber 
\end{eqnarray}
% 
% \begin{equation}
% \label{eq:cs_hs}
%      \sigma=\frac{H_s^2e^{2\{L_s(\alpha_{0s}-1)\}}}{2(\alpha_s'L_s+\mathfrak{g_s})}
%                 +\frac{H_h^2e^{2\{L_h(\alpha_{0h}-1)\}}}{2(\alpha_h'L_h+\mathfrak{g_h})}
%                 +2H_sH_he^{L_s(\alpha_{0s}-1)+L_h(\alpha_{0h}-1)}
%                  \frac{\mathfrak{B}\cos\phi_0+\mathfrak{L}\sin\phi_0} {\mathfrak{B}^2 + \mathfrak{L}^2}.
%   \end{equation}
  with the notation:
 \begin{equation}
\label{eq:H-factors}
    H_s=\frac{A_s}{{\Bigl(1+\frac{\widetilde{Q^2}}{{Q_s^2}}\Bigr)}^{n_s}},\;
    H_h=\frac{A_h\Bigl({\widetilde{Q^2}\over{Q_h^2}}\Bigr)}{{\Bigl(1+\frac{\widetilde{Q^2}}{{Q_h^2}}\Bigr)}^{n_h+1}},
 \end{equation}
  where we use $A_{s,h}=\frac{\sqrt{\pi}}{s_0}\widetilde{A_{\phantom{a}}}_{\!\!\!\!s,h}\\$
\begin{math}
  \begin{array}{l l r}
    L_s=\ln\Big(\frac{s}{s_{0s}}\Big),
    &\mathfrak{g_s}=2\Bigl(\frac{a_s}{\widetilde{Q^2}}+\frac{b_s}{2m_p^2}\Bigr),
    &\alpha_s(t)=\alpha_{0s}+\alpha_s't,
    \\L_h=\ln\Big(\frac{s}{s_{0h}}\Big),
    &\mathfrak{g_h}=2\Bigl(\frac{a_h}{\widetilde{Q^2}}+\frac{b_h}{2m_p^2}\Bigr),
    &\alpha_h(t)=\alpha_{0h}+\alpha_h't,\\
  \end{array}
\end{math}

\begin{math}
  \begin{array}{l}
    \mathfrak{B}=L_s\alpha_s' + L_h\alpha_h'+(\mathfrak{g_s}+\mathfrak{g_h}),
    \\\mathfrak{L}=\frac{\pi}{2}(\alpha_s'-\alpha_h'),
    \\\phi_0=\frac{\pi}{2}(\alpha_{0s}-\alpha_{0h}).
  \end{array}
\end{math}

The parameters of the linear Pomeron trajectories are fixed according to Refs.~\cite{L, DL}:
$$
%\begin{math}
%  \begin{array}{l}
    \alpha_s(t) = 1.08 + 0.25t,\;%\\
    \alpha_h(t) = 1.44 + 0.01t.%\\
%  \end{array}
%\end{math}
$$

\section{A model with $Q^2-$dependent Pomeron trajectory}\label{app:Alternative}
An effective way to account for the ``hardening" of dynamics in the one-component model analyzed in this article is the an introduction of $\widetilde Q^2$-dependent parameters of the Pomeron trajectory. Below we present an example of such a treatment. Although it is not a ``solution" of the problem, but it may provide hints for the expected trends in the $\widetilde Q^2-$dependence of the amplitude.

Limiting ourselves to the DVCS case, consider the amplitude presented in Eq.~(\ref{Amplitude1}):
\begin{equation}
A(Q^2, W, t, M_V^2)=\frac{A_0}{(1+\widetilde Q^2/Q_0^2)^n}e^{-\frac{i\pi}{2}\alpha(t)}(s/s_0)^{\alpha(t)}e^{-2\bigl(
\frac{a}{\widetilde Q^2}+\frac{b}{2m_p^2}\bigr)|t|}.
\end{equation}

Here we fix $Q^2_0=1.0$ GeV$^2$ and $s_0=1.0$ GeV$^2$, so that $A_0$, $n$, $a$  and $b$ remain the only free parameters.

We use a linear Pomeron trajectory, $\alpha(t,Q^2)=\alpha_0(\widetilde Q^2)+\alpha'(\widetilde Q^2)t,$ where the $\widetilde Q^2-$ dependence of its parameters
mimics the ``hardening" of the reactions. 
%In a subsequent paper \cite{FFJS} this ``effective" way of the ``hardening" will be lifted with the introduction of thetwo-term, Pomeron.

We define the intercept of the Pomeron trajectory as 
\begin{equation}
\alpha_0(\widetilde Q^2)=\frac{1}{\ln\bigl(d+\frac{1}{f+\widetilde Q^2})},
\end{equation}
with $d=2.16$ and $f=2.744$ providing for the ``soft" cross section limit 
$$\alpha_0(\widetilde Q^2=0)|_{DVCS}=1/\ln({2.16+1/2.744})=1.08,$$
and the ``hard" cross section limit
$$\alpha_0(\widetilde Q^2\rightarrow \infty)|_{DVCS}=1/\ln(2.16)=1.3.$$

Similarly, we introduce the Ê``soft" and ``hard" limits in the slope of the Pomeron trajectory as
\begin{equation}
\alpha'(\widetilde Q^2)=\ln(1+1/(c+\widetilde Q^2)),
\end{equation}
with $c=8.17$, such that 
$$\alpha'(\widetilde Q^2=0)|_{DVCS}=\ln(1+1/8.17)=1.12,$$ 
and 
$$\alpha(\widetilde Q^2\rightarrow \infty)|_{DVCS}=\ln(1)=0.$$

Using the above formulas for the cross sections and the forward slope (see Eqs.~(\ref{eq:cs}),~(\ref{eq:dcsdt}) and (\ref{eq:Bslope})), with the $\widetilde Q^2-$dependent parameters of Pomeron trajectory, we have fitted this model to the HERA data.

From the fitting of the DVCS forward slope we obtained :
$a=0.27$, $b=1.98$, ${\chi^2}/{d.o.f.}=0.11$.
At this point all parameters: $a, b,  c,  d,  f$ are fixed except for the normalization $A_0$ and the exponent $n$.

Fitting this model to the DVCS elastic cross section, we find:\\
$A_0=16.2,\; n=1.43$ (see also \cite{EDS,Canarias}).

%#############################################################################################

\vfill \eject
\end{document}